\documentclass[british,english,aps,a4paper,showpacs,preprint]{revtex4}
\usepackage[T1]{fontenc}
\usepackage[latin1]{inputenc}
\usepackage{amsmath}
\usepackage{graphicx}
\usepackage{amssymb}

\makeatletter

\providecommand{\tabularnewline}{\\}

\makeatletter

\makeatletter

\makeatletter

\makeatletter

\makeatletter

\makeatletter

\usepackage{geometry}

\makeatletter

\usepackage{epsf}

\usepackage{epsfig}

\usepackage{fixmath}

\usepackage{mathrsfs}

\makeatother
\newcommand{\cut}[1]{{\!\!}}

\makeatother

\makeatother

\makeatother

\makeatother

\makeatother

\makeatother

\usepackage{babel}
\makeatother
\begin{document}
%\maketitle

\author{Juan P. Neirotti and David Saad}

\affiliation{The Neural Computing Research Group, Aston University, Birmingham
B4 7ET, UK.}

\begin{abstract}
An efficient Bayesian inference method for problems that can be mapped
onto dense graphs is presented. The approach is based on message passing
where messages are averaged over a large number of replicated variable
systems exposed to the same evidential nodes. An assumption about
the symmetry of the solutions is required for carrying out the averages;
here we extend the previous derivation based on a replica symmetric
(RS) like structure to include a more complex one-step replica symmetry
breaking (1RSB)-like ansatz. To demonstrate the potential of the approach
it is employed for studying critical properties of the Ising linear
perceptron and for multiuser detection in Code Division Multiple Access
(CDMA) under different noise models. Results obtained under the RS
assumption in the non-critical regime give rise to a highly efficient
signal detection algorithm in the context of CDMA; while in the critical
regime one observes a first order transition line that ends in a continuous
phase transition point. Finite size effects are also observed. While
the 1RSB ansatz is not required for the original problems, it was
applied to the CDMA signal detection problem with a more complex noise
model that exhibits RSB behaviour, resulting in an improvement in
performance. 
\end{abstract}


\title{Inference by replication in densely connected systems}

\maketitle

\section{Introduction}

Efficient inference in large complex systems is a major challenge
with significant implications in science, engineering and computing.
Exact inference is computationally hard in complex systems and a range
of approximation methods have been devised over the years, many of
which have been originated in the physics literature~\cite{MPV}.
A recent review~\cite{MFAbook} highlights the links between the
various approximation methods and their applications.

Approximative Bayesian inference techniques arguably offer the most
principled approach to information extraction, by combining a rigorous
statistical approach with a feasible but systematic approximation.
Although message passing techniques have existed for some time in
the computer science community~\cite{Pearl,Jensen} they have enjoyed
growing popularity in recent years~\cite{macKay}, mainly within
the context of Bayesian networks and the use of Belief Propagation
(BP) for a range of inference applications, from signal extraction
in telecommunication to machine learning.

The main advantage of these techniques is their moderate growth in
computational cost, with respect to the systems size, due to the local
nature of the calculation when applied to sparse graphs. Until recently,
message passing techniques were deemed unsuitable for inference in
densely connected systems due to the inherently high number of short
loops in the corresponding graphical representation, and the large
number of connections per node, which results in a high computational
cost. Both properties are considered prohibitive to the use of conventional
message passing techniques in such problems.

A recently suggested method for message passing in densely connected
systems~\cite{KabashimaCDMA} relies on replacing individual messages
by averages sampled from a Gaussian distribution of some mean and
variance that are modified iteratively. The method has been applied
for the CDMA signal detection inference problem; it successfully finds
optimal solutions where the space of solutions is contiguous but breaks
down when the solution space becomes fragmented, for instance, when
there is a mismatch between the true and assumed noise levels in the
CDMA detection problem. The emergence of competing solutions gives
rise to conflicting messages that result in bungled average messages
and suboptimal performance. In statistical physics terms, it corresponds
to the replica symmetric solution in dense systems~\cite{Nishimoribook}
and gives poor estimates when more complex solution structures are
required.

In the current paper, we methodologically extend the approach of Kabashima~\cite{KabashimaCDMA}
for inference in dense graphs by considering a large (infinite) number
of replicated variable systems, exposed to the same evidential data
(received signals). Each one of the systems represents a pure state
and a possible solution. The pseudo posteriors, that form the basis
for our estimates, are based on averages over the replicated systems.
The method has been employed previously only in the non-critical regime~\cite{neirottisaad},
using the most basic (RS-like) ansatz for the solution structure.
In the current paper we study both critical and non-critical regimes
and extend the solution structure considered to include step replica
symmetry breaking (1RSB) like structures~\cite{footnote}. To demonstrate
the potential of this approach and the performance obtained using
the resulting algorithm we apply the method to two different but related
problems: signal detection in Code Division Multiple Access (CDMA)
and learning in the Ising linear perceptron (ILP). \cut{ to demonstrate
its performance and relevance to general inference tasks.}

We investigate both RS and 1RSB-like structures. The former is applied
to both CDMA and ILP problems and seems to be sufficient for obtaining
optimal performances; the latter is applied to a variant of the CDMA
signal detection problem with a more complex noise model that exhibits
RSB-like behaviour, to demonstrate its efficacy for particularly difficult
inference tasks.

In section~\ref{sec:models} we will introduce the general models
studied, followed by a brief review of message passing techniques
for dense systems in section~\ref{sec:message_passing}. The general
derivation of our approach, for both RS and RSB-like solution structures,
will be presented in section~\ref{sec:General-Formalism}; numerical
studies of both CDMA signal detection and ILP learning will be reported
in section~\ref{sec:CDMA}. To demonstrate the method based on the
more complex 1RSB solution structure, and to examine its efficacy
to problems that require such structures, we will introduce a variant
of the CDMA signal detection problem and study it numerically in section~\ref{sec:CDMA2Gauss}.
We will conclude the presentation with a summary and point to future
research directions. Details of the derivation will be provided in
Appendices~\ref{app:RS}-\ref{app:optimisation}.

\section{Models studied}

\label{sec:models}

Before describing the inference method, the approach taken and the
algorithms derived from it, it would be helpful to briefly describe
the exemplar inference problems tackled in this paper.

We apply the method to two different but related inference problems:
signal detection in CDMA and learning in the Ising linear perceptron
(ILP). Both correspond to inference problems where data points are
noisy representations of sums of binary variables modulated by random
binary values.

\begin{figure}
\begin{centering}\begin{picture}(440,190) \put(0,-4){\epsfxsize=98.5mm
\epsfbox{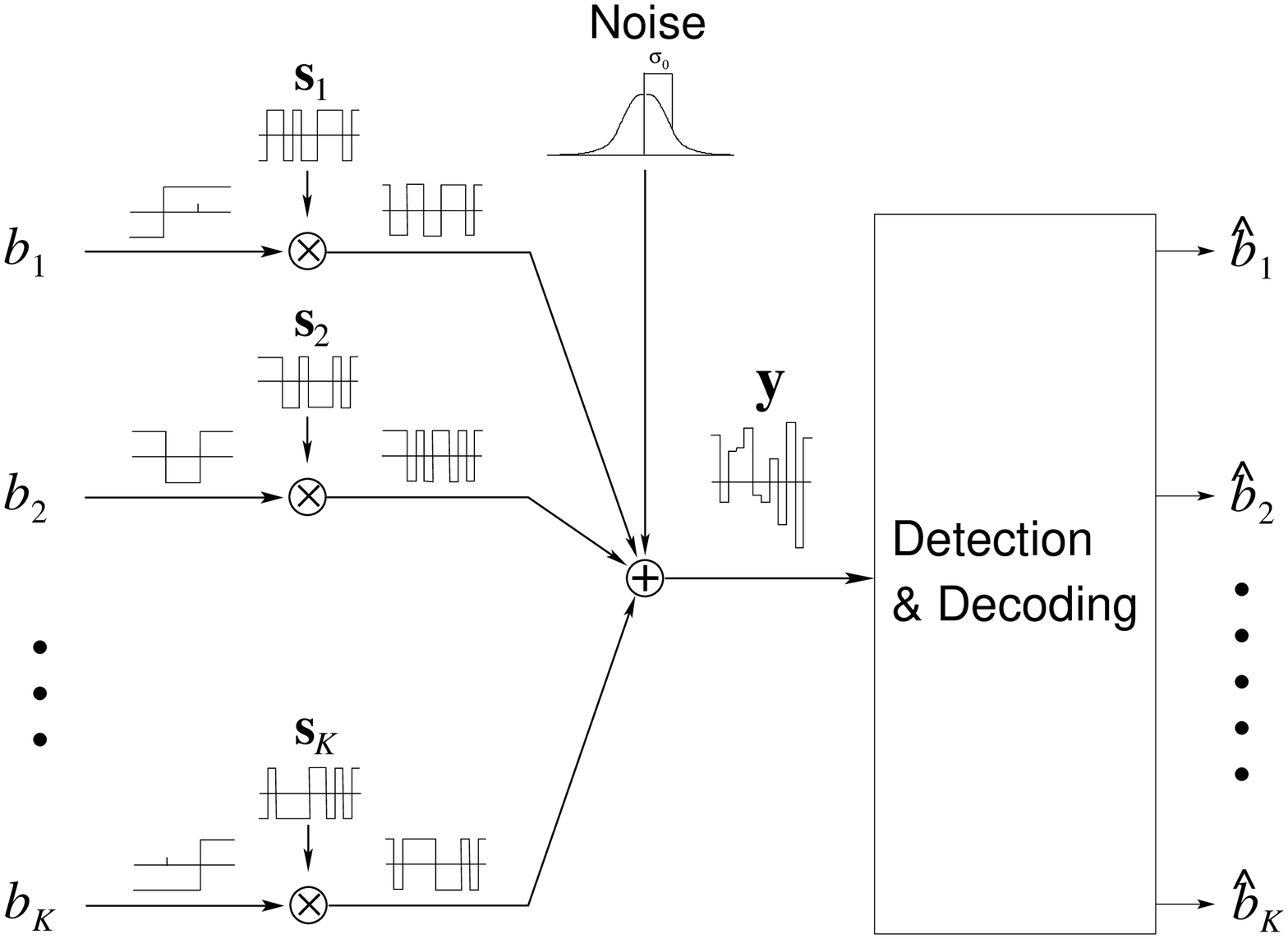}} \put(325,25){\epsfxsize=50.5mm \epsfbox{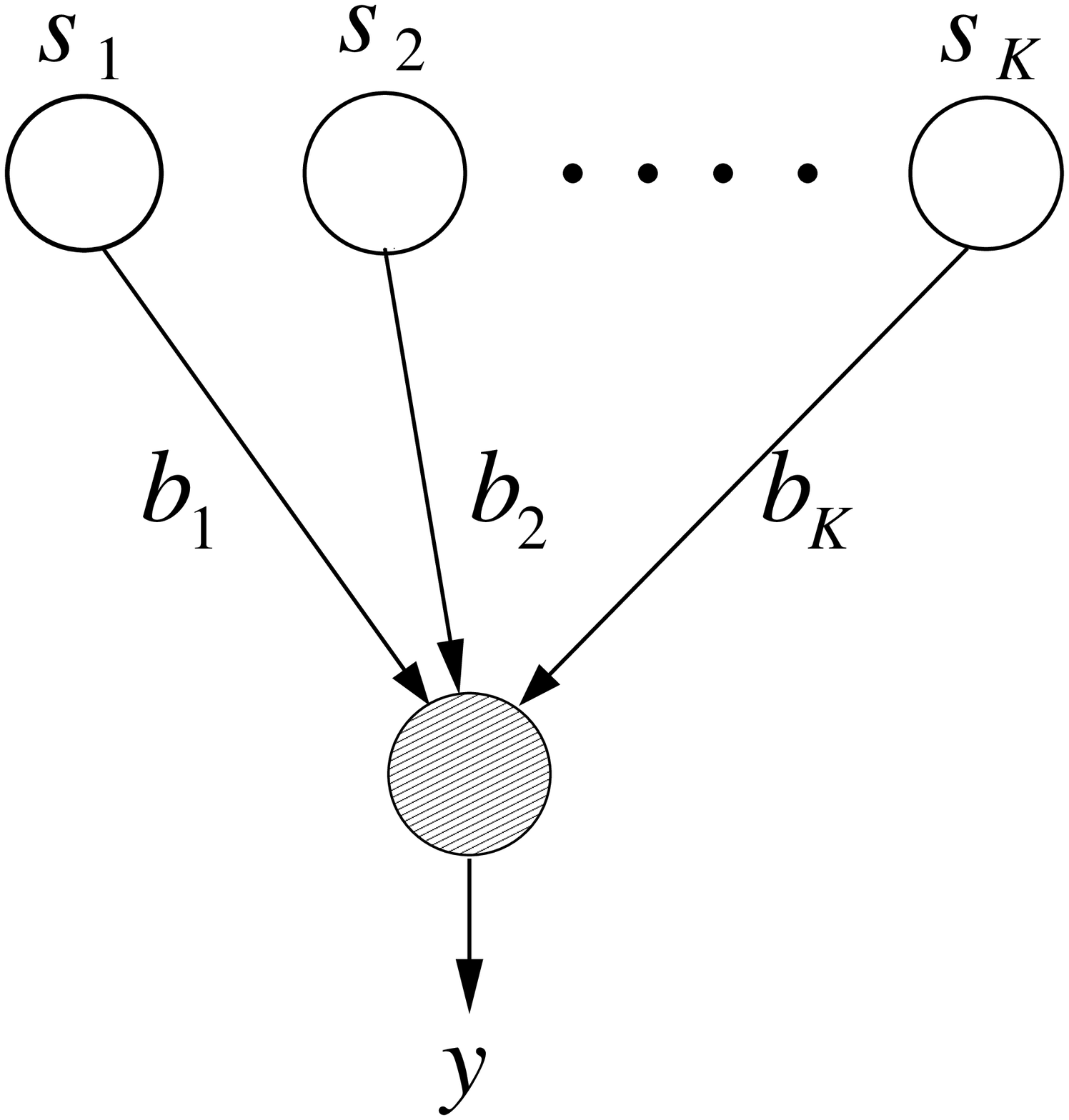}}
\put(0,203){\large (a)} \put(305,203){\large (b)} \end{picture}\par\end{centering}

\caption{Schematic representation of (a) the CDMA system. (b) the ILP.}

\label{cdma-ilp} 
\end{figure}

Multiple access communication refers to the transmission of multiple
messages to a single receiver. The scenario we study here, described
schematically in figure~\ref{cdma-ilp}(a), is that of $K$ users
transmitting independent messages over an additive white Gaussian
noise (AWGN) channel of zero mean and variance $\sigma_{0}^{2}$.
Various methods are in place for separating the messages, in particular
Time, Frequency and Code Division Multiple Access~\cite{CDMAbook}.
The latter, is based on spreading the signal by using $K$ individual
random binary spreading codes of spreading factor $N$. We consider
the large-system limit, in which the number of users $K$ tends to
infinity while the system load $\beta\equiv K/N$ is kept to be $\mathcal{O}(1)$.
We focus on a CDMA system using binary phase shift keying (BPSK) symbols
and will assume the power is completely controlled to unit energy.
The received aggregated, modulated and corrupted signal is of the
form: \begin{equation}
y_{\mu}=\frac{1}{\sqrt{N}}\sum_{k=1}^{K}s_{\mu k}b_{k}+\sigma_{0}n_{\mu}\label{eq:CDMA}\end{equation}
 where $b_{k}$ is the bit transmitted by user $k$, $s_{\mu k}$
is the spreading chip value, $n_{\mu}$ is the Gaussian noise variable
drawn from $\mathcal{N}\left(0,1\right)$, and $y_{\mu}$ the received
message. The task is to infer the original transmission from the set
of received messages. This process is reminiscent of the learning
task performed by a perceptron with binary weights and linear output,
which is the next example considered in this paper.

Learning in neural networks has attracted considerable theoretical
interest. In particular we focus on supervised learning from examples,
which relies on a training set consisting of examples of the target
task~\cite{Seung}. We consider a perceptron, described schematically
in figure~\ref{cdma-ilp}(b), which is a network that sums a single
layer of inputs $s_{\mu k}$ with synaptic weights $b_{k}$ and passes
the result through a transfer function $y_{\mu}$ \begin{equation}
y_{\mu}=g\left(\frac{1}{\sqrt{K}}\sum_{k=1}^{K}s_{\mu k}b_{k}\right)\,,\label{eq:perceptron}\end{equation}
 where $g$ is typically a non-linear sigmoidal function. If $g(x)=x$
the network is termed \emph{linear output perceptron}. If the weights
$b_{k}\in\left\{ \pm1\right\} $ the network is called \emph{Ising
perceptron}. Learning is a search through the weight space for the
perceptron that best approximates a target rule.

The similarity between the linear perceptron of equation~(\ref{eq:perceptron})
and the CDMA detection problem of Eq.(\ref{eq:CDMA}) allows for a
direct relation between the two problems to be established. The main
difference between the problems is the regime of interest. While CDMA
detection applications are of interest mainly for non-critical low
load values, ILP studies focused on the critical regime. We consider
both regimes in this paper. \cut{, but to unify the treatment we
will use the notation and scaling conventions of the CDMA system.}

\section{Message passing for inference in densely connected systems}

\label{sec:message_passing}

Graphical models (Bayes belief networks) provide a powerful framework
for modelling statistical dependencies between variables~\cite{Pearl,Jensen,macKay}.
They play an essential role in devising a principled probabilistic
framework for inference in a broad range of applications.

Message passing techniques are typically used for inference in graphical
models that can be represented by a sparse graph with a few (typically
long) loops. They are aimed at obtaining (pseudo) posterior estimates
for the system's variables by iteratively passing messages (locally
calculated conditional probabilities) between variables. Iterative
message passing of this type is guaranteed to converge to the globally
correct estimate when the system is tree-like; there are no such guarantees
for systems with loops even in the case of large loops and a local
tree-like structure (although message passing techniques have been
used successfully in loopy systems, supported by some limited theory~\cite{weiss}).
A clear link has been established between certain message passing
algorithms and well known methods of statistical mechanics~\cite{MFAbook}
such as the Bethe approximation~\cite{TAPEPL,YFW}.

These inherent limitations seem to prevent the use of message passing
techniques in densely connected systems due to their high connectivity,
implying an exponentially growing cost, and an exponential number
of loops. However, an exciting new approach has been recently suggested~\cite{KabashimaCDMA}
for extending BP techniques~\cite{Pearl,Jensen,macKay} to densely
connected systems. In this approach, messages are grouped together,
giving rise to a macroscopic random variable, drawn from a Gaussian
distribution of varying mean and variance for each of the nodes. The
technique has been successfully applied to CDMA signal detection problems
and the results reported are competitive with those of other state-of-the-art
techniques. However, the current approach has some inherent limitations~\cite{KabashimaCDMA},
presumably due to its similarity to the replica symmetric solution
in the equivalent Ising spin models~\cite{MPV,Nishimoribook}.

In a separate recent development~\cite{MPZ}, the replica-symmetric-equivalent
BP has been extended to Survey Propagation (SP), which corresponds
to one-step replica symmetry breaking in diluted systems. This new
algorithm, motivated by the theoretical physics interpretation of
such problems, has been highly successful in solving hard computational
problems~\cite{MPZ}, far beyond other existing approaches. In addition,
the algorithm facilitated theoretical studies of the corresponding
physical system and contributed to our understanding of it~\cite{MZPRE}.
The SP algorithm has recently been modified to handle Ising and multilayer
perceptrons~\cite{BZ}.

\cut{The approach presented here was inspired by advances made in
BP for densely connected systems and the development of SP. It adopts
the methodology of the former and builds on the main feature of the
latter, namely, the replication of variable nodes and the calculation
of messages averaged over contributions from the replicated systems.}

\section{General Formalism \label{sec:General-Formalism}}

\begin{figure}
\begin{centering}\includegraphics[width=2.55906in]{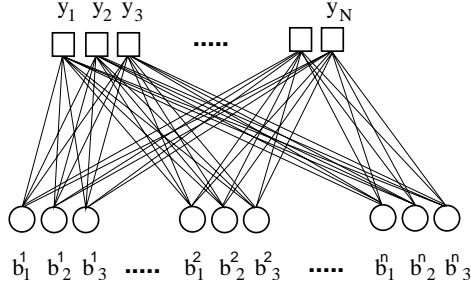}\par\end{centering}

\caption{Replicated solutions $\mathbf{B}\!=\!\left(\mathbf{b}_{1},\,\mathbf{b}_{2},..,\mathbf{b}_{K}\right)$
given data.}

\label{fig1} 
\end{figure}

We recently presented a new approach~\cite{neirottisaad} for inference
in densely connected systems, which was inspired by both the extension
of BP to densely connected graphs and the introduction of SP. The
systems we consider here are characterised by multiplicity of pure
states and a possible fragmentation of the space of solutions. To
address the inference problem in such cases we consider an ensemble
of replicated systems where averages are taken over the ensemble of
potential solutions. This amounts to the presentation of a new graph,
where the observables $y_{\mu}$ are linked to variables in all replicated
systems, namely $\mathbf{B}\!=\!\left(\mathbf{b}^{1},\mathbf{b}^{2},\dots,\mathbf{b}^{n}\right)$;
where $\mathbf{b}^{{\textrm{a}}}\!=\!\left(b_{1}^{{\textrm{a}}},b_{2}^{{\textrm{a}}},\dots,b_{K}^{{\textrm{a}}}\right)^{\textsf{T}}$,
as shown in figure \foreignlanguage{british}{\ref{fig1}}. To
estimate the variables $\mathbf{B}$ given the data $\mathbf{y^{\sf T}}\!=\!\left(y_{1},y_{2},\ldots,y_{N}\right)$,
in a Bayesian framework, we have to maximise the posterior $P\left(\mathbf{B}|\mathbf{y}\right)\!\propto\!\prod_{\mu=1}^{N}P\left(y_{\mu}|\mathbf{B}\right)P\left(\mathbf{B}\right),$
where we have considered independent data, and thus $P\left(\mathbf{y}|\mathbf{B}\right)\!=\!\prod_{\mu=1}^{N}P\left(y_{\mu}|\mathbf{B}\right)$.

The likelihood so defined is of a general form; the explicit expression
depends on the particular problem studied. Here, we are interested
in cases where $\mathbf{b}\!\in\!\left\{ \pm1\right\} ^{K}$ is an
unbiased vector and $P\left(\mathbf{B}\right)\!=\!2^{-Kn}$. The estimate
we would like to obtain is the maximiser of the posterior marginal
(MPM) $\widehat{\mathbf{b}}_{k}\!=\!\mathop\mathrm{argmax}_{\mathbf{b}_{k}\in\left\{ \pm1\right\} ^{n}}\sum_{\left\{ \mathbf{b}_{l\neq k}\right\} }P\left(\mathbf{B}|\mathbf{y}\right)\,\,,$
which is expected to be a vector with equal entries for all replica
$\widehat{b}_{k}^{1}=\widehat{b}_{k}^{2}=\dots=\widehat{b}_{k}^{n}$.
The number of operations required to obtain the full MPM estimator
is of $\mathcal{O}\left(2^{K}\right)$ which is infeasible for large
$K$ values.

To obtain an approximate MPM estimate we apply BP message passing
technique~\cite{Pearl,Jensen,macKay}. In particular we are interested
here in the application of BP to densely connected graphs, similar
to the one presented in~\cite{KabashimaCDMA}. The latter is based
on estimating a single solution and therefore does not converge, as
has been observed, when the solution space becomes fragmented and
multiple solutions emerge. This arguably corresponds to the replica
symmetry breaking phenomena and occurs, for instance, when the noise
level is unknown in the CDMA signal detection case.

A potential algorithmic improvement is achieved by the introduction
of an SP-like approach, based on replicated variable systems, similar
to the approach taken in problems that can be mapped onto sparsely
connected graphs. %

Using Bayes rule one straightforwardly obtains the BP equations: \begin{eqnarray}
P^{t+1}\left(y_{\mu}|\mathbf{b}_{k},\left\{ y_{\nu\neq\mu}\right\} \right) & = & \sum_{\left\{ \mathbf{b}_{l\neq k}\right\} }P\left(y_{\mu}|\mathbf{B}\right)\prod_{l\neq k}P^{t}\left(\mathbf{b}_{l}|\left\{ y_{\nu\neq\mu}\right\} \right)\label{eq:bp1}\\
P^{t}\left(\mathbf{b}_{l}|\left\{ y_{\nu\neq\mu}\right\} \right) & \propto & \prod_{\nu\neq\mu}P^{t}\left(y_{\nu}|\mathbf{b}_{l},\left\{ y_{\sigma\neq\nu}\right\} \right)\,.\label{eq:bp2}\end{eqnarray}
 For calculating the posterior $P\left(\mathbf{y}|\mathbf{B}\right),$
we assume a dependency of the data on the parameters of the form $y_{\mu}=\mathcal{F}\left(\sum_{l=1}^{K}\varepsilon_{\mu l}\mathbf{b}_{l};\mathbold{\gamma}\right)$,
where $\mathcal{F}$ is some general smooth function, $\mathbold{\gamma}$
are model parameters and $\varepsilon_{\mu l}$ are small enough to
ensure that $\sum_{l=1}^{K}\varepsilon_{\mu l}b_{l}^{{\rm a}}\sim\mathcal{O}(1)$.
We define the vector $\mathbf{{\Delta}}_{\mu}\equiv\sum_{l=1}^{K}\varepsilon_{\mu l}\mathbf{b}_{l}=\sum_{l\neq k}\varepsilon_{\mu l}\mathbf{b}_{l}+\varepsilon_{\mu k}\mathbf{b}_{k}=\mathbf{{\Delta}}_{\mu k}+\varepsilon_{\mu k}\mathbf{b}_{k}.$
Thus, using $y_{\mu}=\mathcal{F}\left(\mathbf{{\Delta}}_{\mu k}+\varepsilon_{\mu k}\mathbf{b}_{k};\mathbold{\gamma}\right)$
we can model the likelihood such that \begin{eqnarray}
P\left(y_{\mu}|\mathbf{B}\right) & = & \int{\rm d}\mathbf{{\Delta}}_{\mu k}P\left(y_{\mu},\mathbf{{\Delta}}_{\mu k}|\mathbf{B};\mathbold{\gamma}\right)\nonumber \\
 & = & \int{\rm d}\mathbf{{\Delta}}_{\mu k}P\left(y_{\mu}|\mathbf{{\Delta}}_{\mu k},\mathbf{B};\mathbold{\gamma}\right)P\left(\mathbf{{\Delta}}_{\mu k}|\mathbf{B}\right)\nonumber \\
 & = & \int{\rm d}\mathbf{{\Delta}}_{\mu k}P\left(y_{\mu}|\mathbf{{\Delta}}_{\mu k}+\varepsilon_{\mu k}\mathbf{b}_{k};\mathbold{\gamma}\right)P\left(\mathbf{{\Delta}}_{\mu k}|\mathbf{B}\right)\nonumber \\
 & \simeq & \int{\rm d}\mathbf{{\Delta}}_{\mu k}\left[1+\varepsilon_{\mu k}\mathbf{b}_{k}^{\sf T}\nabla_{\mathbf{{\Delta}}_{\mu k}}\ln P\left(y_{\mu}|\mathbf{{\Delta}}_{\mu k};\mathbold{\gamma}\right)\right]P\left(y_{\mu}|\mathbf{{\Delta}}_{\mu k};\mathbold{\gamma}\right)\, P\left(\mathbf{{\Delta}}_{\mu k}|\mathbf{B}\right)\,,\label{eq:likelihood}\end{eqnarray}
 where we have assumed that $P\left(y_{\mu}|\mathbf{{\Delta}}_{\mu k},\mathbf{B};\mathbold{\gamma}\right)\approx P\left(y_{\mu}|\mathbf{{\Delta}}_{\mu k}+\varepsilon_{\mu k}\mathbf{b}_{k};\mathbold{\gamma}\right)$,

due to the assumed dependence of the observed values $y_{\mu}$ on
$\mathbf{\Delta}_{\mu k}$ and $\mathbf{b}_{k}$.

\subsection{Inter-replica correlations}

An explicit expression for inter-dependence between solutions is required
for obtaining a closed set of update equations. We assume a dependence
of the form \begin{equation}
P^{t}\left(\mathbf{b}_{k}\left|\left\{ y_{\nu\neq\mu}\right\} \right.\right)\propto\exp\left\{ \mathbf{h}_{\mu k}^{t\mathsf{T}}\,\mathbf{b}_{k}+\frac{1}{2}\mathbf{b}_{k}^{\textsf{T}}\mathbf{Q}_{\mu k}^{t}\,\mathbf{b}_{k}\right\} ,\label{eq:ansatz}\end{equation}
 where $\mathbf{h}_{\mu k}^{t}$ is a vector representing an external
field and $\mathbf{Q}_{\mu k}^{t}$ the matrix of cross-replica interaction.
The form of $\mathbf{Q}_{\mu k}^{t}$ depends upon the particular
case considered. We assume one of the following symmetry relation
between the replicated solutions: \begin{eqnarray*}
\left(\mathbf{h}_{\mu k}^{t}\right)^{\ell{\rm a}} & = & h_{\mu k}^{t},~\mbox{~~~~and~~}\\
\left(^{{\rm (RS)}}\mathbf{Q}_{\mu k}^{t}\right)^{\mathrm{a}{\rm a}^{\prime}} & = & \delta^{\mathrm{a}{\rm a}^{\prime}}\, q_{0\mu k}^{t}+\left(1-\delta^{{\rm a}{\rm a}^{\prime}}\right)\, q_{1\mu k}^{t}~\mbox{~~or~~}\\
\left(^{{\rm (1RSB)}}\mathbf{Q}_{\mu k}^{t}\right)^{\ell{\rm a}\;\ell^{\prime}{\rm a}^{\prime}} & = & \delta^{\ell\ell^{\prime}}\left(^{{\rm (RS)}}\mathbf{Q}_{\mu k}^{t}\right)^{\mathrm{a}{\rm a}^{\prime}}+\left(1-\delta^{\ell\ell^{\prime}}\right)q_{2\mu k}^{t}\,,\end{eqnarray*}
 where $\ell$ is a block index that runs from 1 to $L$ and `a' is
a intra-block replica index that runs form 1 to $n$ where $n$ is
the number of variables per block. We also make the following reasonable
assumption $q_{0\mu k}^{t}>q_{1\mu k}^{t}>q_{2\mu k}^{t}>0$, as one
expects correlations to gradually decrease between variables with
non-identical replica and block indices, respectively.

For both types of symmetries considered, the correlation matrix defined
as: \[
\left(\mathbf{\Upsilon}_{\mu k}^{t}\right)^{\sf\mathbf{I}\;\sf\mathbf{I}^{\prime}}\equiv\left\langle \Delta_{\mu k}^{\sf\mathbf{I}}\Delta_{\mu k}^{\sf\mathbf{I}^{\prime}}\right\rangle -\left\langle \Delta_{\mu k}^{\sf\mathbf{I}^{\phantom{\prime}}}\right\rangle \left\langle \Delta_{\mu k}^{\sf\mathbf{I}^{\prime}}\right\rangle \,\]
 where $\sf\mathbf{I}$ is an index or a pair of indices for RS and
1RSB, respectively. The correlation matrix is assumed to be self-averaging,
i.e. $\mathbf{\Upsilon}_{\mu k}^{t}\simeq\mathbf{\Upsilon}^{t}$ and
preserves the symmetry of the matrix $\mathbf{Q}_{\mu k}^{t}$. An
explicit derivation of the entries of $\mathbf{\Upsilon}^{t}$ is
presented in Appendices~\ref{app:RS} and~\ref{app:RSB}, for the
RS and RSB-like correlation structures, respectively; the matrices
take following the general form: \begin{eqnarray*}
\left(^{{\rm (RS)}}\mathbf{\Upsilon}^{t}\right)^{{\rm a}{\rm a}^{\prime}} & = & \delta^{{\rm a}{\rm a}^{\prime}}X^{t}+\left(1-\delta^{{\rm a}{\rm a}^{\prime}}\right)\frac{1}{n}R^{t}\\
\left(^{{\rm (1RSB)}}\mathbf{\Upsilon}^{t}\right)^{{\rm a}\ell\;{\rm a}^{\prime}\ell^{\prime}} & = & \delta^{\ell\ell^{\prime}}\left[\delta^{{\rm a}{\rm a}^{\prime}}X^{t}+\left(1-\delta^{{\rm a}{\rm a}^{\prime}}\right)\frac{1}{n}V^{t}\right]+\left(1-\delta^{\ell\ell^{\prime}}\right)\frac{1}{nL}\left(V^{t}-R^{t}\right)\,.\end{eqnarray*}
 Thus, for the appropriate centre of the distribution $\mathbf{u}_{\mu k}^{t}$
(see equations~ (\ref{eq:urs}) and (\ref{eq:u1rsb})), the probability
of $\mathbf{{\Delta}}_{\mu k}$ can be expressed as: {\footnotesize \begin{eqnarray}
P\left(\mathbf{{\Delta}}_{\mu k}|\mathbf{B}\right) & = & \sqrt{\frac{1}{\left(2\pi\right)^{n}\det\left(\mathbf{\Upsilon}^{t}\right)}}\exp\left\{ -\frac{1}{2}\left(\mathbf{{\Delta}}_{\mu k}-\mathbf{u}_{\mu k}^{t}\right)^{\mathsf{T}}\left(\mathbf{\Upsilon}^{t}\right)^{-1}\left(\mathbf{{\Delta}}_{\mu k}-\mathbf{u}_{\mu k}^{t}\right)\right\} \\
 & \propto & {\small\begin{cases}
\int{\rm d}\vartheta\,\exp\left\{ -n\,{\displaystyle \frac{\left(\vartheta-u_{\mu k}^{t}\right)^{2}}{2R^{t}}}\right\} \prod_{{\rm a}=1}^{n}\exp\left\{ -{\displaystyle \frac{\left(\Delta_{\mu k}^{{\rm a}}-\vartheta\right)^{2}}{2X^{t}}}\right\}  & \mbox{(RS)}\label{eq:pdelrs}\\
\int{\rm d}\mathbf{{\Theta}}\,\prod_{\ell=1}^{L}\exp\left\{ -{\displaystyle
\frac{n}{2}\left[\frac{\left(\vartheta^{0}\right)^{2}}{V^{t}-R^{t}}+\frac{\left(\vartheta^{\ell}\right)^{2}}{V^{t}-L^{-1}\left(V^{t}-R^{t}\right)}\right]}\right\}
\prod_{{\rm a}=1}^{n}\exp\left\{ -{\displaystyle \frac{\left(\Delta_{\mu
k}^{\ell{\rm a}}-\vartheta_{\mu k}^{0\ell
t}\right)^{2}}{2\left(X^{t}-n^{-1}V^{t}\right)}}\right\}  &
\mbox{(RSB)}\end{cases}}\nonumber \end{eqnarray}}
 for the RS and RSB-like correlation matrices, respectively, where
$\vartheta_{\mu k}^{0\ell t}\equiv\vartheta^{0}+\vartheta^{\ell}+u_{\mu k}^{t}$
and $\mathbf{{\Theta}}^{\sf T}=\left(\vartheta^{0},\vartheta^{1},\dots,\vartheta^{L}\right).$

\subsection{Messages}

Having obtained the conditional probability distribution $P\left(\mathbf{{\Delta}}_{\mu k}|\mathbf{B}\right)$
one can then derive explicit expressions for the messages $m_{\mu k}$
(magnetisation) and $\widehat{m}_{\mu k}$ that can be viewed as parameters
in the corresponding marginalised binary distributions $P^{t}\left(y_{\mu}|b_{k},\{ y_{\nu\ne\mu}\}\right)\propto(1+\hat{m}_{\mu k}^{t}b_{k})/2$
and $P^{t}\left(b_{k}|\{ y_{\nu\ne\mu}\}\right)=(1+m_{\mu k}^{t}b_{k})/2$.

The messages from nodes $y_{\mu}$ to nodes $\mathbf{b}_{k}$, as
derived in Appendix~\ref{app:messages}, equations~(\ref{eq:mhat})-(\ref{eq:mm1rsb})
\begin{eqnarray}
\widehat{m}_{\mu k}^{t+1} & = & \begin{cases}
\varepsilon_{\mu k}{\displaystyle \frac{\tilde{\vartheta}_{\mu k}^{t}-u_{\mu k}^{t}}{R^{t}}} & \mbox{(RS)}\\
\varepsilon_{\mu k}{\displaystyle \frac{\tilde{\vartheta}_{\mu k}^{t}-u_{\mu k}^{t}}{2V^{t}-R^{t}}+\frac{\varepsilon_{\mu k}}{2n}\,\frac{\mathcal{P}_{2}V^{t}}{1-\mathcal{P}_{1}V^{t}}\,} & \mbox{(RSB)}\end{cases},\label{eq:m_hat}\end{eqnarray}
 where $\mathcal{P}_{j}=\left.{\displaystyle \frac{\partial^{j}\mathcal{P}}{\partial\vartheta^{j}}}\right|_{\vartheta=\tilde{\vartheta}_{\mu k}^{t}}$,
$\mathcal{P}$ is defined in equation~(\ref{eq:gcal}) and $\tilde{\vartheta}_{\mu k}^{t}$
is obtained from the saddle point equations given by equation~(\ref{eq:wrs})
in the RS case and by equation~(\ref{eq:w1rsb}) in the 1RSB case.
The messages from nodes $\mathbf{b}_{k}$ to $y_{\mu}$ are given
in both cases by the expression $m_{\mu k}^{t}\simeq\tanh\left(\sum_{\nu\neq\mu}\widehat{m}_{\nu k}^{t}\right).$

For the gauged field $b_{k}h_{\mu k}^{t}$ where $h_{\mu k}^{t}\equiv{\rm artanh}\left(m_{\mu k}^{t}\right)=\sum_{\nu\neq\mu}{\rm artanh}\left(\hat{m}_{\nu k}^{t}\right)\simeq\sum_{\nu\neq\mu}\hat{m}_{\nu k}^{t}$.
The distribution of this field is well approximated by a Gaussian
as a result of the central limit theorem. The mean and variance of
the Gaussian are $E^{t}$ and $F^{t}$ respectively: \begin{eqnarray}
E^{t} & = & \frac{1}{K}\sum_{k=1}^{K}\sum_{\mu=1}^{N}b_{k}\hat{m}_{\mu k}^{t}\label{eq:EF}\\
F^{t} & = & \sum_{\mu=1}^{N}\left[\frac{1}{K}\sum_{k=1}^{K}\left(b_{k}\hat{m}_{\mu k}^{t}\right)^{2}-\left(\frac{1}{K}\sum_{k=1}^{K}b_{k}\hat{m}_{\mu k}^{t}\right)^{2}\right]\simeq\frac{1}{K}\sum_{k=1}^{K}\sum_{\mu=1}^{N}\left(\hat{m}_{\mu k}^{t}\right)^{2}\,.\nonumber \end{eqnarray}
 Both $E^{t}$ and $F^{t}$ are assumed to be independent of the index
$\mu$ by virtue of the self-averaging property. For the same reason
we expect the macroscopic variables defined as $M_{\mu}^{t}\equiv\sum_{k=1}^{K}b_{k}m_{\mu k}^{t}/K\simeq\sum_{k=1}^{K}b_{k}m_{k}^{t}/K=M^{t}$
and $N_{\mu}^{t}\equiv\sum_{k=1}^{K}\left(m_{\mu k}^{t}\right)^{2}/K\simeq\sum_{k=1}^{K}\left(m_{k}^{t}\right)^{2}/K=N^{t}$,
where $m_{k}^{t}\simeq\tanh\left(\sum_{\nu=1}^{N}\widehat{m}_{\nu k}^{t}\right)$,
to be independent of the index $\mu.$ Thus, these macroscopic variables
can be evaluated by the following integrals \[
M^{t}=\int\mathcal{D}u\,\tanh\left(\sqrt{F^{t}}u+E^{t}\right)\quad N^{t}=\int\mathcal{D}u\,\tanh^{2}\left(\sqrt{F^{t}}u+E^{t}\right)\,,\]
 where $\mathcal{D}u=\exp\left(-u^{2}/2\right)/\sqrt{2\pi}$.

\subsection{Optimisation}

The structure of the correlation matrix used introduces free variables
in the form of the correlation terms between replicated solutions.
These are used for optimising the estimation provided with respect
to a given performance measure.

Since the MPM estimator is given by $\hat{b}_{k}^{t}={\rm sgn}\left(m_{k}^{t}\right)\simeq{\rm sgn}\left(m_{\mu k}^{t}\right)={\rm sgn}\left(h_{\mu k}^{t}\right)$,
the expression for the error per bit rate takes the form: \begin{equation}
P_{b}^{t}=\frac{1}{2K}\sum_{k=1}^{K}\left(1-{\rm sgn}\left(b_{k}m_{k}^{t}\right)\right)\,,\label{eq:pdet}\end{equation}
 which is minimised when the true message vector \textbf{$\mathbf{b}$}
and the vector of messages $\mathbf{m}^{t}$ are parallel. Therefore,
the error rate per bit decreases as the ratio $M^{t}/\sqrt{N^{t}}=\cos\left(\widehat{\mathbf{b}\,\mathbf{m}^{t}}\right)$
increases. The optimal value is reached when $E^{t}\left(\mathbold\gamma^{c}\right)=F^{t}\left(\mathbold\gamma^{c}\right)$
and $\left.{\displaystyle \frac{\partial E^{t}}{\partial\gamma_{i}}-\frac{1}{2}\,\frac{E^{t}}{F^{t}}\,\frac{\partial F^{t}}{\partial\gamma_{i}}}\right|_{\gamma_{i}^{c}}=0$
as derived in Appendix~\ref{app:optimisation}.

\section{CDMA and linear Ising perceptron\label{sec:CDMA}}

Using this notation one defines $\varepsilon_{\mu k}=s_{\mu k}/\sqrt{N}$
for the CDMA problem and $\varepsilon_{\mu k}=s_{\mu k}/\sqrt{K}$
for the Ising perceptron. The goal is to get an accurate estimate
of the vector $\mathbf{b}$ for all users given the received message
vector $\mathbf{y}$ via a principled approximation of the posterior
$P(\mathbf{b}|\mathbf{y})$. An expression representing the likelihood
is required and is easily derived from the noise model (assuming zero
mean and variance $\sigma^{2}$). If the arithmetic variance over
replicas of the macroscopic message $\Delta_{\mu k}^{{\rm a}}$ is
finite and independent of the sub indexes $\mu$ and $k$, i.e. $\Sigma^{2}\equiv\frac{1}{n}\sum_{{\rm a}}\left(\Delta_{\mu k}^{{\rm a}}\right)^{2}-\left(\frac{1}{n}\sum_{{\rm a}}\Delta_{\mu k}^{{\rm a}}\right)^{2}<\infty\;\forall\,\mu k$,
then $P\left(y_{\mu}|\mathbf{B}\right)$ can be expanded as \begin{eqnarray}
P\left(y_{\mu}|\mathbf{B}\right) & \simeq & \sqrt{\frac{n}{2\pi\sigma^{2}}}{\rm e}^{\frac{\Sigma^{2}}{2\sigma^{2}}}\exp\left\{ -\frac{\left(\mathbf{y}_{\mu}-\mathbf{\Delta}_{\mu k}\right)^{\textsf{T}}\left(\mathbf{y}_{\mu}-\mathbf{\Delta}_{\mu k}\right)}{2\sigma^{2}}\right\} \left[1+\frac{\varepsilon_{\mu k}}{\sigma^{2}}\mathbf{b}_{k}^{\sf T}\left(\mathbf{y}_{\mu}-\mathbf{\Delta}_{\mu k}\right)\right]\,,\label{supp}\end{eqnarray}
 where $\mathbf{y}_{\mu}=y_{\mu}\mathbf{u}$ and $\mathbf{u}^{\textsf{T}}\equiv\stackrel{nL}{\overbrace{\left(1,\,1,\,\cdots,\,1\right)}}$.
The function $\mathcal{P}\left(\vartheta,y_{\mu}\right)$, defined
in equation~(\ref{eq:pcal}), and obtained from this distribution
is linear in $\vartheta$; therefore, the second derivative used for
calculating the messages in equation~(\ref{eq:m_hat}) $\mathcal{P}_{2}=0$
and the corresponding structure of the correlation matrix is RS-like.

To calculate correlations between replica we expand $P\left(y_{\mu}|\mathbf{B}\right)$
in the large \emph{N} limit in~(\ref{supp}), as shown in equation~(\ref{eq:likelihood}).
According to the RS correlation assumption, the macroscopic variables
satisfy the following relation: \begin{eqnarray*}
u_{\mu k}^{t} & = & \frac{1}{\sqrt{{\rm e}_{1}N}}\sum_{l\neq k}s_{\mu l}m_{\mu l}^{t}\\
X^{t} & \simeq & {\rm e}_{2}\left(1-N^{t}\right)\,,\end{eqnarray*}
 where ${\rm e}_{1}=1\,(\beta)$ for the CDMA (ILP) system and ${\rm e}_{2}=\beta\,(1)$
for the CDMA (ILP) systems, respectively, due to the change in scaling.
The saddle point equation~(\ref{eq:hcal1rsb}) provides a dominant
value for the variable~$\vartheta$ \begin{eqnarray*}
\tilde{\vartheta} & = & \frac{R^{t}}{\sigma^{2}+X^{t}+R^{t}}\left(\frac{\sigma^{2}u_{\mu k}^{t}}{X^{t}+R^{t}}+y_{\mu}\right)\,.\end{eqnarray*}

\subsection{Messages}

The message from $y_{\mu}$ to $b_{k}^{{\rm a}}$ at time $t+1$ is
then given by: \begin{equation}
\widehat{m}_{\mu k}^{t+1}=\varepsilon_{\mu k}\frac{y_{\mu}-u_{\mu k}^{t}}{\sigma^{2}+X^{t}+R^{t}}\,.\label{mhat1}\end{equation}

The main difference between equation~(\ref{mhat1}) and the equivalent
equation in~\cite{KabashimaCDMA} is the dependence of the pre-factor
on $R^{t}$, reflecting correlations between different solutions groups
(replica). To determine this term we optimise the choice of $\sigma^{2}$
by applying the condition $E^{t}=F^{t}$. Forcing this condition leads
to a relation between the structure of the space of solutions, represented
by $R^{t}$, and the free parameter of the model $\sigma^{2}$. From
equation~(\ref{mhat1}) and using $E^{t}=F^{t}$ and $M^{t}=N^{t}$
one obtains: \begin{eqnarray*}
E^{t+1} & = & \frac{{\rm e}_{1}^{-1}}{\sigma^{2}+X^{t}+R^{t}}\qquad F^{t+1}={\rm e}_{1}\left[\sigma_{0}^{2}+X^{t}\right]\left(E^{t+1}\right)^{2}\,,\end{eqnarray*}
 which imply, after simplification, that for both cases $R^{t}=\sigma_{0}^{2}-\sigma^{2}$.
Despite the simplicity of this result, the process from which we obtained
it provides us with a practical way to estimate the true noise variance.
Notice that for calculating $E^{t}$ and $F^{t}$ we used the limits
$K,N\to\infty\,{\rm with}\; K/N=\beta$. So that $\sigma_{0}^{2}$,
which appears in the expression for $F^{t}$, can be obtained from
the signal vector of $y_{\mu}$ with an infinite number of entries.
Thus \[
\lim_{N\to\infty}\frac{1}{N}\sum_{\mu=1}^{N}\left(y_{\mu}\right)^{2}={\rm e}_{2}+\sigma_{0}^{2}\,.\]
 Using this expression we can finally express the message as:\begin{eqnarray}
\widehat{m}_{\mu k}^{t+1} & \simeq & \varepsilon_{\mu k}\frac{y_{\mu}-u_{\mu k}^{t}}{{\displaystyle \frac{1}{N}\sum_{\mu=1}^{N}\left(y_{\mu}\right)^{2}}-{\rm e}_{2}N^{t}}\,,\label{mhatfin}\end{eqnarray}
 where no prior belief of $\sigma$ is required.

\subsection{Steady state and critical analysis}

The steady state equations for the macroscopic variables $N^{t}$
and $E^{t}$ are obtained by taken the limit $t\to\infty$. Let us
define $\overline{N}\equiv\lim_{t\to\infty}N^{t}$ and $\overline{E}\equiv\lim_{t\to\infty}E^{t}$.
In the asymptotic regime the following relations hold: \begin{eqnarray}
\overline{N}\left(\sigma_{0}^{2},\beta\right) & = & \int\mathcal{D}u\,\tanh^{2}\left(\sqrt{\overline{E}\left(\sigma_{0}^{2},\beta\right)}u+\overline{E}\left(\sigma_{0}^{2},\beta\right)\right)\label{eq:ebar}\\
\overline{E}\left(\sigma_{0}^{2},\beta\right) & = & \frac{{\rm e}_{1}^{-1}}{\sigma_{0}^{2}+{\rm e}_{2}\left(1-\overline{N}\left(\sigma_{0}^{2},\beta\right)\right)}\nonumber\end{eqnarray}
 and from these expressions one can obtain the full expression for
the error per bit rate: \begin{equation}
\overline{P}_{b}\left(\sigma_{0}^{2},\beta\right)=\frac{1}{2}\left[1+\,{\rm erf}\left(\sqrt{\frac{\overline{E}\left(\sigma_{0}^{2},\beta\right)}{2}}\right)\right]\,.\label{eq:epbbar}\end{equation}

\subsection{CDMA signal detection - numerical results}

The inference algorithm requires an iterative update of equations~(\ref{eq:mmhatapprox},\ref{mhatfin})
and converges to a reliable estimate of the signal, with no need for
prior information of the noise level. The computational complexity
of the algorithm is of ${\mathcal{O}}(K^{2})$. %
\begin{figure}
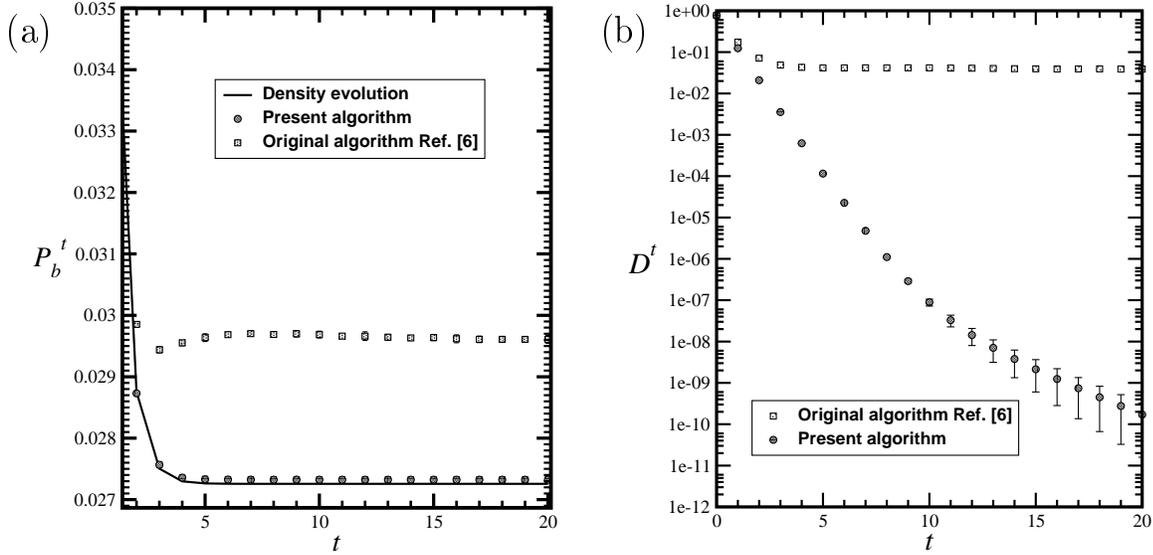

\begin{centering}\begin{picture}(440,190) \put(0,0){\epsfxsize=70mm
\epsfbox{graf1.eps}} \put(225,0){\epsfxsize=70mm \epsfbox{graf2.eps}}
\put(-10,193){\large (a)} \put(215,193){\large (b)} \end{picture}\par\end{centering}

\caption{(a) Error probability of the inferred solution evolving in time.
The system load $\beta=0.25$, true noise level $\sigma_{0}^{2}=0.25$
and estimated noise $\sigma^{2}=0.01$. Squares represent results
of the original algorithm \cite{KabashimaCDMA}, solid line the dynamics
obtained from our equations; circles represent results obtained from
the suggested practical algorithm. Variances are smaller than the
symbol size. (b) $D^{t}$, a measure of convergence for the obtained
solutions, as a function of time; symbols are as in the main figure.}

\label{fig2} 
\end{figure}

To test the performance of our algorithm we carried out a set of experiments
of the CDMA signal detection problem under typical conditions. Error
probability of the inferred signals was calculated for a system load
of $\beta\!=\!0.25$, where the true noise level is $\sigma_{0}^{2}\!=\!0.25$
and the estimated noise is $\sigma^{2}\!=\!0.01$, as shown in figure~\ref{fig2}(a).
The solid line represents the expected theoretical results (density
evolution), knowing the exact values of $\sigma_{0}^{2}$ and $\sigma^{2}$,
while circles represent simulation results obtained via the suggested
\emph{practical} algorithm, where no such knowledge is assumed. The
results presented are based on $10^{5}$ trials per point and a system
size $N\!=\!2000$ and are superior to those obtained using the original
algorithm~\cite{KabashimaCDMA}.

Another performance measure one should consider is \[
D^{t}\equiv\frac{1}{K}\left(\mathbf{m}^{t}-\mathbf{m}^{t-1}\right)\cdot\left(\mathbf{m}^{t}-\mathbf{m}^{t-1}\right),\]
 that provides an indication to the stability of the solutions obtained.
In figure~\ref{fig2}(b) we see that results obtained from our algorithm
show convergence to a reliable solution in contrast to the original
algorithm~\cite{KabashimaCDMA}. The physical interpretation of the
difference between the two results is assumed to be related to a replica
symmetry breaking phenomenon.

\subsection{Ising linear perceptron - numerical results}

\begin{figure}
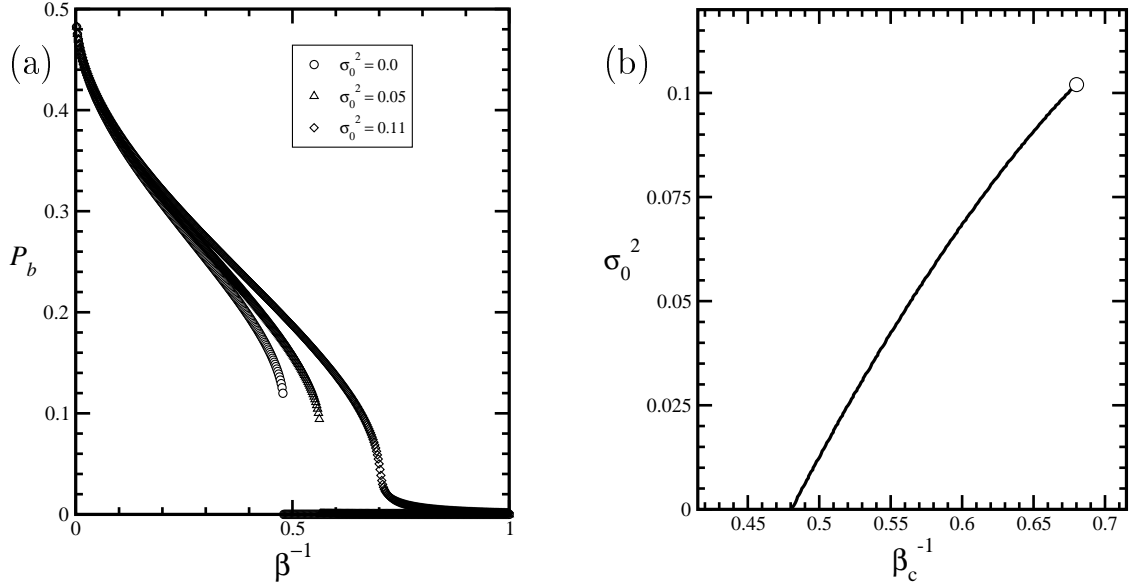

\begin{centering}\begin{picture}(440,190) \put(0,0){\epsfxsize=67.5mm
\epsfbox{4a.eps}} \put(225,0){\epsfxsize=70mm \epsfbox{4b.eps}}
\put(0,193){\large (a)} \put(225,193){\large (b)} \end{picture}\par\end{centering}

\caption{(a) The error probability $\overline{P}_{b}$ at the steady state,
equation~(\ref{eq:epbbar}), as a function of $\beta^{-1}$ for different
values of the noise parameter. For values of $\sigma_{0}^{2}$ below
0.1025 the curves show discontinuity at certain $\beta$ values, which
becomes continuous but non-analytic at $\sigma_{0}^{2}=0.1025$ around
$\beta^{-1}\simeq0.68$. For noise variance values above $\sigma_{0}^{2}=0.1025$
the curves become analytical. (b) Position of the non analyticity
of the error rate curve $\beta_{C}^{-1}$ as a function of the noise
parameter $\sigma_{0}^{2}$. This first order phase transition-like
curve ends in a second order phase transition-like point marked by
($\circ$).}

\label{fig-crit-1-2} 
\end{figure}

For the ILP, the $K>N$ regime of high interest as the system develops
a critical behaviour for a range of $\sigma_{0}^{2}$ values. We carried
out a set of experiments for this system based on density evolution.
In figure~\ref{fig-crit-1-2}(a) we present curves of the bit error
probability $\overline{P}_{b}$, defined in equation~(\ref{eq:epbbar}),
as a function of the inverse load $\beta^{-1}$ for different values
of $\sigma_{0}^{2}$. Three different regimes have been observed:
For $\sigma_{0}^{2}<0.1025$ the curves exhibit a discontinuity at
a value of $\beta$ that varies with $\sigma_{0}^{2}$ (first order
phase transition-like behaviour). At $\sigma_{0}^{2}=0.1025$ the
curve becomes continuous but its slope diverges (second order phase
transition-like behaviour). The $\overline{P}_{b}$ curves show analytical
behaviour for noise values above 0.1025. Figure~\ref{fig-crit-1-2}(b)
exhibits a phase diagram of the ILP system; it shows the dependency
of the critical load $\beta_{C}^{-1}$ as a function of the noise
parameter. The first order transition line ends in a second order
transition point marked by a circle. The results obtained, and in
particular the critical $\beta$ value, are consistent with those
derived using the replica symmetric statistical mechanics-based analysis
of the problem~\cite{Seung}.

\begin{figure}
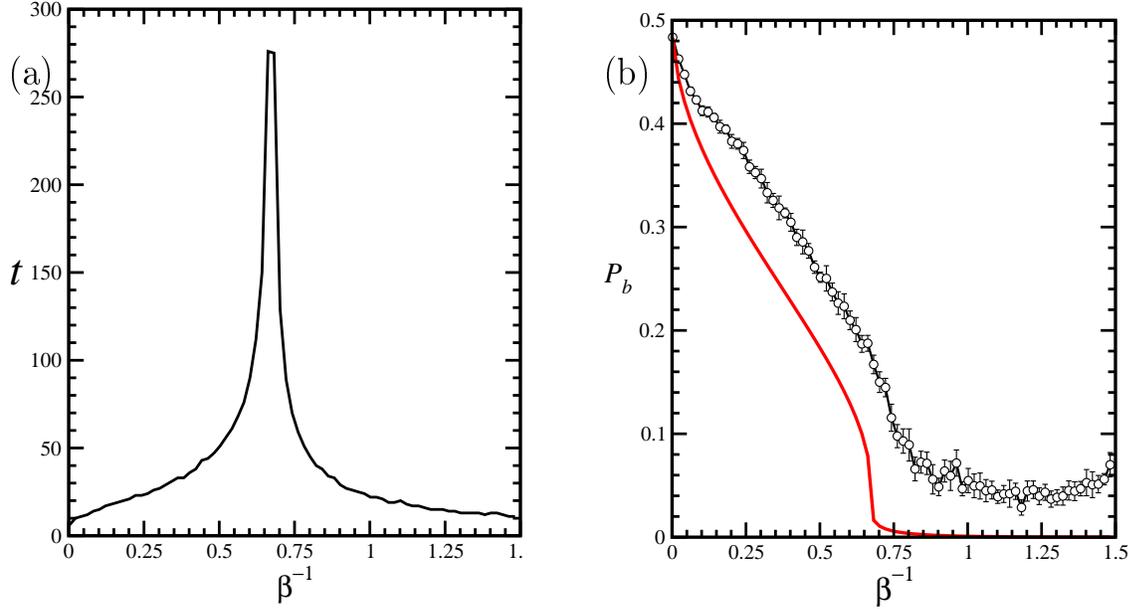

\begin{centering}\begin{picture}(440,190) \put(0,-4){\epsfxsize=70mm
\epsfbox{5a.eps}} \put(225,-5){\epsfxsize=70mm \epsfbox{5b.eps}}
\put(0,193){\large (a)} \put(225,193){\large (b)} \end{picture}\par\end{centering}

\caption{(a) Number of iterations of equation~(\ref{eq:ebar}) required for
convergence as a function of $\beta$, for $\sigma_{0}^{2}=0.10$;
one clearly identifies the $\beta$ value where the error rate curve
exhibits a discontinuity. (b) Finite size effects are observed at
all $\beta$ values. The noise level used is $\sigma_{0}^{2}=0.10$
with $K=500$. The curves provide mean values and error-bars over
1000 experiments. The solid curve obtained from the iteration of the
steady state equations is presented as a reference.}

\label{fig-crit-3-4} 
\end{figure}

Another indication for the critical behaviour is the number of steps
required for the recursive update of equation~(\ref{eq:ebar}) to
convergence. In figure~\ref{fig-crit-3-4}(a) we present the number
of iterations required to reach a steady state as a function of $\beta^{-1}$
when the noise parameter is set to $\sigma_{0}^{2}=0.1$. The number
of iterations diverges when the critical value of $\beta$ is reached.

Finally, we wish to explore the efficiency of the algorithm as a function
of the system size. In figure~\ref{fig-crit-3-4}(b) we present the
result of iterating equations~(\ref{eq:mmhatapprox}) and (\ref{mhatfin})
for a system size of \emph{K}=500. The curve presents mean values
and error bars over 1000 experiments. There is a strong dependency
of the error per bit rate on the size of the system, which is expected
to converge to the asymptotic limit (infinite system size) represented
by the solid line.

\section{CDMA signal detection with dual-peaked Gaussian noise}

\label{sec:CDMA2Gauss}

To demonstrate the suitability of the method for more complex inference
problems that require a system with 1RSB-like structures, we will
consider the CDMA signal of equation~(\ref{eq:CDMA}) where the noise
$n_{\mu}$ is drawn from a bi-Gaussian distribution: \begin{equation}
P\left(n_{\mu}\right)=\frac{1-r_{0}}{2}\frac{1}{\sqrt{2\pi}}\exp\left\{ -\frac{\left(n_{\mu}+\varepsilon_{0}/\sigma_{0}\right)^{2}}{2}\right\} +\frac{1+r_{0}}{2}\frac{1}{\sqrt{2\pi}}\exp\left\{ -\frac{\left(n_{\mu}-\varepsilon_{0}/\sigma_{0}\right)^{2}}{2}\right\} \,,\label{eq:2gruido}\end{equation}
 where $r_{0}\in\left(-1,1\right)$ represents the bias and $\pm\varepsilon_{0}/\sigma_{0}$
the positions of the Gaussian peaks. We consider the particular case
where $\left|\varepsilon_{0}/\sigma_{0}\right|\ll1$, so that the
Gaussian peaks are slightly off centre. For this model the likelihood
expression takes the form: \[
P\left(y_{\mu}|\Delta_{\mu};r,\varepsilon,\sigma^{2}\right)\propto\prod_{\ell=1}^{L}\prod_{{\rm a}=1}^{n}\left\{ \frac{1-r}{2}\exp\left[-\frac{\left(y_{\mu}-\Delta_{\mu}^{\ell{\rm a}}+\varepsilon\right)^{2}}{2\sigma^{2}}\right]+\frac{1+r}{2}\exp\left[-\frac{\left(y_{\mu}-\Delta_{\mu}^{\ell{\rm a}}-\varepsilon\right)^{2}}{2\sigma^{2}}\right]\right\} \,,\]
 where \emph{r}, $\varepsilon$ and $\sigma^{2}$ are estimates of
the true parameters $r_{0}$, $\varepsilon_{0}$ and $\sigma_{0}^{2}$.

To derive the messages in this case we first calculate the function
$\mathcal{P}\left(\vartheta,y_{\mu}\right)$ of equation~(\ref{eq:pcal}),
which has the form: \[
\mathcal{P}\left(\vartheta,y_{\mu}\right)=\frac{y_{\mu}-\vartheta}{\sigma^{2}+X^{t}}-\frac{\varepsilon}{\sigma^{2}+X^{t}}\tanh\left(\varepsilon\frac{y_{\mu}-\vartheta}{\sigma^{2}+X^{t}}+{\rm arctanh}(r)\right)\,,\]
 where $X^{t}=\beta\left(1-N^{t}\right).$

Following the derivation of Appendix~\ref{app:messages}, the saddle
point equations~(\ref{eq:wrs}) and (\ref{eq:w1rsb}) can be expressed
as: \begin{eqnarray*}
\tilde{\vartheta}_{\mu k}^{t} & = & u_{\mu k}^{t}+W^{t}\mathcal{P}\left(\tilde{\vartheta}_{\mu k}^{t},y_{\mu}\right)\\
y_{\mu}-\tilde{\vartheta}_{\mu k}^{t} & = & y_{\mu}-u_{\mu k}^{t}-W^{t}\,\frac{y_{\mu}-\tilde{\vartheta}_{\mu k}^{t}}{\sigma^{2}+X^{t}}+\varepsilon\,\frac{W^{t}}{\sigma^{2}+X^{t}}\tanh\left(\varepsilon\frac{y_{\mu}-\tilde{\vartheta}_{\mu k}^{t}}{\sigma^{2}+X^{t}}+{\rm arctanh}(r)\right)\\
\frac{y_{\mu}-\tilde{\vartheta}_{\mu k}^{t}}{\sigma^{2}+X^{t}} & = & \frac{y_{\mu}-u_{\mu k}^{t}}{\sigma^{2}+X^{t}+W^{t}}+\frac{\varepsilon}{\sigma^{2}+X^{t}}\,\frac{W^{t}}{\sigma^{2}+X^{t}+W^{t}}\tanh\left(\varepsilon\frac{y_{\mu}-\tilde{\vartheta}_{\mu k}^{t}}{\sigma^{2}+X^{t}}+{\rm arctanh}(r)\right)\\
z & = & \rho_{W}\left(y_{\mu}-u_{\mu k}^{t}\right)+\varepsilon\left(\rho_{0}-\rho_{W}\right)\tanh\left(\varepsilon z+{\rm arctanh(r)}\right)\\
 & \simeq & z_{0}+r\,\upDelta\rho_{W}\,\varepsilon+\left(1-r^{2}\right)\upDelta\rho_{W}\, z\,\varepsilon^{2}\\
 &  & -r\left(1-r^{2}\right)\upDelta\rho_{W}\, z^{2}\varepsilon^{3}-\frac{1}{3}\left(1-r^{2}\right)\left(1-3r^{2}\right)\upDelta\rho_{W}\, z^{3}\varepsilon^{4}\,,\end{eqnarray*}
 where we denote $W^{t}=R^{t}$ for the RS case and $W^{t}=2V^{t}-R^{t}$
for the 1RSB case, $z\equiv\frac{y_{\mu}-\tilde{\vartheta}_{\mu k}^{t}}{\sigma^{2}+X^{t}}$,
$\rho_{A}\equiv\left(\sigma^{2}+X^{t}+A\right)^{-1}$, $z_{0}\equiv\rho_{W}\left(y_{\mu}-u_{\mu k}^{t}\right)$
and $\upDelta\rho_{W}\equiv\rho_{0}-\rho_{W}$.

The solution of this equation provides, up to order $\mathcal{O}\left(\varepsilon^{4}\right)$,
\begin{eqnarray*}
z\left(\varepsilon\right) & \simeq & z_{0}+r\,\upDelta\rho_{W}\,\varepsilon+\left(1-r^{2}\right)\upDelta\rho_{W}\left[\, z_{0}\varepsilon^{2}+r\,\left(\upDelta\rho_{W}-z_{0}^{2}\right)\varepsilon^{3}+\left(1-3r^{2}\right)\, z_{0}\left(\upDelta\rho_{W}-\frac{1}{3}z_{0}^{2}\right)\varepsilon^{4}\right].\end{eqnarray*}
 The function $\mathcal{P}$ and its two first derivatives at the
saddle point value are: \begin{eqnarray*}
\mathcal{P}_{0} & = & -r\left[1+\left(1-r^{2}\right)\upDelta\rho_{W}\,\varepsilon^{2}\right]\rho_{W}\,\varepsilon+\\
 &  & +\left[1-\left(1-r^{2}\right)\rho_{W}^{2}\,\varepsilon^{2}-\left(1-r^{2}\right)\left(1-3r^{2}\right)\upDelta\rho_{W}\,\rho_{W}^{2}\,\varepsilon^{4}\right]\left(y_{\mu}-u_{\mu k}^{t}\right)+\\
 &  & +r\left(1-r^{2}\right)\rho_{W}^{3}\left(y_{\mu}-u_{\mu k}^{t}\right)^{2}\varepsilon^{3}+\frac{1}{3}\left(1-r^{2}\right)\left(1-3r^{2}\right)\rho_{W}^{4}\left(y_{\mu}-u_{\mu k}^{t}\right)^{3}\varepsilon^{4}\\
\mathcal{P}_{1} & \simeq & -\rho_{0}+\mathcal{O}\left(\varepsilon^{2}\right)\\
\mathcal{P}_{2} & = & 2\rho_{0}^{3}\left(1-r^{2}\right)\left[r\varepsilon^{3}+\left(1-3r^{2}\right)\rho_{W}\left(y_{\mu}-u_{\mu k}^{t}\right)\varepsilon^{4}\right]\,,\end{eqnarray*}
 therefore, one can obtain the following expression, required for
calculating the messages in the 1RSB case (\ref{eq:mm1rsb}) \[
\frac{1}{2}\,\frac{\mathcal{P}_{2}V^{t}}{1-\mathcal{P}_{1}V^{t}}=\left(1-r^{2}\right)\rho_{0}\upDelta\rho_{V}\left[r\,\varepsilon^{3}+\left(1-3r^{2}\right)\rho_{W}\left(y_{\mu}-u_{\mu k}^{t}\right)\varepsilon^{4}\right]\,,\]
 where $\upDelta\rho_{V}\equiv\rho_{0}-\rho_{V}$. This straightforwardly
leads to the following expression for the message: \begin{eqnarray}
^{{\textrm{(1RSB)}}}\widehat{m}_{\mu k}^{t+1} & = & \frac{s_{\mu k}}{\sqrt{N}}\left\{ -\left[\rho_{W}+\left(1-r^{2}\right)\left(\Upsilon_{n}-\rho_{W}^{2}\right)\varepsilon^{2}\right]r\,\varepsilon+\right.\nonumber \\
 &  & +\rho_{W}\left[1-\left(1-r^{2}\right)\rho_{W}\,\varepsilon^{2}-\left(1-r^{2}\right)\left(1-3r^{2}\right)\left(\Upsilon_{n}-\rho_{W}^{2}\right)\varepsilon^{4}\right]\,\left(y_{\mu}-u_{\mu k}^{t}\right)+\nonumber \\
 &  & \left.+r\left(1-r^{2}\right)\rho_{W}^{3}\,\varepsilon^{3}\left(y_{\mu}-u_{\mu k}^{t}\right)^{2}+\frac{1}{3}\left(1-r^{2}\right)\left(1-3r^{2}\right)\rho_{W}^{4}\,\varepsilon^{4}\left(y_{\mu}-u_{\mu k}^{t}\right)^{3}\right\} \,,\label{eq:2gauss1rsb}\end{eqnarray}
 where $\Upsilon_{n}\equiv\rho_{0}\left(\rho_{W}-\frac{1}{n}\upDelta\rho_{V}\right)$.
The expression for the message in the RS case is recovered from equation~(\ref{eq:2gauss1rsb})
in the limit $n\to\infty.$

\subsection{Optimisation and messages}

Calculating the expressions for the macroscopic variables $E^{t+1}$
and $F^{t+1}$, used in the optimisation process, requires performing
the following sums, in the limit of $K,N\to\infty$ with $K/N=\beta<\infty$:
\begin{eqnarray*}
A_{j} & \equiv & \lim_{K,N\to\infty}\sum_{\mu}^{N}\frac{1}{K}\sum_{k=1}^{K}\frac{s_{\mu k}b_{k}}{\sqrt{N}}\left(y_{\mu}-u_{\mu k}^{t}\right)^{j}\\
B_{l} & \equiv & \lim_{K,N\to\infty}\frac{1}{N}\sum_{\mu}^{N}\frac{1}{K}\sum_{k=1}^{K}\left(y_{\mu}-u_{\mu k}^{t}\right)^{l}\,,\end{eqnarray*}
 where $j=0,\dots,3$ and $l=0,\dots,4$. From the definition of the
signal $y_{\mu}$ (\ref{eq:CDMA}) and the expression for the noise
(\ref{eq:2gruido}) we find that $A_{0}=0$, $A_{1}=1$, $A_{2}=2B_{1},$
$A_{3}=3B_{2},$ $B_{0}=1$, $B_{1}=r_{0}\varepsilon_{0}$, $B_{2}=\beta\left(1-2M^{t}+N^{t}\right)+\sigma_{0}^{2}+\varepsilon_{0}^{2}$,
$B_{3}=B_{1}\left(3B_{2}-2\varepsilon_{0}^{2}\right)$ and $B_{4}=3B_{2}^{2}-2\varepsilon_{0}^{4}.$
The explicit expressions derived for the macroscopic variables are:
\begin{eqnarray*}
E^{t+1} & = & \rho_{W}-\left(1-r^{2}\right)\rho_{W}^{2}\,\varepsilon^{2}+2r\left(1-r^{2}\right)B_{1}\rho_{W}^{3}\,\varepsilon^{3}-\left(1-r^{2}\right)\left(1-3r^{2}\right)\left[\Upsilon_{n}-\left(1+B_{2}\rho_{r}\right)\rho_{W}^{2}\right]\rho_{W}\,\varepsilon^{4}\\
F^{t+1} & = & B_{2}\rho_{W}^{2}-2rB_{1}\rho_{W}^{2}\,\varepsilon\\
 &  & +\left[r^{2}-2\left(1-r^{2}\right)B_{2}\rho_{W}\right]\rho_{W}^{2}\,\varepsilon^{2}-2r\left(1-r^{2}\right)B_{1}\left[\Upsilon_{n}-\left(2+3B_{2}\rho_{W}\right)\rho_{W}^{2}\right]\rho_{W}\,\varepsilon^{3}+\\
 &  & +\left(1-r^{2}\right)\left[2r^{2}\left(\Upsilon_{n}-\rho_{W}^{2}\right)\rho_{W}+\left(1-3r^{2}\right)B_{2}\left(3\rho_{W}^{2}+2B_{2}\rho_{W}^{3}-2\Upsilon_{n}\right)\rho_{W}^{2}\right]\varepsilon^{4}\,.\end{eqnarray*}
 Applying the optimisation conditions of Appendix~\ref{app:optimisation},
$E^{t}\left(\mathbold\gamma^{c}\right)=F^{t}\left(\mathbold\gamma^{c}\right)$
and $\left.{\displaystyle \frac{\partial E^{t}}{\partial\gamma_{i}}-\frac{1}{2}\,\frac{E^{t}}{F^{t}}\,\frac{\partial F^{t}}{\partial\gamma_{i}}}\right|_{\gamma_{i}^{c}}=0$,
where $\mathbold\gamma^{\sf T}=\left(r,\varepsilon,\sigma^{2},\frac{1}{n}\right)$
one obtain the following conditions: \begin{eqnarray}
\rho_{W} & = & \frac{1}{B_{2}}+\frac{\varepsilon^{2}}{B_{2}^{2}}-\frac{\varepsilon^{4}}{B_{2}^{3}}+\left(1-r^{2}\right)^{2}\,\frac{1-B_{2}\rho_{0}\left(1-\frac{1}{n}B_{2}\upDelta\rho_{V}\right)}{B_{2}^{3}}\,\varepsilon^{4}\label{eq:cond1}\\
r\,\varepsilon & = & B_{1}+r\left(1-r^{2}\right)\,\frac{1-B_{2}\rho_{0}\left(1-\frac{1}{n}B_{2}\upDelta\rho_{V}\right)}{B_{2}}\,\varepsilon^{3}\,.\label{eq:cond2}\end{eqnarray}

In the 1RSB case one can further simplify these expressions by a suitable
choice of $V^{t}$ and the number of replicas per block \emph{n}.
Optimisation with respect to the latter results in \begin{equation}
1=B_{2}\rho_{0}\left(1-\frac{1}{n}B_{2}\upDelta\rho_{V}\right),\label{eq:cond3}\end{equation}
 which implies \[
V^{t}=\frac{\left(X^{t}+\sigma^{2}\right)^{2}\left(\sigma_{0}^{2}-\sigma^{2}\right)}{{\displaystyle \frac{1}{n}\,\left(X^{t}+\sigma_{0}^{2}\right)^{2}-\left(X^{t}+\sigma^{2}\right)\left(\sigma_{0}^{2}-\sigma^{2}\right)}}\,,\]
 that by definition is larger than zero. This condition is satisfied
if our estimate for the noise variance is smaller than the true parameter
$\left(\sigma^{2}<\sigma_{0}^{2}\right)$. In this case the number
of replicas per block has to satisfy the condition \[
1\leq n\leq f\left(X^{t};\sigma_{0}^{2},\sigma^{2}\right)\equiv\frac{\left(X^{t}+\sigma_{0}^{2}\right)^{2}}{\left(X^{t}+\sigma^{2}\right)\left(\sigma_{0}^{2}-\sigma^{2}\right)}.\]
 Interestingly this ties the noise level mismatch to the number of
replicas, thus giving further insight to the role played by the structure
of the inter-replica correlation matrix.

For $0\leq X^{t}$, the minimum value of $f\left(X^{t};\sigma_{0}^{2},\sigma^{2}\right)$
is reached at $X_{min}=\max\left(0,\sigma_{0}^{2}-2\sigma^{2}\right)$.
It is also possible to prove that $4\leq f\left(X_{min};\sigma_{0}^{2},\sigma^{2}\right).$
Although $V^{t}$ and $n$ will not be explicitly used in the following
expressions, the correct choice of the value for these parameters
allows one to use equations~(\ref{eq:cond1}) and (\ref{eq:cond2})
in order to find the final expression for the macroscopic variable
$E^{t+1}$, where no estimates are needed for the noise parameters:
\begin{eqnarray*}
^{{\textrm{(1RSB)}}}E^{t+1} & = & \frac{1}{B_{2}-B_{1}^{2}}.\end{eqnarray*}

Note that in the RS case we do not have the freedom to choose the
number of replicas per block, given that this case is equivalent to
take $n\to\infty$ in the absence of the additional replica $l=1,\ldots,L$.
For this reason equations~(\ref{eq:cond1}) and (\ref{eq:cond2})
and (\ref{eq:cond2}) take the form: \begin{eqnarray}
\rho_{W} & = & \frac{1}{B_{2}}+\frac{\varepsilon^{2}}{B_{2}^{2}}-\frac{\varepsilon^{4}}{B_{2}^{3}}+\left(1-r^{2}\right)^{2}\,\frac{1-B_{2}\rho_{0}}{B_{2}^{3}}\,\varepsilon^{4}\label{eq:cond1rs}\\
r\,\varepsilon & = & B_{1}+r\left(1-r^{2}\right)\,\frac{1-B_{2}\rho_{0}}{B_{2}}\,\varepsilon^{3}\,,\label{eq:cond2rs}\end{eqnarray}
 and the macroscopic variable \begin{eqnarray*}
^{{\textrm{(RS)}}}E^{t+1} & = & ^{{\textrm{(1RSB)}}}E^{t+1}+\frac{2B_{1}^{2}\left(\varepsilon^{2}-B_{1}^{2}\right)}{B_{2}^{3}}\left(\frac{B_{2}}{X^{t}+\sigma^{2}}-1\right)\,,\end{eqnarray*}
 which depends on both estimates of the noise variance $\sigma^{2}$
and bias $\varepsilon.$

Given that the algorithm deals with finite signal vectors $\left(N<\infty\right)$,
the quantities $B_{1}$ and $B_{2}$ have to be approximated by the
correspondent finite sums. Therefore, we have: \begin{eqnarray}
B_{1} & = & \lim_{N,K\to\infty}\frac{1}{N}\sum_{\mu=1}^{N}\frac{1}{K}\sum_{k=1}^{K}\left(y_{\mu}-u_{\mu k}^{t}\right)\approx\frac{1}{N}\sum_{\mu=1}^{N}y_{\mu}\equiv\overline{B}_{1}\label{eq:estB1B2}\\
B_{2} & = & \lim_{N,K\to\infty}\frac{1}{N}\sum_{\mu=1}^{N}\frac{1}{K}\sum_{k=1}^{K}\left(y_{\mu}-u_{\mu k}^{t}\right)^{2}\approx\frac{1}{N}\sum_{\mu=1}^{N}y_{\mu}^{2}+\beta N^{t}\equiv\overline{B}_{2}\,,\nonumber \end{eqnarray}
 where we used the fact that $\lim_{N,K\to\infty}\frac{1}{NK}\sum_{\mu,k}u_{\mu k}^{t}=0$.
Observe that no information about the true noise has been used to
derive these expressions.

Having the estimates~(\ref{eq:estB1B2}) we can write down the messages
explicitly: \begin{eqnarray*}
^{{\textrm{(1RSB)}}}\widehat{m}_{\mu k}^{t+1} & = & \frac{s_{\mu k}}{\sqrt{N}}\,\left\{ -\frac{\overline{B}_{1}}{\overline{B}_{2}}+\frac{\overline{B}_{1}}{\overline{B}_{2}^{2}}\varepsilon^{2}+\left(\frac{1}{\overline{B}_{2}}+\frac{\overline{B}_{1}^{2}}{\overline{B}_{2}^{2}}-\frac{3\varepsilon^{2}-2\overline{B}_{1}^{2}}{\overline{B}_{2}^{3}}\varepsilon^{2}\right)\,\left(y_{\mu}-u_{\mu k}^{t}\right)+\right.\\
 &  & \left.+\frac{\overline{B}_{1}\left(\varepsilon^{2}-\overline{B}_{1}^{2}\right)}{\overline{B}_{2}^{3}}\,\left(y_{\mu}-u_{\mu k}^{t}\right)^{2}+\frac{1}{3}\frac{\left(\varepsilon^{2}-\overline{B}_{1}^{2}\right)\left(\varepsilon^{2}-3\overline{B}_{1}^{2}\right)}{\overline{B}_{2}^{4}}\,\left(y_{\mu}-u_{\mu k}^{t}\right)^{3}\right\} \\
^{{\textrm{(RS)}}}\widehat{m}_{\mu k}^{t+1} & = & ^{{\textrm{(1RSB)}}}\widehat{m}_{\mu k}^{t+1}+\frac{s_{\mu k}}{\sqrt{N}}\left(1-\frac{\overline{B}_{2}}{X^{t}+\sigma^{2}}\right)\frac{\varepsilon^{2}-\overline{B}_{1}^{2}}{\overline{B}_{2}^{2}}\left[\overline{B}_{1}+2\frac{\varepsilon^{2}-2\overline{B}_{1}^{2}}{\overline{B}_{2}^{2}}\,\left(y_{\mu}-u_{\mu k}^{t}\right)\right]\,,\end{eqnarray*}
 which can be now used recursively for obtaining the inferred solutions
for this problem. Notice that an estimate of \emph{both} $\varepsilon$
and $\sigma$ in required in the RS case.

\subsection{Numerical results}

\begin{figure}
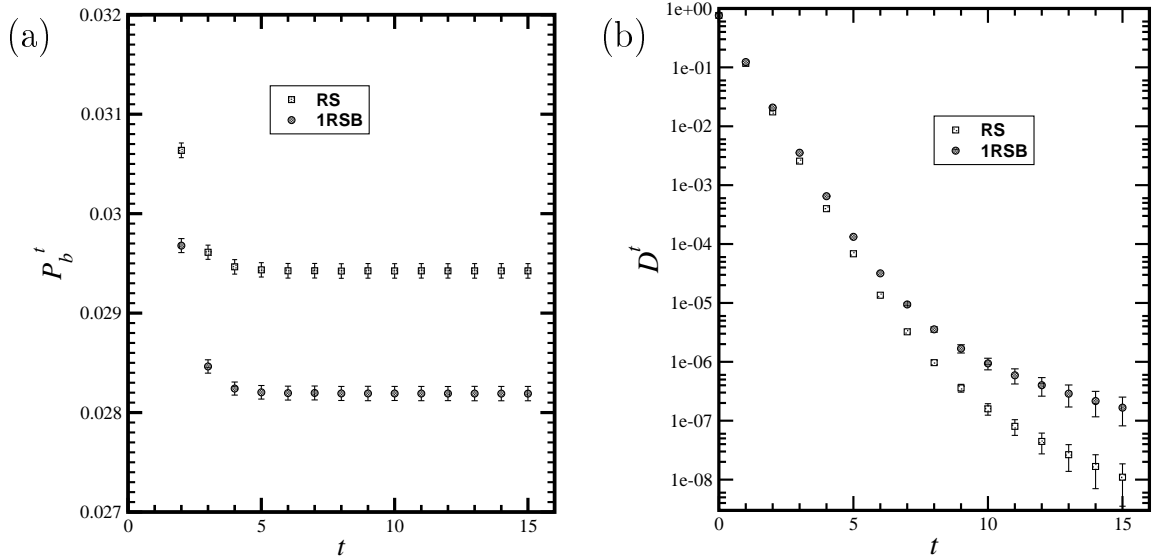

\begin{centering}\begin{picture}(440,190) \put(0,0){\epsfxsize=70mm
\epsfbox{graf3.eps}} \put(225,0){\epsfxsize=70mm \epsfbox{graf4.eps}}
\put(-10,193){\large (a)} \put(215,193){\large (b)} \end{picture} \par\end{centering}

\caption{(a) Error probability of the inferred solution evolving in time,
for the bi-Gaussian noise case. The system load $\beta=0.25$, true
noise level $\sigma_{0}^{2}=0.25$ and estimated noise $\sigma^{2}=0.01$.
Squares represent results of the RS algorithm and circles represent
results obtained from the 1RSB algorithm. (b) $D^{t}$, a measure
of convergence in the obtained solutions, as a function of time; symbols
are as in the main figure.}

\label{fign} 
\end{figure}

To test the performance of the 1RSB algorithm we carried out a set
of experiments of the CDMA signal detection problem with bi-Gaussian
noise. The results shown in figure~\ref{fign}(a) describe the error
probability of the inferred signals as a function of the number of
iterations has been calculated using both RS and 1RSB-like correlation
matrices for the case of parameters mismatch. The system load used
in the simulations was $\beta\!=\!0.25$, the true noise level $\sigma_{0}^{2}\!=\!0.25$,
Gaussian bias of $\varepsilon_{0}=0.06$ and weight $r_{0}=$0.6.
The estimated noise parameters are $\sigma^{2}\!=\!0.01$ and $\varepsilon=0.2$.
The circles represent simulation results obtained via the 1RSB algorithm
while the squares correspond to the RS-like structure. The results
presented are based on $10^{5}$ trials per point and a system size
$N\!=\!1000$; error-bars are also provided. The results obtained
using the 1RSB-like structure are superior to those obtained using
the RS algorithm. As shown in figure~\ref{fign}(b) using the stability
measure $D^{t}$, both RS and 1RSB-based algorithms converge to reliable
solutions; the 1RSB-based algorithm is slightly slower to converge,
presumably due to the more complex message passing scheme.

\section{Conclusions}

We present and methodologically develop a new algorithm for using
BP in densely connected systems that enables one to obtain reliable
solutions even when the solution space is fragmented. The algorithm
relies on the introduction of a large number of replicated variable
systems exposed to the same evidential nodes. Messages are obtained
by averaging over all replicated systems leading to pseudoposterior
that is then used to infer the variable nodes most probable values.
This is done with no actual replication, by introducing an assumption
about correlations between the replicated variables and exploiting
the high number of replicated systems. The algorithm was developed
in a systematic manner to accommodate more complex correlation matrices.
It was successfully applied to the CDMA signal detection and ILP learning
problems, using the RS-like correlation matrix, and to the CDMA inference
problem with bi-modal Gaussian noise model in the 1RSB-like correlation
matrix. The algorithm provides superior results to other existing
algorithms~\cite{KabashimaCDMA,Kabashimanew} and a systematic improvement
where more complex correlation matrices are introduced, where required.

Further research is required to fully determine the potential of the
new algorithm. Two particular areas which we consider as particularly
promising are inference problems characterised by discrete data variables
and noise model and problems that can be mapped onto \emph{sparse}
graphs. Both activities are currently underway.

\begin{acknowledgments}
Support from EVERGROW IP No.~1935 of the EU FP-6 is gratefully acknowledged. 
\end{acknowledgments}
\appendix
%dummy comment inserted by tex2lyx to ensure that this paragraph is not empty
%dummy comment inserted by tex2lyx to ensure that this paragraph is not empty
%dummy comment inserted by tex2lyx to ensure that this paragraph is not empty
%dummy comment inserted by tex2lyx to ensure that this paragraph is not empty
%dummy comment inserted by tex2lyx to ensure that this paragraph is not empty
%dummy comment inserted by tex2lyx to ensure that this paragraph is not empty
%dummy comment inserted by tex2lyx to ensure that this paragraph is not empty

\section{The Replica Symmetric (RS) Ansatz \label{app:RS}}

Within the RS setting, the interaction term in equation~(\ref{eq:ansatz})
is: \[
\mathbf{b}_{k}^{\sf T}\mathbf{Q}_{\mu k}^{t}\mathbf{b}_{k}=n\left(q_{0\mu k}^{t}-q_{1\mu k}^{t}\right)+q_{1\mu k}^{t}\left(\sum_{{\rm a}=1}^{n}b_{k}^{{\rm a}}\right)^{2}\,,\]

A simplified expression for equation~(\ref{eq:ansatz}) immediately
follows \begin{eqnarray*}
P^{t}\left(\mathbf{b}_{k}|\left\{ y_{\nu\neq\mu}\right\} \right) & = & [\mathcal{Z}_{\mu k}^{t}]^{-1}\exp\left\{ h_{\mu k}^{t}\sum_{\mathrm{a}=1}^{n}b_{k}^{\mathrm{a}}+\frac{1}{2}q_{1\mu k}^{t}\left(\sum_{\mathrm{a}=1}^{n}b_{k}^{\mathrm{a}}\right)^{2}\right\} \\
 & = & [\mathcal{Z}_{\mu k}^{t}]^{-1}{\displaystyle \int_{-\infty}^{\infty}\mathrm{d}x\,\exp\left\{ -\frac{x^{2}}{2q_{1\mu k}^{t}}+\left(x+h_{\mu k}^{t}\right)\sum_{\mathrm{a}=1}^{n}b_{k}^{\mathrm{a}}\right\} }\end{eqnarray*}
 where $\mathcal{Z}_{\mu k}^{t}$ is a normalisation constant. The
diagonal elements $q_{0\mu k}^{t}$ only affect the normalisation
term and can therefore be taken to zero with no loss of generality.

We expect the logarithm of the normalisation term $\mathcal{Z}_{\mu k}^{t}$
(linked to the free energy), obtained from the well behaved distribution
$P^{t}$, to be self-averaging. We therefore expect \[
\lim_{n\rightarrow\infty}\frac{1}{n}\log\left(\overline{\mathcal{Z}_{\mu k}^{t}}\right)=\lim_{n\rightarrow\infty}\frac{1}{n}\log\left(\mathcal{Z}_{\mu k}^{t}\left(\hat{h},\hat{q}_{1}\right)\right),\]
 where $\hat{h}$ and $\hat{q}_{1}$ are the mean values of the parameters
drawn for some suitable distributions and the over-line represents
the mean value of the partition function over these distributions.

In the following we will drop the upper-index \emph{t} and the sub-indices
$\mu$ and $k$ for brevity. To obtain the scaling behaviour of the
various parameters one calculates $\mathcal{Z}\left(h,q_{1}\right)$
explicitly, assuming the parameter $q_{1}$ is taken from a normal
distribution $\mathcal{N}\left(\hat{q}_{1},\sigma_{q}^{2}\right)$.
The partition function takes the form : \begin{equation}
\mathcal{Z}\left(h,q_{1}\right)=\int_{-\infty}^{\infty}\frac{\mathrm{d}x}{\sqrt{2\pi q_{1}}}\,\exp\left(-\frac{\left(x-h\right)^{2}}{2q_{1}}+n\ln\left(2\cosh(x)\right)\right).\label{eq:z}\end{equation}
 Thus, the mean value of the partition function over the set of parameters
is: \[
\overline{\mathcal{Z}\left(h,q_{1}\right)}=\int\mathcal{D}_{q_{1}}\,\mathcal{Z}\left(h,q_{1}\right),\]
 where $\mathcal{D}_{q_{1}}=\mathrm{d}q_{1}\,\mathcal{N}\left(\hat{q}_{1},\sigma_{q_{1}}^{2}\right).$
The normalisation can be expressed as: \begin{eqnarray*}
\overline{\mathcal{Z}\left(h,q_{1}\right)} & = & \sum_{\textrm{a}=0}^{n}\binom{n}{{\rm a}}\,\exp\left\{ n\left[h\left(1-\frac{2\textrm{a}}{n}\right)+\frac{\hat{q}_{1}}{2}\left(1-\frac{2\textrm{a}}{n}\right)^{2}n+\frac{\sigma_{q_{1}}^{2}}{8}\left(1-\frac{2\textrm{a}}{n}\right)^{4}n^{3}\right]\right\} \\
 & = & \mathcal{A}(n)\,(n+1)\binom{n}{n/2}\,\exp\left\{ n\left[\left|h\right|+n\frac{\hat{q}_{1}}{2}+n^{3}\frac{\sigma_{q_{1}}^{2}}{8}\right]\right\} \\
 & \simeq & \sqrt{\frac{2}{\pi}}\mathcal{A}(n)\,\exp\left\{ n\left[\ln(2)+\left|h\right|+n\frac{\hat{q}_{1}}{2}+n^{3}\frac{\sigma_{q_{1}}^{2}}{8}\right]\right\} ,\end{eqnarray*}
 where $\mathcal{A}(n)\sim\mathcal{O}(1)$. Thus, $h\sim\mathcal{O}\left(1\right)$,
$\hat{q}_{1}\sim\mathcal{O}\left(n^{-1}\right)$ and $\sigma_{q_{1}}^{2}\sim\mathcal{O}\left(n^{-3}\right)$.
>From now on we will take the off-diagonal elements of the RS matrix
$\mathbf{Q}_{\mu k}^{t}$ equal to $g_{1\mu k}^{t}/n$, where $g_{1\mu k}^{t}\sim\mathcal{O}\left(1\right)$.

The form of the marginalised posterior at time \emph{t} is then: \begin{equation}
P^{t}\left(\mathbf{b}_{k}|\left\{ y_{\nu\neq\mu}\right\} \right)=\frac{{\displaystyle {\displaystyle {\displaystyle {\displaystyle \int_{-\infty}^{\infty}\mathrm{d}x\,\exp{\textstyle {\displaystyle \left\{ -n\frac{\left(x-h_{\mu k}^{t}\right)^{2}}{2g_{1\mu k}^{t}}+x\sum_{\textrm{a}=1}^{n}b_{k}^{\textrm{a}}\right\} }}}}}}}{{\displaystyle {\displaystyle {\displaystyle \int_{-\infty}^{\infty}\mathrm{d}x\,\exp\left\{ -n\Phi\left(x;h_{\mu k}^{t},g_{1\mu k}^{t}\right)\right\} }}}},\label{pp}\end{equation}
 where \[
\Phi\left(x;h_{\mu k}^{t},g_{1\mu k}^{t}\right)=\frac{\left(x-h_{\mu k}^{t}\right)^{2}}{2g_{1\mu k}^{t}}-\ln\left(2\cosh(x)\right).\]
 The function $\Phi\left(x;h,g_{1}\right)$ presents one or two minima
according to the following table:\\

\begin{center}\begin{tabular}{|c|c|c|}
\hline 
$h$&
$g_{1}$&
Number of minima\tabularnewline
\hline 
$h\in\mathbb{R}$&
$0<g_{1}\leq1$&
one min.\tabularnewline
\hline 
$|h|=h_{c}$&
$g_{1}>1$&
one min. and one hump\tabularnewline
\hline 
$|h|<h_{c}$&
$g_{1}>1$&
two min. \tabularnewline
\hline
\end{tabular}\par\end{center}

\noindent where $h_{c}=\sqrt{g_{1}(g_{1}-1)}-\cosh^{-1}\left(\sqrt{g_{1}}\right)$;
the coefficient $g_{1}$ plays the role of the inverse temperature.
Below the critical value $g_{1c}=1$ a spontaneous magnetisation appears.

\begin{figure}
\begin{centering}\includegraphics[width=2.55906in]{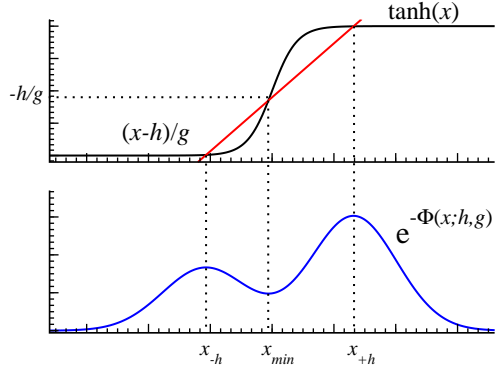} \par\end{centering}

\caption{Solutions for the mean field equation~(\ref{eq:derivada}) with
two maxima and one minimum for a positive value of the field $h$.\label{2peaks}}
\end{figure}

\noindent This results from analysing the equation: \begin{eqnarray}
\frac{\partial\Phi\left(x;h,g_{1}\right)}{\partial x} & = & \frac{x-h}{g_{1}}-\tanh(x)=0.\label{eq:derivada}\end{eqnarray}
 The case of two maxima is presented in figure~\ref{2peaks}.

\noindent %

We define the mean values from the distribution equation~(\ref{pp}).
If the field $h$ is not zero, as shown in figure~\ref{2peaks},
$\left[\exp\left(-\Phi\right)\right]^{n}$ develops one dominant maximum
as $n\to\infty$. For large enough $n$, only this maximum contributes
to the integrals (\ref{pp}) and the algorithm obtained from this
assumption turns out to be the same as the one presented in ~\cite{KabashimaCDMA}.
However, if the field is sufficiently small it gives rise to a new
regime where the two maxima contribute. At the same time, it is important
to note that a small, non zero field favours the solution of Eq.(\ref{eq:derivada})
that satisfies ${\rm sgn}(x)={\rm sgn}(h).$ To analyse the behaviour
of the field, we will explore the solutions of Eq.(\ref{eq:derivada})
in the regime $0\lesssim\left|h\right|\ll1$. With this aim, suppose
that the solutions for the Eq.(\ref{eq:derivada}) at zero field are
$x_{0}=\pm g_{1}\left|m\right|$ where $m\equiv\tanh\left(x_{0}\right)$
and ${\rm sgn}(m)={\rm sgn}(h)$. If the field is sufficiently small
one can expand the solutions of equation~(\ref{eq:derivada}) as
$x_{\pm h}=\pm g_{1}m+\xi(m,g_{1})h$ where $\xi(m,g_{1})h$ is expected
to be small and satisfies ${\rm sgn}\left(\xi(m,g_{1})h\right)={\rm sgn}(h)$.
Observe that if the field is positive (negative), both roots are displaced
to the right (left) with respect to the zero field solutions. Using
this expression for the roots in Eq.(\ref{eq:derivada}) and disregarding
terms of $\mathcal{O}\left(h^{2}\right)$ one finds that \begin{equation}
\xi(m,g_{1})=\frac{1}{1-g_{1}\left(1-m^{2}\right)}\,.\label{eq:xih}\end{equation}
 The expression for the exponent $\Phi$ near the roots and in the
$0\lesssim\left|h\right|\ll1$ regime is then $\Phi\left(x_{\pm h};h\to0,g_{1}\right)\simeq\Phi\left(x_{0};0,g_{1}\right)\mp mh=\Phi_{0}\mp mh\,,$
and, by the definition of the $m$, the product $mh$ is positively
defined.

Let us define $\beta_{\pm h}\left(m,g_{1}\right)\equiv\left(1-m^{2}\right)\left[1\mp2\xi\left(m,g_{1}\right)mh\right]$.
We expect that, for large $n$ the following approximation to be valid:
\begin{eqnarray}
\exp\left\{ -n\Phi\left(x;h\to0,g_{1}\right)\right\}  & \simeq & {\rm e}^{-n\Phi_{0}}\left\{ {\rm e}^{nmh}\exp\left\{ -\frac{n}{2}\left[g_{1}^{-1}-\beta_{h}\left(m,g_{1}\right)\right]\left(x-x_{h}\right)^{2}\right\} \right.\nonumber \\
 &  & \left.\qquad+{\rm e}^{-nmh}\exp\left\{ -\frac{n}{2}\left[g_{1}^{-1}-\beta_{-h}\left(m,g_{1}\right)\right]\left(x-x_{-h}\right)^{2}\right\} \right\} \,.\label{exph0}\end{eqnarray}
 Using equation~(\ref{exph0}) one can calculate the normalisation
in equation~(\ref{eq:z}) \begin{eqnarray}
\mathcal{Z}\left(h\to0,g_{1}\right) & \simeq & {\rm e}^{-n\left(\Phi_{0}-mh\right)}\int{\rm d}x\,\exp\left\{ -\frac{n}{2}\left[g_{1}^{-1}-\beta_{h}\left(m,g_{1}\right)\right]\left(x-x_{h}\right)^{2}\right\} \nonumber \\
 &  & +{\rm e}^{-n\left(\Phi_{0}+mh\right)}\int{\rm d}x\,\exp\left\{ -\frac{n}{2}\left[g_{1}^{-1}-\beta_{-h}\left(m,g_{1}\right)\right]\left(x-x_{-h}\right)^{2}\right\} \nonumber \\
 & \simeq & \sqrt{\frac{2\pi g_{1}\xi\left(m,g_{1}\right)}{n}}\!\,{\rm e}^{-n\Phi_{0}}\!\left\{ \,{\rm e}^{nmh}\!\left(1-g_{1}\left(1-m^{2}\right)\xi^{2}\left(m,g_{1}\right)\, mh\right)\right.\nonumber \\
 &  & \qquad\qquad\qquad\left.+\,\,{\rm e}^{-nmh}\!\left(1+g_{1}\left(1-m^{2}\right)\xi^{2}\left(m,g_{1}\right)\, mh\right)\right\} \,\,.\label{eq:zzz}\end{eqnarray}
 The mean value of a given function $f(x)$ with respect to the conditional
probability distribution defined in equation~(\ref{pp}) is then:\begin{eqnarray*}
\left\langle f(x)|h\to0,g_{1}\right\rangle  & \simeq & \mathcal{Z}^{-1}{\rm e}^{-n\left(\Phi_{0}-mh\right)}\int\!\!{\rm d}x\,\exp\left\{ -\frac{n}{2}\left[g_{1}^{-1}\!-\!\left(1\!-\! m^{2}\right)\!\left(1\!-\!2\xi\left(m,g_{1}\right)\, mh\right)\right]\!\left(x\!-\! x_{h}\right)^{2}\right\} \\
 &  & \qquad\qquad\qquad\qquad\qquad\left[f\left(x_{h}\right)+\left(x-x_{h}\right)f^{\prime}\left(x_{h}\right)+\frac{1}{2}\left(x-x_{h}\right)^{2}f^{\prime\prime}\left(x_{h}\right)\right]\\
 &  & +\mathcal{Z}^{-1}{\rm e}^{-n\left(\Phi_{0}+mh\right)}\int\!\!{\rm d}x\,\exp\left\{ -\frac{n}{2}\left[g_{1}^{-1}\!-\!\left(1\!-\! m^{2}\right)\!\left(1\!+\!2\xi\left(m,g_{1}\right)\, mh\right)\right]\!\left(x\!-\! x_{-h}\right)^{2}\right\} \\
 &  & \qquad\qquad\qquad\qquad\qquad\left[f\left(x_{-h}\right)+\left(x-x_{-h}\right)f^{\prime}\left(x_{-h}\right)+\frac{1}{2}\left(x-x_{-h}\right)^{2}f^{\prime\prime}\left(x_{-h}\right)\right]\,,\end{eqnarray*}
 which implies, considering that the integrals of the linear terms
are zero and keeping only the leading terms in the expansions, that
the expectation values takes the form: \begin{eqnarray*}
\left\langle f(x)|h\to0,g_{1}\right\rangle  & \simeq & \left[1-{\rm e}^{-2nmh}\left(1+2\xi^{2}\left(m,g_{1}\right)\, mh\right)\right]\left\{ f\left(x_{h}\right)+\frac{g_{1}}{2n}\xi\left(m,g_{1}\right)\, f^{\prime\prime}\left(x_{h}\right)\right\} \\
 &  & +{\rm e}^{-2nmh}\left(1+2\xi^{2}\left(m,g_{1}\right)\, mh\right)f\left(x_{-h}\right)\,.\end{eqnarray*}
 Considering the expansion of $f\left(x_{\pm h}\right)\simeq f\left(\pm g_{1}m+\xi\left(m,g_{1}\right)\, h\right)\simeq f\left(\pm g_{1}m\right)+\xi\left(m,g_{1}\right)\, f^{\prime}\left(\pm g_{1}m\right)\, h$
and disregarding terms of $\mathcal{O}\left(h{\rm e}^{-2nmh}\right)$,
one can write: \begin{equation}
\left\langle f(x)|h\!\to\!0,g_{1}\right\rangle \simeq f\left(mg_{1}\right)\!+\!\frac{g_{1}}{2n}\xi\left(m,g_{1}\right)f^{\prime\prime}\left(mg_{1}\right)\!-\!{\rm e}^{-2nmh}\!\left[f\left(mg_{1}\right)\!-\! f\left(\!-\! mg_{1}\right)\right]\!+\! f^{\prime}\left(mg_{1}\right)\!\xi\left(m,g_{1}\right)h.\label{eq:meanrs}\end{equation}
 The resulting one and two variable expectation values become \begin{eqnarray*}
\left\langle b_{k}^{\textrm{a}}|h_{\mu k}^{t}\to0,g_{\mu k}^{t}\right\rangle  & = & \sum_{\left\{ \mathbf{b}_{k}\right\} }P^{t}\left(\mathbf{b}_{k}|\left\{ y_{\nu\neq\mu}\right\} \right)b_{k}^{\textrm{a}}=\left\langle \tanh(x)|h_{\mu k}^{t}\to0,g_{1\mu k}^{t}\right\rangle \\
 & \simeq & \left[1-\frac{g_{1\mu k}^{t}}{n}\left[1-\left(m_{\mu k}^{t}\right)^{2}\right]\xi\left(m_{\mu k}^{t},g_{1\mu k}^{t}\right)-2{\rm e}^{-2nm_{\mu k}^{t}h_{\mu k}^{t}}\right]m_{\mu k}^{t}\\
 &  & \qquad\qquad+\xi\left(m_{\mu k}^{t},g_{1\mu k}^{t}\right)\,\left[1-\left(m_{\mu k}^{t}\right)^{2}\right]h_{\mu k}^{t}\end{eqnarray*}

and\begin{eqnarray*}
\left\langle b_{k}^{\textrm{a}}b_{k}^{\textrm{b}}|h_{\mu k}^{t}\to0,g_{1\mu k}^{t}\right\rangle  & = & P^{t}\left(\mathbf{b}_{k}|\left\{ y_{\nu\neq\mu}\right\} \right)b_{k}^{\textrm{a}}b_{k}^{\textrm{b}}=\delta^{\textrm{ab}}+\left(1-\delta^{\textrm{ab}}\right)\left\langle \tanh^{2}(x)|h_{\mu k}^{t}\to0,g_{1\mu k}^{t}\right\rangle \,,\end{eqnarray*}
 where \[
\left\langle \tanh^{2}(x)|h_{\mu k}^{t}\to0,g_{1\mu k}^{t}\right\rangle =\left(m_{\mu k}^{t}\right)^{2}+\xi\left(m_{\mu k}^{t},g_{1\mu k}^{t}\right)\left[1-\left(m_{\mu k}^{t}\right)^{2}\right]\,\left\{ \frac{g_{1\mu k}^{t}}{n}\,\left[1-3\left(m_{\mu k}^{t}\right)^{2}\right]+2m_{\mu k}^{t}h_{\mu k}^{t}\right\} \,,\]
 and \[
\left\langle b_{k}^{\textrm{a}}b_{l}^{\textrm{b}}|h_{\mu k}^{t}\to0,g_{1\mu k}^{t}\right\rangle =\left\langle b_{k}^{\textrm{a}}|h_{\mu k}^{t}\to0,g_{1\mu k}^{t}\right\rangle \left\langle b_{l}^{\textrm{b}}|h_{\mu k}^{t}\to0,g_{1\mu k}^{t}\right\rangle .\]

Thus, the leading terms for the covariance matrix of the replicated
variables are: \begin{eqnarray*}
\left(\mathbf{{\Psi}}_{\mu kl}^{t}\right)^{\mathrm{ab}} & \equiv & \left\langle b_{k}^{\textrm{a}}b_{l}^{\textrm{b}}|h_{\mu k}^{t}\to0,g_{1\mu k}^{t};h_{\mu l}^{t}\to0,g_{1\mu l}^{t}\right\rangle -\left\langle b_{k}^{\textrm{a}}|h_{\mu k}^{t}\to0,g_{1\mu k}^{t}\right\rangle \left\langle b_{l}^{\textrm{b}}|h_{\mu l}^{t}\to0,g_{1\mu l}^{t}\right\rangle =\delta_{kl}\left(\mathbf{{\Psi}}_{\mu k}^{t}\right)^{\mathrm{ab}}\\
\left(\mathbf{{\Psi}}_{\mu k}^{t}\right)^{\mathrm{ab}} & \simeq & \delta^{{\rm ab}}\left[1-\left(m_{\mu k}^{t}\right)^{2}\right]\\
 &  & +\left(1-\delta^{{\rm ab}}\right)\left\{ \frac{g_{1\mu k}^{t}}{n}\xi\left(m_{\mu k}^{t},g_{1\mu k}^{t}\right)\left[1-\left(m_{\mu k}^{t}\right)^{2}\right]^{2}+4{\rm e}^{-2nm_{\mu k}^{t}h_{\mu k}^{t}}\left(1-{\rm e}^{-2nm_{\mu k}^{t}h_{\mu k}^{t}}\right)\left(m_{\mu k}^{t}\right)^{2}\right\} \,.\end{eqnarray*}
 If one requires the non-diagonal elements of this covariance matrix
to have the same scaling as the inter-replica interaction matrix,
the field has to behave in such a way that the exponential term contributes
at most in $\mathcal{O}\left(n^{-1}\right).$ One thus expects the
field to obey $m_{\mu k}^{t}h_{\mu k}^{t}<{\displaystyle \frac{1}{n}}\ln\left|{\displaystyle \frac{2n}{n_{\mu k}^{t}}}\right|$,
where the $n_{\mu k}^{t}$ are appropriate constants. With this asymptotic
behaviour, the expression for the entries in the covariance matrix
is \[
\left(\mathbf{{\Psi}}_{\mu k}^{t}\right)^{\mathrm{ab}}\simeq\delta^{{\rm ab}}\left[1-\left(m_{\mu k}^{t}\right)^{2}\right]+\left(1-\delta^{{\rm ab}}\right)\,\frac{g_{1\mu k}^{t}\xi\left(m_{\mu k}^{t},g_{1\mu k}^{t}\right)}{n}\,\left[1-\left(m_{\mu k}^{t}\right)^{2}\right]^{2}\,,\]
 which serves to define the probability distribution for the macroscopic
variable $\Delta_{\mu k}^{\mathrm{a}}=\sum_{l\neq k}\varepsilon_{\mu l}b_{l}^{\mathrm{a}}$.

As $\varepsilon_{\mu k}$ and $b_{k}^{\mathrm{a}}$ are unbiased variables,
the variable $\Delta_{\mu k}^{\mathrm{a}}$, by virtue of the central
limit theorem, obeys a normal distribution, with mean value and covariance
matrix given by (to highest order) \begin{eqnarray}
\left(\mathbf{u}_{\mu k}^{t}\right)^{\mathrm{a}}\equiv\left\langle \Delta_{\mu k}^{\mathrm{a}}\right\rangle  & = & \sum_{\left\{ \mathbf{b}_{l\neq k}\right\} }\prod_{l\neq k}P^{t}\left(\mathbf{b}_{l}|\left\{ y_{\nu\neq\mu}\right\} \right)\sum_{l\neq k}\varepsilon_{\mu l}b_{l}^{\mathrm{a}}=\sum_{l\neq k}\varepsilon_{\mu l}m_{\mu l}^{t}\label{eq:urs}\\
\left(\mathbf{{\Upsilon}}_{\mu k}^{t}\right)^{\mathrm{ab}}\equiv\left\langle \Delta_{\mu k}^{\textrm{a}}\Delta_{\mu k}^{\textrm{b}}\right\rangle -\left\langle \Delta_{k}^{\textrm{a}\phantom{b}}\right\rangle \left\langle \Delta_{k}^{\textrm{b}}\right\rangle  & = & \sum_{\left\{ \mathbf{b}_{l\neq k}\right\} }\prod_{l\neq k}P^{t}\left(\mathbf{b}_{l}|\left\{ y_{\nu\neq\mu}\right\} \right)\sum_{\substack{l\neq k\\
j\neq k}
}\varepsilon_{\mu l}\varepsilon_{\mu j}b_{l}^{\mathrm{a}}b_{j}^{\mathrm{b}}-\left(\sum_{l\neq k}\varepsilon_{\mu l}m_{\mu l}^{t}\right)^{2}\nonumber \\
 & = & \sum_{l\neq k}\varepsilon_{\mu l}^{2}\left(\mathbf{{\Psi}}_{\mu lj}^{t}\right)^{\mathrm{ab}}=\delta^{\mathrm{ab}}X_{\mu k}+\left(1-\delta^{\mathrm{ab}}\right)\frac{1}{n}R_{\mu k}^{t},\nonumber \end{eqnarray}
 where \begin{eqnarray}
X_{\mu k}^{t} & \equiv & \sum_{l\neq k}\varepsilon_{\mu l}^{2}\,\left[1-\left(m_{\mu l}^{t}\right)^{2}\right]\mbox{~~~and}\label{eq:xxx}\\
R_{\mu k}^{t} & \equiv & \sum_{l\neq k}\varepsilon_{\mu l}^{2}\, g_{1\mu l}^{t}\,\xi\left(m_{\mu l}^{t},g_{1\mu l}^{t}\right)\left[1-\left(m_{\mu l}^{t}\right)^{2}\right]^{2}\,,\nonumber \end{eqnarray}
 are macroscopic variables of $\mathcal{O}(1)$. In particular, $R_{\mu k}^{t}$
is a free variable that can be used later on to optimise a given performance
measure. This variables have the property of being self-averaging,
therefore we can drop the sub-indices $\mu$ and \emph{k. }

\section{The One Step Replica Symmetry Breaking (1RSB) Ansatz \label{app:RSB}}

%DS - is this consistent with the result discussed in the
%DS - paragraph below (19) (page 18)????
Under a solution correlation matrix that resembles the 1RSB structure,
the system comprises $nL$ variables, where both the number of blocks
\emph{L} and the number of variables per block \emph{n} are considered
large. As before we are interested in the regime where $L$ and $n\to\infty.$

With this setting, the interaction term in equation~(\ref{eq:ansatz})
is now:\[
\mathbf{b}_{k}^{\sf T}\mathbf{Q}_{\mu k}^{t}\mathbf{b}_{k}=-q_{1\mu k}^{t}nL+\left(q_{1\mu k}^{t}-q_{2\mu k}^{t}\right)\sum_{\ell=1}^{L}\left(\sum_{{\rm a}=1}^{n}b_{k}^{\ell{\rm a}}\right)^{2}+q_{2\mu k}^{t}\left(\sum_{\ell=1}^{L}\sum_{{\rm a}=1}^{n}b_{k}^{\ell{\rm a}}\right)^{2}\,,\]
 thus we have now $L+1$ squared sums in the exponent that can be
replaced by integrals:\begin{eqnarray*}
P^{t}\left(\mathbf{b}_{k}|\left\{ y_{\nu\neq\mu}\right\} \right) & = & [\mathcal{Z}_{\mu k}^{t}]^{-1}{\displaystyle \int\mathrm{d}\mathbold{x}}\,\exp\left\{ -\frac{x_{0}^{2}}{2q_{2\mu k}^{t}}-\sum_{\ell=1}^{L}\frac{x_{\ell}^{2}}{2\upDelta q_{\mu k}^{t}}+\sum_{\ell=1}^{L}\left(x_{0}+x_{\ell}+h_{\mu k}^{t}\right)\sum_{\mathrm{a}=1}^{n}b_{k}^{\ell\mathrm{a}}\right\} \,,\end{eqnarray*}
 where $\upDelta q_{\mu k}^{t}\equiv q_{1\mu k}^{t}-q_{2\mu k}^{t}>0$
and $\mathbold x^{\sf T}=\left(x_{0},x_{1},\dots,x_{L}\right)$. Also
here we expect the logarithm of the normalisation term (linked to
the free energy) obtained from the well behaved distribution $P^{t}$
to be self-averaging, thus:\[
\lim_{n\to\infty}\lim_{L\to\infty}\frac{1}{nL}\log\left(\overline{\mathcal{Z}_{\mu k}^{t}}\right)=\lim_{n\to\infty}\lim_{L\to\infty}\frac{1}{nL}\log\left(\mathcal{Z}_{\mu k}^{t}\left(h_{\mu k}^{t},q_{1\mu k}^{t},q_{2\mu k}^{t}\right)\right),\]
 which is satisfied if the entries behave like $q_{2\mu k}^{t}\sim g_{2\mu k}^{t}/nL$
and $\upDelta q_{\mu k}^{t}\sim g_{1\mu k}^{t}/n,$ where $g_{1\mu k}^{t}$
and $g_{2\mu k}^{t}\sim\mathcal{O}(1)$. Using this new scaled parameters,
the expression for the normalisation is $\mathcal{Z}_{\mu k}^{t}=\int{\rm d}\mathbf{x}\,\exp\left\{ -nL\Phi\left(\mathbf{x};h_{\mu k}^{t},g_{1\mu k}^{t},g_{2\mu k}^{t}\right)\right\} $
where\[
\Phi\left(\mathbf{x};h_{\mu k}^{t},g_{1\mu k}^{t},g_{2\mu k}^{t}\right)\equiv\frac{x_{0}^{2}}{2g_{1\mu k}^{t}}+\frac{1}{L}\sum_{\ell=1}^{L}\frac{x_{\ell}^{2}}{2g_{2\mu k}^{t}}-\frac{1}{L}\sum_{\ell=1}^{L}\log\left[2\cosh\left(x_{0}+x_{\ell}+h_{\mu k}^{t}\right)\right]\,.\]

As before, we drop the indexes $\mu$, \emph{k}, and \emph{t} for
brevity. The critical points of the function $\Phi\left(\mathbf{x};h,g_{1},g_{2}\right)$
satisfy the following set of equations: \begin{eqnarray*}
\frac{\partial\Phi}{\partial x_{0}} & = & \frac{x_{0}}{g_{1}}-\frac{1}{L}\sum_{\ell=1}^{L}\tanh\left(x_{0}+x_{\ell}+h\right)=0\\
\frac{\partial\Phi}{\partial x_{\ell}} & = & \frac{1}{L}\left(\frac{x_{\ell}}{g_{2}}-\tanh\left(x_{0}+x_{\ell}+h\right)\right)=0\,,\end{eqnarray*}
 which are satisfied for the following values: \begin{eqnarray}
x_{0}^{*} & = & \frac{g_{1}}{g_{2}}\,\frac{1}{L}\sum_{\ell=1}^{L}x_{\ell}^{*}=\frac{g_{1}}{g_{2}}x^{*}\nonumber \\
\frac{x_{\ell}^{*}}{g_{2}} & = & \tanh\left(x_{\ell}^{*}+\frac{g_{1}}{g_{2}}x^{*}+h\right)\,,\label{eq:set}\end{eqnarray}
 where $x^{*}\equiv\frac{1}{L}\sum_{\ell=1}^{L}x_{\ell}^{*}$. The
second equation in the set, equation~(\ref{eq:set}), has the same
form for all $\ell=1,\dots,L$ and in the small field regime it has
at most three different solutions. From the three possible solutions,
one is a local maximum; of the other two, the one that has the same
sign as \emph{h} is dominant. Thus we can expect, for all $\ell$,
$x_{\ell}^{*}=x^{*}$. This reduces the set of $L+1$ equations to
one \[
\frac{x^{*}}{g_{2}}=\tanh\left(\frac{G}{g_{2}}x^{*}+h\right)\,,\]
 where $G\equiv g_{1}+g_{2}$. With the substitution $u=\left(G/g_{2}\right)x^{*}$
the equation has the same form as equation~(\ref{eq:derivada}),
i.e. $u=G\,\tanh\left(u+h\right).$ If one considers again the field
\emph{h} to be small, the solutions can be expressed as an expansion
of the zero field solutions $u_{\pm h}\simeq\pm Gm+\xi(m,G)h$, where
$\xi(m,G)$ is given by equation~(\ref{eq:xih}), and ${\rm sgn}(m)={\rm sgn}(h).$
Using these expansions the critical values are given by: $x_{0,\pm h}^{*}\simeq g_{1}\left[\pm m+G^{-1}\xi(m,G)h\right]$
and $x_{\ell,\pm h}^{*}\simeq g_{2}\left[\pm m+G^{-1}\xi(m,G)h\right]$
for all $\ell=1,\dots,L.$

As in the RS case, the expansion of $\Phi$ around the critical points
in the small field regime is $\Phi\left(\mathbf{x}_{\pm h}^{*};h\to0,g_{1},g_{2}\right)\simeq\Phi\left(\mathbf{x}_{0}^{*};0,g_{1},g_{2}\right)\mp mh=\Phi_{0}\mp mh$.
So the dominant solution is the one that shares the sign with the
field.

For a sufficiently large system with $nL$ variables, one expects
the following expansion to be valid:{\small \begin{eqnarray}
\exp\left\{ -nL\Phi\left(\mathbf{x};h\to0,g_{1},g_{2}\right)\right\}  & \simeq & {\rm e}^{-nL\Phi_{0}}\left\{ {\rm e}^{nLmh}\exp\left[-\frac{nL}{2}\left(\mathbf{x}-\mathbf{x}_{h}^{*}\right)^{\sf T}\mathbf{H}_{\Phi,h}\left(\mathbf{x}-\mathbf{x}_{h}^{*}\right)\right]\right.\nonumber \\
 &  & \left.\qquad\quad+{\rm e}^{-nLmh}\exp\left[-\frac{nL}{2}\left(\mathbf{x}\!-\!\mathbf{x}_{-h}^{*}\right)^{\sf T}\mathbf{H}_{\Phi,-h}\left(\mathbf{x}\!-\!\mathbf{x}_{-h}^{*}\right)\right]\right\} \,,\label{eq:expmultivar}\end{eqnarray}}
 where $\mathbf{H}_{\Phi,\pm h}$ is the Hessian of $\Phi$ in $\mathbf{x}_{\pm h}^{*}$.

Defining $\beta_{\pm h}\equiv\left(1-m^{2}\right)\left\{ 1\mp2\left[\xi(m,G)+1\right]\, mh\right\} $,
the entries of the Hessian become \begin{eqnarray*}
\left.\frac{\partial^{2}\Phi}{\partial x_{0}^{2}}\right|_{\mathbf{x}_{\pm h}^{*}} & \simeq & g_{1}^{-1}-\beta_{\pm h}\equiv\alpha_{\pm h}\\
\left.\frac{\partial^{2}\Phi}{\partial x_{\ell}^{2}}\right|_{\mathbf{x}_{\pm h}^{*}} & \simeq & \frac{1}{L}\left(g_{2}^{-1}-\beta_{\pm h}\right)\equiv\frac{1}{L}\gamma_{\pm h}\\
\left.\frac{\partial^{2}\Phi}{\partial x_{0}\partial x_{\ell}}\right|_{\mathbf{x}_{\pm h}^{*}} & \simeq & -\frac{1}{L}\beta_{\pm h}\\
\left.\frac{\partial^{2}\Phi}{\partial x_{\ell}\partial x_{\ell^{\prime}}}\right|_{\mathbf{x}_{\pm h}^{*}} & = & 0~.\end{eqnarray*}
 The corresponding characteristic equation is: \[
\det\left(\mathbf{H}_{\Phi,\pm h}-\lambda\openone\right)=\left(\frac{1}{L}\gamma_{\pm h}-\lambda\right)^{L-1}\left\{ \left(\alpha_{\pm h}-\lambda\right)\left(\frac{1}{L}\gamma_{\pm h}-\lambda\right)-\frac{1}{L}\beta_{\pm h}^{2}\right\} =0\,.\]
 The solutions for this equation, disregarding terms of $\mathcal{O}\left(L^{-2}\right)$
and $\mathcal{O}\left(hL^{-1}\right)$, are: \begin{eqnarray}
\lambda_{0,\pm h} & = & \alpha_{\pm h}+\frac{1}{L}\frac{\beta_{\pm h}^{2}}{\alpha_{\pm h}}\nonumber \\
 & \simeq & \lambda_{0}\left\{ 1\pm2\left[\xi(m,g_{1})-1\right]\left[\xi(m,G)+1\right]\left(1-m^{2}\right)\, mh\right\}\nonumber \\
\lambda_{1,\pm h} & = & \frac{1}{L}\left(\gamma_{\pm h}-\frac{\beta_{\pm h}^{2}}{\alpha_{\pm h}}\right)\label{eq:eigenvalues}\\
 & \simeq & \lambda_{1}\left\{ 1\pm2\frac{\xi(m,g_{1})^{2}\left[\xi(m,g_{2})-1\right]}{1-\left[\xi(m,g_{1})-1\right]\left[\xi(m,g_{2})-1\right]}\left[\xi(m,G)+1\right]\left(1-m^{2}\right)\, mh\right\}\nonumber \\
\lambda_{\ell,\pm h} & = & \frac{1}{L}\gamma_{\pm h}\nonumber \\
 & \simeq & \lambda_{\ell}\left\{ 1\pm2\left[\xi(m,g_{2})-1\right]\left[\xi(m,G)+1\right]\left(1-m^{2}\right)\, mh\right\} \quad\forall\,\ell=2,\dots,L\,,\nonumber\end{eqnarray}
 where $\lambda_{0}\equiv\alpha_{0}+\frac{1}{L}\,\frac{\beta_{0}^{2}}{\alpha_{0}}$,
$\lambda_{1}\equiv\frac{1}{L}\left(\gamma_{0}-\frac{\beta_{0}^{2}}{\alpha_{0}}\right)$
and $\lambda_{\ell}\equiv\frac{1}{L}\gamma_{0}$. The corresponding
eigenvectors, up to order $L^{-1}$, are: \begin{eqnarray}
\mathbf{u}_{0,\pm h} & = & \left(1,\stackrel{L\;{\rm times}}{\overbrace{-\frac{1}{L}\,\frac{\beta_{\pm h}}{\alpha_{\pm h}},-\frac{1}{L}\,\frac{\beta_{\pm h}}{\alpha_{\pm h}},\dots,-\frac{1}{L}\,\frac{\beta_{\pm h}}{\alpha_{\pm h}}}}\right)^{\sf T}\nonumber \\
\mathbf{u}_{1,\pm h} & = & \frac{1}{\sqrt{L}}\left(\frac{\beta_{\pm h}}{\alpha_{\pm h}},\stackrel{L\;{\rm times}}{\overbrace{1,1,\dots,1}}\right)^{\sf T}\label{eq:eigenvectors}\\
\mathbf{u}_{\ell,\pm h} & = & \frac{1}{\sqrt{\ell\left(\ell-1\right)}}\left(0,\stackrel{\ell-1\;{\rm times}}{\overbrace{1,1,\dots,1}},-\left(\ell-1\right),\stackrel{L-\ell\;{\rm times}}{\overbrace{0,0,\dots,0}}\right)^{\sf T}\quad\forall\,\ell=2,\dots,L\,.\nonumber \end{eqnarray}
 These vectors satisfy the normalisation condition $\mathbf{u}_{\ell,\pm h}^{\sf T}\mathbf{u}_{\ell^{\prime},\pm h}=\delta^{\ell\ell^{\prime}}\left[1+\mathcal{O}\left(L^{-1}\right)\right]\;\;\forall\ell,\ell^{\prime}=0,1,\dots,L.$
The linear transformation from the canonical basis to the basis of
eigenvectors is then represented by a matrix with the entries \begin{eqnarray}
\left(\mathbf{U}_{\pm h}\right)_{ij}\simeq\left(\mathbf{U}_{0}\right)_{ij} & = & \delta_{0i}\delta_{0j}+\frac{1}{\sqrt{j(j-1)}}\left[\sum_{k=1}^{j-1}\delta_{ki}-\left(1-\delta_{0j}\right)\left(1-\delta_{1j}\right)\delta_{ij}(j-1)\right]\nonumber \\
 &  & +\frac{1}{\sqrt{L}}\,\delta_{1j}\left[\delta_{0i}\,\frac{\beta_{0}}{\alpha_{0}}+\left(1-\delta_{0i}\right)\right]-\frac{1}{L}\,\delta_{0j}\left(1-\delta_{0i}\right)\,\frac{\beta_{0}}{\alpha_{0}}\,,\label{eq:rot}\end{eqnarray}
 ignoring terms of $\mathcal{O}\left(hL^{-1/2}\right)$. Because this
transformation is a rigid rotation, the following properties are satisfied:
$\left|\det\left(\mathbf{U}_{\pm}\right)\right|=1$ and $\mathbf{U}_{\pm h}^{\sf T}\mathbf{U}_{\pm h}=\mathbf{U}_{\pm h}\mathbf{U}_{\pm h}^{\sf T}=\openone.$

Second order terms in equation~(\ref{eq:expmultivar}) can be re-written
using the diagonal representation of the Hessian. Therefore, keeping
only terms of order $\mathcal{O}\left(L^{-1}\right)$ we have that:
$\left(\mathbf{x}-\mathbf{x}_{\pm h}\right)^{\sf T}\mathbf{H}_{\Phi,\pm h}\left(\mathbf{x}-\mathbf{x}_{\pm h}\right)=\left(\mathbf{x}-\mathbf{x}_{\pm h}\right)^{\sf T}\mathbf{U}_{0}\mathbf{U}_{0}^{\sf T}\mathbf{H}_{\Phi,\pm h}\mathbf{U}_{0}\mathbf{U}_{0}^{\sf T}\left(\mathbf{x}-\mathbf{x}_{\pm h}\right)=\left(\mathbf{y}-\mathbf{y}_{\pm h}\right)^{\sf T}\mathbf{H}_{\Phi,\pm h}^{\prime}\left(\mathbf{y}-\mathbf{y}_{\pm h}\right)$,
where $\mathbf{y}\equiv\mathbf{U}_{0}^{\sf T}\mathbf{x}$ and $\mathbf{H}_{\Phi,\pm h}^{\prime}\equiv\mathbf{U}_{0}^{\sf T}\mathbf{H}_{\Phi,\pm h}\mathbf{U}_{0}$
is the diagonal representation of the Hessian, i.e. $\left(\mathbf{H}_{\Phi,\pm h}^{\prime}\right)_{ij}=\delta_{ij}\lambda_{i,\pm h}$.
Using the diagonal representation in conjunction with equation~(\ref{eq:expmultivar})
one obtains an expression for the normalisation term \begin{eqnarray*}
\mathcal{Z}\left(h\to0,g_{1},g_{2}\right) & \simeq & {\rm e}^{-nL\left(\Phi_{0}-mh\right)}\int{\rm d}\mathbf{y}\,\exp\left[-\frac{nL}{2}\left(\mathbf{y}-\mathbf{y}_{h}^{*}\right)^{\sf T}\mathbf{H}_{\Phi,h}^{\prime}\left(\mathbf{y}-\mathbf{y}_{h}^{*}\right)\right]\\
 &  & +{\rm e}^{-nL\left(\Phi_{0}+mh\right)}\int{\rm d}\mathbf{y}\,\exp\left[-\frac{nL}{2}\left(\mathbf{y}-\mathbf{y}_{-h}^{*}\right)^{\sf T}\mathbf{H}_{\Phi,-h}^{\prime}\left(\mathbf{y}-\mathbf{y}_{-h}^{*}\right)\right]\\
 & \simeq & {\rm e}^{-nL\Phi_{0}}\left(\frac{2\pi}{nL}\right)^{\frac{L+1}{2}}\left[{\rm e}^{nLmh}\prod_{\ell=0}^{L}\lambda_{\ell,h}^{-\frac{1}{2}}+{\rm e}^{-nLmh}\prod_{\ell=0}^{L}\lambda_{\ell,-h}^{-\frac{1}{2}}\right]\,.\end{eqnarray*}
 For a small field, the product of the eigenvalues can be approximated
by \begin{eqnarray*}
\prod_{\ell=0}^{L}\lambda_{\ell,\pm h}^{-\frac{1}{2}} & \simeq & \left\{ 1\mp\left[\xi(m,g_{2})-1\right]\left[\xi(m,G)+1\right]Lmh\right\} \prod_{\ell=0}^{L}\lambda_{\ell}^{-\frac{1}{2}}\,.\end{eqnarray*}
 Thus, the expression for $\mathcal{Z}$ reduces to \begin{eqnarray*}
\mathcal{Z}\left(h\to0,g_{1},g_{2}\right) & \simeq & {\rm e}^{-nL\Phi_{0}}\left(\frac{2\pi}{nL}\right)^{\frac{L+1}{2}}\prod_{\ell=0}^{L}\lambda_{\ell}^{-\frac{1}{2}}\left\{ {\rm e}^{nLmh}\left\{ 1-\left[\xi(m,g_{2})-1\right]\left[\xi(m,G)+1\right]Lmh\right\} \right.\\
 &  & \qquad\qquad\qquad+\left.{\rm e}^{-nLmh}\left\{ 1+\left[\xi(m,g_{2})-1\right]\left[\xi(m,G)+1\right]Lmh\right\} \right\} \,.\end{eqnarray*}
 The mean value of a given function $f(\mathbf{x})$ is then given
by \begin{eqnarray*}
\left\langle f(\mathbf{x})|h\to0,g_{1},g_{2}\right\rangle  & \simeq & \mathcal{Z}^{-1}{\rm e}^{-nL\left(\Phi_{0}-mh\right)}\int{\rm d}\mathbf{y}\,\exp\left\{ -\frac{nL}{2}\left(\mathbf{y}-\mathbf{y}_{h}^{*}\right)^{\sf T}\mathbf{H}_{\Phi,h}^{\prime}\left(\mathbf{y}-\mathbf{y}_{h}^{*}\right)\right\} \\
 &  & \qquad\qquad\qquad\qquad\qquad\left[f\left(\mathbf{x}_{h}\right)+\frac{1}{2}\left(\mathbf{y}-\mathbf{y}_{h}^{*}\right)^{\sf T}\mathbf{H}_{f,h}^{\prime}\left(\mathbf{y}-\mathbf{y}_{h}^{*}\right)\right]\\
 &  & +\mathcal{Z}^{-1}{\rm e}^{-nL\left(\Phi_{0}+mh\right)}\int{\rm d}\mathbf{y}\,\exp\left\{ -\frac{nL}{2}\left(\mathbf{y}-\mathbf{y}_{-h}^{*}\right)^{\sf T}\mathbf{H}_{\Phi,-h}^{\prime}\left(\mathbf{y}-\mathbf{y}_{-h}^{*}\right)\right\} \\
 &  & \qquad\qquad\qquad\qquad\qquad\left[f\left(\mathbf{x}_{-h}\right)+\frac{1}{2}\left(\mathbf{y}-\mathbf{y}_{-h}^{*}\right)^{\sf T}\mathbf{H}_{f,-h}^{\prime}\left(\mathbf{y}-\mathbf{y}_{-h}^{*}\right)\right]\,,\end{eqnarray*}
 where $\mathbf{H}_{f,\pm h}^{\prime}$ is the Hessian of the function
$f$ in the basis of eigenvectors of $\mathbf{H}_{\Phi\pm h}$, evaluated
at the critical points. The linear terms in the expansion of $f\left(\mathbf{x}\right)$
do not contribute to the expectation value. The Gaussian integral
of the cross products of the type $\left(y_{i}-y_{i,\pm h}^{*}\right)\left(y_{j}-y_{j,\pm h}^{*}\right)$
with $i\neq j$ are zero, thus the Gaussian integral of the second
term in the expansion of $f\left(\mathbf{x}\right)$ becomes: \begin{eqnarray}
I_{\pm} & = & \frac{1}{2}\mathcal{Z}^{-1}{\rm e}^{-nL\left(\Phi_{0}\mp mh\right)}\int{\rm d}\mathbf{y}\,\exp\left[-\frac{nL}{2}\left(\mathbf{y}-\mathbf{y}_{\pm h}^{*}\right)^{\sf T}\mathbf{H}_{\Phi,\pm h}^{\prime}\left(\mathbf{y}-\mathbf{y}_{\pm h}^{*}\right)\right]\left(\mathbf{y}-\mathbf{y}_{\pm h}^{*}\right)^{\sf T}\mathbf{H}_{f,\pm h}^{\prime}\left(\mathbf{y}-\mathbf{y}_{\pm h}^{*}\right)\nonumber \\
I_{+} & \simeq & \frac{1}{2}\,\left\{ 1-{\rm e}^{-2nLmh}\left\{ 1+2\left[\xi(m,g_{2})-1\right]\left[\xi(m,G)+1\right]Lmh\right\} \right\} \;\frac{1}{nL}\,\sum_{\ell=0}^{L}\lambda_{\ell,h}^{-1}\left(\mathbf{H}_{f,h}^{\prime}\right)_{\ell\ell}\nonumber \\
I_{-} & \simeq & \frac{1}{2}\,{\rm e}^{-2nLmh}\left\{ 1+2\left[\xi(m,g_{2})-1\right]\left[\xi(m,G)+1\right]Lmh\right\} \;\frac{1}{nL}\,\sum_{\ell=0}^{L}\lambda_{\ell,-h}^{-1}\left(\mathbf{H}_{f,-h}^{\prime}\right)_{\ell\ell}\,.\label{b6}\end{eqnarray}
 Using the expansion $f\left(\mathbf{x_{\pm}}\right)\simeq f\left(\pm\mathbf{x}_{0}+h\,\mathbold{\xi}(m,G)\right)\simeq f\left(\pm\mathbf{x}_{0}\right)+h\,\mathbold\xi^{\sf T}(m,G)\nabla f\left(\pm\mathbf{x}_{0}\right)=f\left(\pm\mathbf{x}_{0}\right)+\delta f\left(\pm\mathbf{x}_{0}\right)\, h$
where $\mathbf{x}_{0}^{\sf T}=m\left(g_{1},\stackrel{L\;{\rm times}}{\overbrace{g_{2},g_{2},\dots,g_{2}}}\right)$
and $\mathbold\xi^{\sf T}(m,G)=G^{-1}\xi(m,G)\left(g_{1},\stackrel{L\;{\rm times}}{\overbrace{g_{2},g_{2},\dots,g_{2}}}\right)$,
the diagonal entries of the transformed Hessian are: \begin{eqnarray}
\left(\mathbf{H}_{f,\pm h}^{\prime}\right)_{\ell\ell} & = & \sum_{i,j=0}^{L}\left(\mathbf{U}_{\pm h}\right)_{\ell i}\left(\mathbf{U}_{\pm h}\right)_{\ell j}\left(\mathbf{H}_{f,\pm h}\right)_{ij}\nonumber \\
 & = & \sum_{i,j=0}^{L}\left(\mathbf{U}_{\pm h}\right)_{\ell i}\left(\mathbf{U}_{\pm h}\right)_{\ell j}\,\left.\frac{\partial^{2}f}{\partial x_{i}\partial x_{j}}\right|_{\mathbf{x}_{\pm}}\nonumber \\
 & \simeq & \sum_{i,j=0}^{L}\left(\mathbf{U}_{0}\right)_{\ell i}\left(\mathbf{U}_{0}\right)_{\ell j}\,\left(\left.\frac{\partial^{2}f}{\partial x_{i}\partial x_{j}}\right|_{\pm\mathbf{x}_{0}}+h\left.\mathbold\xi^{\sf T}(m,G)\nabla\frac{\partial^{2}f}{\partial x_{i}\partial x_{j}}\right|_{\pm\mathbf{x}_{0}}\right)\label{b7}\\
 & \simeq & \left(\left.\mathbf{H}_{f}^{\prime}\right|_{\pm\mathbf{x}_{0}}\right)_{\ell\ell}+\left(\left.\delta\mathbf{H}_{f}^{\prime}\right|_{\pm\mathbf{x}_{0}}\right)_{\ell\ell}h\,,\nonumber \end{eqnarray}
 with $\left(\left.\delta\mathbf{H}_{f}^{\prime}\right|_{\pm\mathbf{x}_{0}}\right)_{\ell\ell}$
defined by the second term in (\ref{b7}). Using the entries of the
diagonalised Hessian, the last term in the integrals (\ref{b6}) becomes
\begin{eqnarray*}
\frac{1}{nL}\,\sum_{\ell=0}^{L}\lambda_{\ell,-h}^{-1}\left(\mathbf{H}_{f,\pm h}^{\prime}\right)_{\ell\ell} & \simeq & \frac{1}{nL}\,\frac{1}{\alpha_{0}}\,\left(\left.\mathbf{H}_{f}^{\prime}\right|_{\pm\mathbf{x}_{0}}\right)_{00}+\frac{1}{n}\,\frac{\alpha_{0}}{\left(\alpha_{0}\gamma_{0}-\beta_{0}^{2}\right)}\,\left(\left.\mathbf{H}_{f}^{\prime}\right|_{\pm\mathbf{x}_{0}}\right)_{11}+\frac{1}{n}\,\frac{1}{\gamma_{0}}\,\sum_{\ell=2}^{L}\left(\left.\mathbf{H}_{f}^{\prime}\right|_{\pm\mathbf{x}_{0}}\right)_{\ell\ell}\,,\end{eqnarray*}
 disregarding terms of $\mathcal{O}\left(\frac{h}{n}+\frac{h}{L}\right).$
The expectation value of an arbitrary function $f$ can then be approximated
by \begin{eqnarray}
\left\langle f(\mathbf{x})|h\!\to\!0^{+},g_{1},g_{2}\right\rangle  & \simeq\! & f\left(\mathbf{x}_{0}\right)\!+\!\frac{1}{2}\frac{1}{nL}\sum_{\ell=0}^{L}\lambda_{\ell}^{-1}\left(\left.\mathbf{H}_{f}^{\prime}\right|_{\mathbf{x}_{0}}\right)_{\ell\ell}\!-\!{\rm e}^{-2nLmh}\!\left[f\left(\mathbf{x}_{0}\right)\!-\! f\left(-\!\mathbf{x}_{0}\right)\right]\!+\!\delta\! f\!\left(\mathbf{x}_{0}\right)\! h,\label{eq:mean1rsb}\end{eqnarray}
 where we have disregarded terms of $\mathcal{O}\left(\frac{h}{n}+\frac{h}{L}\right)$,$\mathcal{O}\left(\frac{1}{n}{\rm e}^{-2nLmh}\right)$
and $\mathcal{O}\left(Lh{\rm e}^{-2nLmh}\right)$. By simple inspection,
equation~(\ref{eq:mean1rsb}) is equivalent to the RS mean value
equation~(\ref{eq:meanrs}).

The single variable mean value is then: \begin{eqnarray*}
\left\langle b_{k}^{\ell\textrm{a}}|h_{\mu k}^{t}\to0,g_{1\mu k}^{t},g_{2\mu k}^{t}\right\rangle  & = & {\displaystyle \sum_{\left\{ \mathbf{b}_{k}\right\} }}P^{t}\left(\mathbf{b}_{k}|\left\{ y_{\nu\neq\mu}\right\} \right)b_{k}^{\ell\textrm{a}}=\left\langle \tanh\left(x_{0}+x_{\ell}+h_{\mu k}^{t}\right)|h_{\mu k}^{t}\to0,g_{1\mu k}^{t},g_{2\mu k}^{t}\right\rangle \,.\end{eqnarray*}
 The expansion for $f(\mathbf{x})=\tanh\left(x_{0}+x_{\ell}+h_{\mu k}^{t}\right)$
is \[
f(\mathbf{x})\simeq m_{\mu k}^{t}+\left[1-\left(m_{\mu k}^{t}\right)^{2}\right]\left(1,\stackrel{\ell-1\;{\rm times}}{\overbrace{0,0,\dots,0}},1,\stackrel{L-\ell\;{\rm times}}{\overbrace{0,0,\dots,0}}\right)^{\sf T}\mathbold{\xi}\left(m_{\mu k}^{t},G_{\mu k}^{t}\right)h_{\mu k}^{t}\,,\]
 which results in the following expression for the single variable
mean value \begin{eqnarray*}
\left\langle b_{k}^{\ell\textrm{a}}|h_{\mu k}^{t}\to0,g_{1\mu k}^{t},g_{2\mu k}^{t}\right\rangle  & \simeq & \left(1-2{\rm e}^{-2nLm_{\mu k}^{t}h_{\mu k}^{t}}\right)m_{\mu k}^{t}+\xi\left(m_{\mu k}^{t},G_{\mu k}^{t}\right)\,\left[1-\left(m_{\mu k}^{t}\right)^{2}\right]h_{\mu k}^{t}\\
 &  & -m_{\mu k}^{t}\left[1-\left(m_{\mu k}^{t}\right)^{2}\right]\frac{1}{nL}\sum_{k=0}^{L}\lambda_{k}^{-1}\left(\mathbf{M^{\prime}}_{0\ell}\right)_{kk}\,,\end{eqnarray*}
 where $\left(\mathbf{M}_{0\ell}\right)_{ij}=\delta_{0i}\delta_{0j}+\delta_{0i}\delta_{\ell j}+\delta_{\ell i}\delta_{0j}+\delta_{\ell i}\delta_{\ell j}$
is a matrix such that $\left.\mathbf{H}_{\tanh\left(x_{0}+x_{\ell}\right)}\right|_{\mathbf{x}_{0}}=-2m_{\mu k}^{t}\left[1-\left(m_{\mu k}^{t}\right)^{2}\right]\mathbf{M}_{0\ell}$.
In the basis of the $\mathbf{H}_{\Phi}$ eigenvalues, the expressions
for the diagonal elements of this matrix are \begin{eqnarray}
\left(\mathbf{M}_{0\ell}^{\prime}\right)_{kk} & = & \sum_{i,j=0}^{L}\left(\mathbf{U}_{\pm h}\right)_{ik}\left(\mathbf{U}_{\pm h}\right)_{jk}\left(\delta_{0i}\delta_{0j}+\delta_{0i}\delta_{\ell j}+\delta_{\ell i}\delta_{0j}+\delta_{\ell i}\delta_{\ell j}\right)\nonumber \\
 & = & \left(\left(\mathbf{U}_{\pm h}\right)_{0k}+\left(\mathbf{U}_{\pm h}\right)_{\ell k}\right)^{2}\nonumber \\
\left(\mathbf{M}_{0\ell}^{\prime}\right)_{00} & \simeq & 1-\frac{2}{L}\,\frac{\beta_{0}}{\alpha_{0}}\label{eq:diagm}\\
\left(\mathbf{M}_{0\ell}^{\prime}\right)_{11} & \simeq & \frac{1}{L}\,\left(\frac{\alpha_{0}+\beta_{0}}{\alpha_{0}}\right)^{2}\nonumber \\
\left(\mathbf{M}_{0\ell}^{\prime}\right)_{kk} & = & \delta_{k\ell}{\displaystyle \frac{\ell-1}{\ell}}+\Theta\left(k-\ell-1\right)\frac{1}{k\left(k-1\right)}\quad\forall\;\ell=2,\dots,L\,,\nonumber \end{eqnarray}
 where $\Theta(n)=1$ if $n>0$ and 0 otherwise. The sum of the eigenvalues'
inverse times the diagonal elements equation~(\ref{eq:diagm}) results
in \begin{eqnarray*}
\frac{1}{nL}\,\sum_{k=0}^{L}\lambda_{k}^{-1}\left(\mathbf{M}_{0\ell}^{\prime}\right)_{kk} & \simeq & \frac{1}{n}\,\frac{1}{\gamma_{0}}\,\sum_{k=2}^{L}\left[\delta_{k\ell}{\displaystyle \frac{\ell-1}{\ell}}+\Theta\left(k-\ell-1\right)\frac{1}{k\left(k-1\right)}\right]+\frac{1}{nL}\,\frac{1}{\alpha_{0}}\left[1+\frac{\left(\alpha_{0}+\beta_{0}\right)^{2}}{\alpha_{0}\gamma_{0}-\beta_{0}^{2}}\right]\\
 & = & \frac{1}{n}\,\frac{1}{\gamma_{0}}\,\left[\sum_{k=2}^{L}\frac{1}{k\left(k-1\right)}\right]+\frac{1}{nL}\,\frac{1}{\alpha_{0}}\left[1+\frac{\left(\alpha_{0}+\beta_{0}\right)^{2}}{\alpha_{0}\gamma-\beta_{0}^{2}}\right]\\
 & = & \frac{1}{n}\,\frac{1}{\gamma_{0}}+\frac{1}{nL}\,\frac{1}{\alpha_{0}}\left[1+\frac{\left(\alpha_{0}+\beta_{0}\right)^{2}}{\alpha_{0}\gamma_{0}-\beta_{0}^{2}}-\frac{\alpha_{0}}{\gamma_{0}}\right]\\
 & = & \frac{1}{n}\, g_{2\mu k}^{t}\xi\left(m_{\mu k}^{t},g_{2\mu k}^{t}\right)+\frac{1}{nL}\,\left[G_{\mu k}^{t}\xi\left(m_{\mu k}^{t},G_{\mu k}^{t}\right)-g_{2\mu k}^{t}\xi\left(m_{\mu k}^{t},g_{2\mu k}^{t}\right)\right]\end{eqnarray*}
 where we have used that $\sum_{k=2}^{L}\left[k\left(k-1\right)\right]^{-1}=(L-1)/L,$
$\gamma_{0}^{-1}=g_{2\mu k}^{t}\xi\left(m_{\mu k}^{t},g_{2\mu k}^{t}\right)$
and ${\displaystyle \frac{1}{\alpha_{0}}}\left[1+{\displaystyle \frac{\left(\alpha_{0}+\beta_{0}\right)^{2}}{\alpha_{0}\gamma_{0}-\beta_{0}^{2}}}-{\displaystyle \frac{\alpha_{0}}{\gamma_{0}}}\right]=G_{\mu k}^{t}\xi\left(m_{\mu k}^{t},G_{\mu k}^{t}\right)-g_{2\mu k}^{t}\xi\left(m_{\mu k}^{t},g_{2\mu k}^{t}\right).$
The final expression for the expectation value of a single variable
is \begin{eqnarray}
\left\langle b_{k}^{\ell\textrm{a}}|h_{\mu k}^{t}\to0,g_{1\mu k}^{t},g_{2\mu k}^{t}\right\rangle  & \simeq & \left(1-2{\rm e}^{-2nm_{\mu k}^{t}h_{\mu k}^{t}}\right)m_{\mu k}^{t}-\frac{g_{2\mu k}^{t}\xi\left(m_{\mu k}^{t},g_{2\mu k}^{t}\right)}{n}\left[1-\left(m_{\mu k}^{t}\right)^{2}\right]\, m_{\mu k}^{t}\nonumber \\
 &  & -\frac{G_{\mu k}^{t}\xi\left(m_{\mu k}^{t},G_{\mu k}^{t}\right)-g_{2\mu k}^{t}\xi\left(m_{\mu k}^{t},g_{2\mu k}^{t}\right)}{nL}\left[1-\left(m_{\mu k}^{t}\right)^{2}\right]\, m_{\mu k}^{t}\nonumber \\
 &  & +\xi\left(m_{\mu k}^{t},G_{\mu k}^{t}\right)\left[1-\left(m_{\mu k}^{t}\right)^{2}\right]\, h_{\mu k}^{t}\,.\label{eq:meanb1rsb}\end{eqnarray}
 To calculate $\left\langle b_{k}^{\ell\textrm{a}}b_{k}^{\ell{\rm a}^{\prime}}|h_{\mu k}^{t}\to0,g_{\mu k}^{t},\Delta g_{\mu k}^{t}\right\rangle $,
an off-diagonal element ($\textrm{a}\neq\textrm{a}^{\prime}$) in
the same block~$\ell$, we can apply the equation~(\ref{eq:mean1rsb})
with $f(\mathbf{x})=\tanh^{2}\left(x_{0}+x_{\ell}+h_{\mu k}^{t}\right)$,
thus the Hessian matrix is $\left.\mathbf{H}_{\tanh^{2}\left(x_{0}+x_{\ell}\right)}\right|_{\mathbf{x}_{0}}=2\left[1-\left(m_{\mu k}^{t}\right)^{2}\right]\left[1-3\left(m_{\mu k}^{t}\right)^{2}\right]\mathbf{M}_{0\ell}$,
thus: \begin{eqnarray}
\left\langle b_{k}^{\ell\textrm{a}}b_{k}^{\ell{\rm a}^{\prime}}|h_{\mu k}^{t}\to0,g_{1\mu k}^{t},g_{2\mu k}^{t}\right\rangle  & \simeq & \left(m_{\mu k}^{t}\right)^{2}+\frac{g_{2\mu k}^{t}\xi\left(m_{\mu k}^{t},g_{2\mu k}^{t}\right)}{n}\left[1-\left(m_{\mu k}^{t}\right)^{2}\right]\left[1-3\left(m_{\mu k}^{t}\right)^{2}\right]\nonumber \\
 &  & +\frac{G_{\mu k}^{t}\xi\left(m_{\mu k}^{t},G_{\mu k}^{t}\right)-g_{2\mu k}^{t}\xi\left(m_{\mu k}^{t},g_{2\mu k}^{t}\right)}{nL}\left[1-\left(m_{\mu k}^{t}\right)^{2}\right]\left[1-3\left(m_{\mu k}^{t}\right)^{2}\right]\nonumber \\
 &  & +2\xi\left(m_{\mu k}^{t},G_{\mu k}^{t}\right)\left[1-\left(m_{\mu k}^{t}\right)^{2}\right]\, m_{\mu k}^{t}h_{\mu k}^{t}.\label{eq:meanbblock1rsb}\end{eqnarray}
 Finally, to calculate the expectation value for the product of two
variables belonging to different blocks $\ell\neq\ell^{\prime}$ (the
sub-block index $\textrm{a}$ is insignificant in this case), $\left\langle b_{k}^{\ell\textrm{a}}b_{k}^{\ell^{\prime}{\rm a}}|h_{\mu k}^{t}\to0,g_{1\mu k}^{t},g_{2\mu k}^{t}\right\rangle $.
We set $f(\mathbf{x})=\tanh\left(x_{0}+x_{\ell}+h_{\mu k}^{t}\right)\tanh\left(x_{0}+x_{\ell^{\prime}}+h_{\mu k}^{t}\right)$,
thus the Hessian matrix \begin{eqnarray*}
\left(\left.\mathbf{H}_{\tanh\left(x_{0}+x_{\ell}\right)\tanh\left(x_{0}+x_{\ell^{\prime}}\right)}\right|_{\mathbf{x}_{0}}\right)_{ij} & = & \mathcal{M}_{0}\left(m_{\mu k}^{t}\right)\left(2\delta_{i0}\delta_{j0}+\delta_{i0}\delta_{j\ell}+\delta_{i\ell}\delta_{j0}+\delta_{i0}\delta_{j\ell^{\prime}}+\delta_{i\ell^{\prime}}\delta_{j0}\right)\\
 &  & +\mathcal{M}_{1}\left(m_{\mu k}^{t}\right)\left(\delta_{i\ell}\delta_{j\ell^{\prime}}+\delta_{i\ell^{\prime}}\delta_{j\ell}\right)-2\mathcal{M}_{2}\left(m_{\mu k}^{t}\right)\left(\delta_{i\ell}\delta_{j\ell}+\delta_{i\ell^{\prime}}\delta_{j\ell^{\prime}}\right)\,,\end{eqnarray*}
 where $\mathcal{M}_{0}\left(m_{\mu k}^{t}\right)\equiv\left[1-\left(m_{\mu k}^{t}\right)^{2}\right]\left[1-3\left(m_{\mu k}^{t}\right)^{2}\right]$,
$\mathcal{M}_{1}\left(m_{\mu k}^{t}\right)\equiv\left[1-\left(m_{\mu k}^{t}\right)^{2}\right]^{2}$
and $\mathcal{M}_{2}\left(m_{\mu k}^{t}\right)\equiv\left(m_{\mu k}^{t}\right)^{2}\left[1-\left(m_{\mu k}^{t}\right)^{2}\right]$.
The diagonal elements $\mathcal{K}_{\ell\ell^{\prime};k}\equiv\left(\left.\mathbf{H}_{\tanh\left(x_{0}+x_{\ell}\right)\tanh\left(x_{0}+x_{\ell^{\prime}}\right)}^{\prime}\right|_{\mathbf{x}_{0}}\right)_{kk}$
in the basis of eigenvectors of $\mathbf{H}_{\Phi}$ are \begin{eqnarray*}
\mathcal{K}_{\ell\ell^{\prime};0} & \simeq & 2\mathcal{M}_{0}\left(m_{\mu k}^{t}\right)\\
\mathcal{K}_{\ell\ell^{\prime};1} & \simeq & \frac{2\mathcal{M}_{0}\left(m_{\mu k}^{t}\right)}{L}\left[\frac{\beta_{0}}{\alpha_{0}}\left(\frac{\beta_{0}+2\alpha_{0}}{\alpha_{0}}\right)+1\right]\\
\mathcal{K}_{\ell\ell^{\prime};j} & = & -2\delta_{j\ell}\mathcal{M}_{2}\left(m_{\mu k}^{t}\right)\frac{\ell-1}{\ell}-2\left[\Theta\left(j-\ell\right)\Theta\left(\ell^{\prime}-j\right)\right]\frac{\mathcal{M}_{2}\left(m_{\mu k}^{t}\right)}{j(j-1)}\\
 &  & -2\delta_{j\ell^{\prime}}\left[\frac{\mathcal{M}_{1}\left(m_{\mu k}^{t}\right)}{\ell^{\prime}}+\frac{\mathcal{M}_{2}\left(m_{\mu k}^{t}\right)}{\ell^{\prime}}\left(\ell^{\prime}-1+\frac{1}{\ell^{\prime}-1}\right)\right]+2\Theta\left(j-\ell^{\prime}\right)\frac{\mathcal{M}_{0}\left(m_{\mu k}^{t}\right)}{j(j-1)}\,,\end{eqnarray*}
 thus, the sum of the diagonal elements is: \begin{eqnarray*}
\frac{1}{2}\,\frac{1}{nL}\,\sum_{k=0}^{L}\lambda_{k,-\pm h}^{-1}\mathcal{K}_{\ell\ell^{\prime};k} & \simeq & \frac{1}{nL}\frac{\mathcal{M}_{0}\left(m_{\mu k}^{t}\right)}{\alpha_{0}}\left[1\!+\!\frac{\left(\beta_{0}+\alpha_{0}\right)^{2}}{\left(\alpha_{0}\gamma_{0}-\beta_{0}^{2}\right)}\right]\!-\!\frac{1}{n}\frac{1}{\gamma_{0}}\left\{ \mathcal{M}_{2}\left(m_{\mu k}^{t}\right)\left[\frac{\ell-1}{\ell}\!+\!\sum_{j=\ell+1}^{\ell^{\prime}-1}\frac{1}{j(j-1)}\right]\right.\\
 &  & +\left.\frac{\mathcal{M}_{1}\left(m_{\mu k}^{t}\right)}{\ell^{\prime}}+\frac{\mathcal{M}_{2}\left(m_{\mu k}^{t}\right)}{\ell^{\prime}}\left(\ell^{\prime}-1+\frac{1}{\ell^{\prime}-1}\right)-\mathcal{M}_{0}\left(m_{\mu k}^{t}\right)\sum_{j=\ell^{\prime}+1}^{L}\frac{1}{j(j-1)}\right\} \\
 & = & -\frac{g_{2\mu k}^{t}\xi\left(m_{\mu k}^{t},g_{2\mu k}^{t}\right)}{n}\,\left[\mathcal{M}_{1}\left(m_{\mu k}^{t}\right)-\mathcal{M}_{0}\left(m_{\mu k}^{t}\right)\right]\\
 &  & \qquad+\frac{G_{\mu k}^{t}\xi\left(m_{\mu k}^{t},G_{\mu k}^{t}\right)-g_{2\mu k}^{t}\xi\left(m_{\mu k}^{t},g_{2\mu k}^{t}\right)}{nL}\mathcal{M}_{0}\left(m_{\mu k}^{t}\right)\,.\end{eqnarray*}
 Using the sum of diagonal terms one then derives the expected correlation
for variables belonging to two different blocks \begin{eqnarray}
\left\langle b_{k}^{\ell\textrm{a}}b_{k}^{\ell^{\prime}{\rm a}}|h_{\mu k}^{t}\to0,g_{\mu k}^{t},\Delta g_{\mu k}^{t}\right\rangle  & \simeq & \left(m_{\mu k}^{t}\right)^{2}-2\,\frac{g_{2\mu k}^{t}\xi\left(m_{\mu k}^{t},g_{2\mu k}^{t}\right)}{n}\,\left(m_{\mu k}^{t}\right)^{2}\left[1-\left(m_{\mu k}^{t}\right)^{2}\right]\nonumber \\
 &  & +\frac{G_{\mu k}^{t}\xi\left(m_{\mu k}^{t},G_{\mu k}^{t}\right)-g_{2\mu k}^{t}\xi\left(m_{\mu k}^{t},g_{2\mu k}^{t}\right)}{nL}\left[1-3\left(m_{\mu k}^{t}\right)^{2}\right]\left[1-\left(m_{\mu k}^{t}\right)^{2}\right]\nonumber \\
 &  & +2\xi\left(m_{\mu k}^{t},G_{\mu k}^{t}\right)\left[1-\left(m_{\mu k}^{t}\right)^{2}\right]\, m_{\mu k}^{t}h_{\mu k}^{t}\,.\label{eq:covllprime}\end{eqnarray}
 Keeping in mind that $\left\langle b_{k}^{\ell\textrm{a}}b_{k}^{\ell\textrm{a}}|h_{\mu k}^{t},g_{1\mu k}^{t},g_{2\mu k}^{t}\right\rangle =1$
and using equations~(\ref{eq:meanb1rsb})-(\ref{eq:covllprime}),
the covariance matrix entries can be then calculated: \begin{eqnarray*}
\left(\mathbf{{\Psi}}_{\mu kl}^{t}\right)^{\ell\mathrm{a}\,\ell^{\prime}{\rm a}^{\prime}} & = & \left\langle b_{k}^{\ell\textrm{a}}b_{l}^{\ell^{\prime}\textrm{a}^{\prime}}|h_{\mu k}^{t}\to0,g_{1\mu k}^{t},g_{2\mu k}^{t};h_{\mu l}^{t}\to0,g_{1\mu l}^{t},g_{2\mu l}^{t}\right\rangle \\
 &  & -\left\langle b_{k}^{\ell^{\phantom{\prime}}\textrm{a}}|h_{\mu k}^{t}\to0,g_{1\mu k}^{t},g_{2\mu k}^{t}\right\rangle \left\langle b_{l}^{\ell^{\prime}\textrm{a}^{\prime}}|h_{\mu l}^{t}\to0,g_{1\mu l}^{t},g_{2\mu l}^{t}\right\rangle =\delta_{kl}\left(\mathbf{{\Psi}}_{\mu kk}^{t}\right)^{\ell\mathrm{a}\,\ell^{\prime}{\rm a}^{\prime}}\\
\left(\mathbf{{\Psi}}_{\mu kk}^{t}\right)^{\ell\mathrm{a}\,\ell^{\prime}{\rm a}^{\prime}} & \simeq & \delta^{\ell\ell^{\prime}}\delta^{{\rm a}{\rm a}^{\prime}}\left[1-\left(m_{\mu k}^{t}\right)^{2}\right]+\delta^{\ell\ell^{\prime}}\left(1-\delta^{{\rm a}{\rm a}^{\prime}}\right)\,\frac{g_{2\mu k}^{t}\xi\left(m_{\mu k}^{t},g_{2\mu k}^{t}\right)}{n}\left[1-\left(m_{\mu k}^{t}\right)^{2}\right]^{2}\\
 &  & +\left(1-\delta^{\ell\ell^{\prime}}\right)\,\frac{G_{\mu k}^{t}\xi\left(m_{\mu k}^{t},G_{\mu k}^{t}\right)-g_{2\mu k}^{t}\xi\left(m_{\mu k}^{t},g_{2\mu k}^{t}\right)}{nL}\left[1-\left(m_{\mu k}^{t}\right)^{2}\right]^{2}\,,\end{eqnarray*}
 where we have kept only the dominant terms at each entry, disregarding
terms of order $\mathcal{O}\left({\rm e}^{-2nm_{\mu k}^{t}h_{\mu k}^{t}}+\frac{h}{n}+\frac{h}{L}\right)$.

If the $\varepsilon_{\mu k}$ and $b_{k}^{\mathrm{a}}$ are unbiased
variables, the variable $\Delta_{\mu k}^{\mathrm{a}}=\sum_{l\neq k}\varepsilon_{\mu l}b_{l}^{\mathrm{a}}$,
by virtue of the central limit theorem, obeys a normal distribution,
with mean value and covariance matrix that can be obtained by employing
the expressions derived for $\mathbf{{\Psi}}$ \begin{eqnarray}
\left(\mathbf{u}_{\mu k}^{t}\right)^{\ell\mathrm{a}} & \equiv & \left\langle \Delta_{\mu k}^{\ell\textrm{a}}\right\rangle =\sum_{\left\{ \mathbf{b}_{l\neq k}\right\} }\prod_{l\neq k}P^{t}\left(\mathbf{b}_{l}|\left\{ y_{\nu\neq\mu}\right\} \right)\sum_{l\neq k}\varepsilon_{\mu l}b_{l}^{\ell\mathrm{a}}=\sum_{l\neq k}\varepsilon_{\mu l}m_{\mu l}^{t}\label{eq:u1rsb}\\
\left(\mathbf{{\Upsilon}}_{\mu k}^{t}\right)^{\ell\mathrm{a}\,\ell^{\prime}{\rm a}^{\prime}} & \equiv & \left\langle \Delta_{\mu k}^{\ell\textrm{a}}\Delta_{\mu k}^{\ell^{\prime}\textrm{a}^{\prime}}\right\rangle -\left\langle \Delta_{\mu k}^{\ell\textrm{a}}\right\rangle \left\langle \Delta_{\mu k}^{\ell^{\prime}\textrm{a}^{\prime}}\right\rangle \nonumber \\
 & = & \sum_{\left\{ \mathbf{b}_{l\neq k}\right\} }\prod_{l\neq k}P^{t}\left(\mathbf{b}_{l}|\left\{ y_{\nu\neq\mu}\right\} \right)\sum_{\substack{l\neq k\\
j\neq k}
}\varepsilon_{\mu l}\varepsilon_{\mu j}b_{l}^{\ell\mathrm{a}}b_{j}^{\ell^{\prime}\mathrm{a}^{\prime}}-\left(\sum_{l\neq k}\varepsilon_{\mu l}m_{\mu l}^{t}\right)^{2}\nonumber \\
 & = & \sum_{l\neq k}\varepsilon_{\mu l}^{2}\left(\mathbf{{\Psi}}_{\mu lj}^{t}\right)^{\ell\mathrm{a}\,\ell^{\prime}{\rm a}^{\prime}}=\delta^{\ell\ell^{\prime}}\delta^{{\rm a}{\rm a}^{\prime}}X_{\mu k}^{t}+\delta^{\ell\ell^{\prime}}\left(1-\delta^{{\rm a}{\rm a}^{\prime}}\right)\frac{1}{n}R_{\mu k}^{t}+\left(1-\delta^{\ell\ell^{\prime}}\right)\,\frac{1}{nL}\,\left(V_{\mu k}^{t}-R_{\mu k}^{t}\right),\nonumber \end{eqnarray}
 where $X_{\mu k}^{t}$ is given by equations~(\ref{eq:xxx}) and
\begin{eqnarray*}
R_{\mu k}^{t} & \equiv & \sum_{l\neq k}\varepsilon_{\mu l}^{2}g_{2\mu l}^{t}\xi\left(m_{\mu l}^{t},g_{2\mu l}^{t}\right)\left[1-\left(m_{\mu l}^{t}\right)^{2}\right]^{2}\\
V_{\mu k}^{t} & \equiv & \sum_{l\neq k}\varepsilon_{\mu l}^{2}G_{\mu l}^{t}\xi\left(m_{\mu l}^{t},G_{\mu l}^{t}\right)\left[1-\left(m_{\mu l}^{t}\right)^{2}\right]^{2}\end{eqnarray*}
 are macroscopic variables of $\mathcal{O}(1)$. In particular, $R_{\mu k}^{t}$
and $V_{\mu k}^{t}$ are free variables that can be used to optimise
a given performance measure. \cut{These variables have self averaging,
therefore we can drop the sub-indexes $\mu$ and \emph{k.} The eigenvalues
and eigenvectors of the matrix $\mathbf{{\Upsilon}}^{t}$ satisfy
the following relation: \[
X^{t}v^{\ell{\rm a}}-\frac{1}{n}R^{t}\sum_{{\rm b}\neq{\rm a}}v^{\ell{\rm b}}+\frac{1}{nL}\,\left(V_{\mu k}^{t}-R_{\mu k}^{t}\right)\sum_{\ell^{\prime}\neq\ell}\sum_{{\rm b}=1}^{n}v^{\ell{\rm b}}=\lambda v^{\ell{\rm a}}.\]
 If $\sum_{{\rm b}=1}^{n}v^{\ell{\rm b}}$ is independent of $\ell$,
then for two different entries of the eigenvector one obtains \[
\left(X^{t}-\frac{1}{n}R^{t}\right)\left(v^{\ell{\rm a}}-v^{\ell^{\prime}{\rm a}^{\prime}}\right)=\lambda\left(v^{\ell{\rm a}}-v^{\ell^{\prime}{\rm a}^{\prime}}\right),\]
 which implies that either $v^{\ell{\rm a}}=v^{\ell^{\prime}{\rm a}^{\prime}}\;\forall\ell,\,\ell^{\prime},\,{\rm a},\,{\rm and}\,{\rm a}^{\prime}$
or $\lambda_{1}=X^{t}-\frac{1}{n}R^{t}$. The former leads to the
eigenvalue $\lambda_{0}\simeq X^{t}+V^{t}+\mathcal{O}\left(n^{-1}\right)$
and $\mathbf{v}_{0}^{\sf T}=\mathbf{v}_{1,1}^{\sf T}=\stackrel{nL}{\left(\overbrace{1,\dots,1}\right)}/\sqrt{nL}$.
$\lambda_{1}$ is $\left(nL-1\right)$-fold degenerated, and their
eigenvectors are: \begin{eqnarray}
\mathbf{v}_{1,{\rm a}>1}^{\sf T} & = & \frac{1}{\sqrt{L}}\stackrel{L}{\left(\overbrace{\mathbf{w}_{{\rm a}}^{\sf T},\dots,\mathbf{w}_{{\rm a}}^{\sf T}}\right)}\label{eq:v1rsb0}\\
\mathbf{v}_{\ell>1,{\rm a}}^{\sf T} & = & \frac{1}{\sqrt{\ell\left(\ell-1\right)}}\left(\stackrel{\ell-1}{\overbrace{\mathbf{w}_{{\rm a}}^{\sf T},\dots,\mathbf{w}_{{\rm a}}^{\sf T}}},-\left({\rm \ell}-1\right)\mathbf{w}_{{\rm a}}^{\sf T},\stackrel{L-\ell}{\overbrace{\mathbf{0}^{\sf T},\dots,\mathbf{0}^{\sf T}}}\right)\label{eq:v1rsb1}\end{eqnarray}
 where $\mathbf{0}^{\sf T}=\stackrel{n}{\left(\overbrace{0,\dots,0}\right)}$
and \begin{eqnarray*}
\mathbf{w}_{1}^{\sf T} & = & \frac{1}{\sqrt{n}}\stackrel{n}{\left(\overbrace{1,\dots,1}\right)}\\
\mathbf{w}_{{\rm a}>1}^{\sf T} & = & \frac{1}{\sqrt{{\rm a}\left({\rm a}-1\right)}}\left(\stackrel{{\rm a}-1}{\overbrace{1,1,\dots,1}},-\left({\rm a}-1\right),\stackrel{n-{\rm a}}{\overbrace{0,0,\dots,0}}\right)\end{eqnarray*}
 are the eigenvectors of the RS matrix. Observe that the number of
eigenvectors that satisfy equation~(\ref{eq:v1rsb0}) is $n-1$ (there
are \emph{n} different $\mathbf{w}_{{\rm a}}$), and the number of
eigenvectors that satisfy Eq.(\ref{eq:v1rsb1}) is $\left(L-1\right)n$
which gives us a total of $nL-1$ eigenvectors to the $\lambda_{1}$
eigenvalue. }

\section{The messages \label{app:messages}}

From the conditional probabilities of equations~(\ref{eq:bp1}) and
(\ref{eq:bp2}) and with the application of the probability distributions
$P\left(\mathbf{\Delta}_{\mu k}|\mathbf{B}\right)$ of equation~(\ref{eq:pdelrs})
in~(\ref{eq:likelihood}) we can express the message from nodes $y_{\mu}$
to nodes $b_{k}^{{\rm a}}$ at time $t+1$ as: \begin{eqnarray}
\widehat{m}_{\mu k}^{t+1} & = & \frac{{\displaystyle \sum_{\left\{ \mathbf{B}\right\} }}{\displaystyle {\displaystyle b_{k}^{{\rm a}^{\prime}}\,\prod_{{\rm a}=1}^{n}P\left(y_{\mu}|\mathbf{b}^{{\rm a}}\right)P\left(\mathbf{b}^{{\rm a}}\right)}\prod_{l\neq k}P\left(\mathbf{b}_{l}|\left\{ y_{\nu\neq\mu}\right\} \right)}}{{\displaystyle \sum_{\left\{ \mathbf{B}\right\} }}\,{\displaystyle \prod_{{\rm a}=1}^{n}P\left(y_{\mu}|\mathbf{b}^{{\rm a}}\right)P\left(\mathbf{b}^{{\rm a}}\right)\prod_{l\neq k}P\left(\mathbf{b}_{l}|\left\{ y_{\nu\neq\mu}\right\} \right)}}\label{eq:mhat}\\
 & = & \frac{{\displaystyle \int{\rm d}\mathbf{\Delta}_{\mu k}}P\left(\mathbf{\Delta}_{\mu k}|\mathbf{B}\right){\displaystyle \sum_{\left\{ \mathbf{b}_{k}\right\} }}b_{k}^{{\rm a}^{\prime}}\, P\left(y_{\mu}|\mathbf{{\Delta}}_{\mu k};\mathbold{\gamma}\right)\left[1+\varepsilon_{\mu k}\mathbf{b}_{k}^{\sf T}\nabla_{\mathbf{{\Delta}}_{\mu k}}\ln P\left(y_{\mu}|\mathbf{{\Delta}}_{\mu k};\mathbold{\gamma}\right)\right]}{{\displaystyle \int{\rm d}\mathbf{\Delta}_{\mu k}}P\left(\mathbf{\Delta}_{\mu k}|\mathbf{B}\right){\displaystyle \sum_{\left\{ \mathbf{b}_{k}\right\} }}P\left(y_{\mu}|\mathbf{{\Delta}}_{\mu k};\mathbold{\gamma}\right)\left[1+\varepsilon_{\mu k}\mathbf{b}_{k}^{\sf T}\nabla_{\mathbf{{\Delta}}_{\mu k}}\ln P\left(y_{\mu}|\mathbf{{\Delta}}_{\mu k};\mathbold{\gamma}\right)\right]}\,.\nonumber \end{eqnarray}
 If $P\left(y_{\mu}|\mathbf{{\Delta}}_{\mu k};\mathbold{\gamma}\right)=\prod_{{\rm a}=1}^{n}P\left(y_{\mu}|\Delta_{\mu k}^{{\rm a}};\mathbold{\gamma}\right)$,
and ignoring $\mathcal{O}(\varepsilon_{\mu k}^{2})$ terms, the traces
on $\mathbf{b}_{k}$ can be written as \begin{eqnarray*}
{\displaystyle \sum_{\left\{ \mathbf{b}_{k}\right\} }}P\left(y_{\mu}|\mathbf{{\Delta}}_{\mu k};\mathbold{\gamma}\right)\left[1+\varepsilon_{\mu k}\mathbf{b}_{k}^{\sf T}\nabla_{\mathbf{{\Delta}}_{\mu k}}\ln P\left(y_{\mu}|\mathbf{{\Delta}}_{\mu k};\mathbold{\gamma}\right)\right] & = & 2^{n}P\left(y_{\mu}|\mathbf{{\Delta}}_{\mu k};\mathbold{\gamma}\right)\\
{\displaystyle \sum_{\left\{ \mathbf{b}_{k}\right\} }}b_{k}^{{\rm a}^{\prime}}\, P\left(y_{\mu}|\mathbf{{\Delta}}_{\mu k};\mathbold{\gamma}\right)\left[1+\varepsilon_{\mu k}\mathbf{b}_{k}^{\sf T}\nabla_{\mathbf{{\Delta}}_{\mu k}}\ln P\left(y_{\mu}|\mathbf{{\Delta}}_{\mu k};\mathbold{\gamma}\right)\right] & = & 2^{n}\varepsilon_{\mu k}P\left(y_{\mu}|\mathbf{{\Delta}}_{\mu k};\mathbold{\gamma}\right)\frac{\partial}{\partial\Delta_{\mu k}^{{\rm a}^{\prime}}}\ln P\left(y_{\mu}|\Delta_{\mu k}^{\tilde{{\rm a}}};\mathbold{\gamma}\right)\,,\end{eqnarray*}
 thus, following from (\ref{eq:mhat}) and neglecting $\mathcal{O}(1/n)$
terms \begin{eqnarray}
^{{\rm (RS)}}\widehat{m}_{\mu k}^{t+1} & \simeq & \varepsilon_{\mu k}\frac{{\displaystyle {\displaystyle \int{\rm d}\mathbf{\Delta}_{\mu k}}P\left(\mathbf{\Delta}_{\mu k}|\mathbf{B}\right)\prod_{{\rm a}=1}^{n}P\left(y_{\mu}|\Delta_{\mu k}^{{\rm a}};\mathbold{\gamma}\right)\frac{\partial}{\partial\Delta_{\mu k}^{{\rm a}^{\prime}}}\ln P\left(y_{\mu}|\Delta_{\mu k}^{{\rm a}^{\prime}};\mathbold{\gamma}\right)}}{{\displaystyle \int{\rm d}\mathbf{\Delta}_{\mu k}}P\left(\mathbf{\Delta}_{\mu k}|\mathbf{B}\right){\displaystyle \prod_{{\rm a}=1}^{n}}P\left(y_{\mu}|\Delta_{\mu k}^{{\rm a}};\mathbold{\gamma}\right)}\label{eq:rsmhatfirst}\\
 & = & \frac{\varepsilon_{\mu k}}{^{{\rm (RS)}}\mathscr N_{\mu k}^{t}}{\displaystyle \int{\rm d}\vartheta}\,\exp\left\{ {\displaystyle -n\,\frac{\left(\vartheta-u_{\mu k}^{t}\right)^{2}}{2R^{t}}}\right\} \nonumber \\
 &  & \qquad\times{\displaystyle \left[\int{\rm d}\Delta\exp\left\{ -\frac{\left(\Delta-\vartheta\right)^{2}}{2X^{t}}+\ln P\left(y_{\mu}|\Delta;\mathbold{\gamma}\right)\right\} \right]^{n-1}}\nonumber \\
 &  & \qquad\times\int{\rm d}\Delta\exp\left\{ -\frac{\left(\Delta-\vartheta\right)^{2}}{2X^{t}}\right\} \frac{\partial}{\partial\Delta}P\left(y_{\mu}|\Delta;\mathbold{\gamma}\right)\,,\nonumber \end{eqnarray}
 and \begin{eqnarray*}
^{{\rm (1RSB)}}\widehat{m}_{\mu k}^{t+1} & \simeq & \varepsilon_{\mu k}\frac{{\displaystyle {\displaystyle \int{\rm d}\mathbf{\Delta}_{\mu k}}P\left(\mathbf{\Delta}_{\mu k}|\mathbf{B}\right)\prod_{\ell=1}^{L}\prod_{{\rm a}=1}^{n}P\left(y_{\mu}|\Delta_{\mu k}^{\ell{\rm a}};\mathbold{\gamma}\right)\frac{\partial}{\partial\Delta_{\mu k}^{\ell^{\prime}{\rm a}^{\prime}}}\ln P\left(y_{\mu}|\Delta_{\mu k}^{\ell^{\prime}{\rm a}^{\prime}};\mathbold{\gamma}\right)}}{{\displaystyle \int{\rm d}\mathbf{\Delta}_{\mu k}}P\left(\mathbf{\Delta}_{\mu k}|\mathbf{B}\right){\displaystyle \prod_{{\rm a}=1}^{n}}P\left(y_{\mu}|\Delta_{\mu k}^{{\rm a}};\mathbold{\gamma}\right)}\\
 & = & \frac{\varepsilon_{\mu k}}{^{{\rm (1RSB)}}\mathscr N_{\mu k}^{t}}\int{\rm d}\mathbf{{\Theta}}\,\prod_{\ell=1}^{L}\exp\left\{ -\frac{n}{2}\left[\frac{\left(\vartheta^{0}\right)^{2}}{V^{t}-R^{t}}+\frac{\left(\vartheta^{\ell}\right)^{2}}{V^{t}-L^{-1}\left(V^{t}-R^{t}\right)}\right]\right\} \\
 &  & \qquad\times\prod_{\ell\neq\ell^{\prime}}{\displaystyle \left[\int{\rm d}\Delta\exp\left\{ -\frac{\left(\Delta-\vartheta_{\mu k}^{0\ell t}\right)^{2}}{2X^{t}}+\ln P\left(y_{\mu}|\Delta;\mathbold{\gamma}\right)\right\} \right]^{n}}\\
 &  & \qquad\times{\displaystyle \left[\int{\rm d}\Delta\exp\left\{ -\frac{\left(\Delta-\vartheta_{\mu k}^{0\ell^{\prime}t}\right)^{2}}{2X^{t}}+\ln P\left(y_{\mu}|\Delta;\mathbold{\gamma}\right)\right\} \right]^{n-1}}\\
 &  & \qquad\times\int{\rm d}\Delta\exp\left\{ -\frac{\left(\Delta-\vartheta_{\mu k}^{0\ell^{\prime}t}\right)^{2}}{2X^{t}}\right\} \frac{\partial}{\partial\Delta}P\left(y_{\mu}|\Delta;\mathbold{\gamma}\right)\,,\end{eqnarray*}
 where $^{{\rm (RS)}}\mathscr N_{\mu k}^{t}$ and $^{{\rm (1RSB)}}\mathscr N_{\mu k}^{t}$
are suitable normalisation constants and $\vartheta_{\mu k}^{0\ell t}\equiv\vartheta^{0}+\vartheta^{\ell}+u_{\mu k}^{t}$.
One can then define: \begin{eqnarray}
\mathcal{G}\left(y_{\mu},\vartheta\right) & \equiv & \int{\rm d}\Delta\exp\left\{ -\frac{\left(\Delta-\vartheta\right)^{2}}{2X^{t}}\right\} P\left(y_{\mu}|\Delta;\mathbold{\gamma}\right)\label{eq:gcal}\\
\mathcal{P}\left(y_{\mu},\vartheta\right) & \equiv & \left[\mathcal{G}\left(y_{\mu},\vartheta\right)\right]^{-1}\int{\rm d}\Delta\exp\left\{ -{\displaystyle \frac{\left(\Delta-\vartheta\right)^{2}}{2X^{t}}}\right\} {\displaystyle \frac{\partial}{\partial\Delta}P\left(y_{\mu}|\Delta;\mathbold{\gamma}\right)}\nonumber \\
 & = & \left[\mathcal{G}\left(y_{\mu},\vartheta\right)\right]^{-1}\int{\rm d}\Delta\exp\left\{ -{\displaystyle \frac{\left(\Delta-\vartheta\right)^{2}}{2X^{t}}}\right\} \frac{\Delta-\vartheta}{X^{t}}{\displaystyle P\left(y_{\mu}|\Delta;\mathbold{\gamma}\right)}\nonumber \\
 & = & \frac{\partial}{\partial\vartheta}\ln\mathcal{G}\left(y_{\mu},\vartheta\right)\label{eq:pcal}\\
^{{\rm (RS)}}\mathcal{H}\left(\vartheta,y_{\mu}\right) & \equiv & \frac{\left(\vartheta-u_{\mu k}^{t}\right)^{2}}{2R_{\mu k}^{t}}-\ln\mathcal{G}\left(y_{\mu},\vartheta\right)\label{eq:hcalrs}\\
^{{\rm (1RSB)}}\mathcal{H}\left(\vartheta^{0},\vartheta^{\ell},y_{\mu}\right) & \equiv & \frac{1}{2}\,\left[\frac{\left(\vartheta^{0}\right)^{2}}{V^{t}-R^{t}}+\frac{\left(\vartheta^{\ell}\right)^{2}}{V^{t}-L^{-1}\left(V^{t}-R^{t}\right)}\right]-\ln\mathcal{G}\left(y_{\mu,}\vartheta_{\mu k}^{0\ell t}\right)\,.\label{eq:hcal1rsb}\end{eqnarray}
 Thus the expression for the RS message is: \begin{eqnarray*}
^{{\rm (RS)}}\widehat{m}_{\mu k}^{t+1} & = & \varepsilon_{\mu k}{\displaystyle \frac{{\displaystyle \int{\rm d}\vartheta\,\exp\left\{ -n{}^{{\rm (RS)}}\mathcal{H}\left(\vartheta,y_{\mu}\right)\right\} \mathcal{P}}\left(y_{\mu,}\vartheta\right)}{{\displaystyle \int{\rm d}\vartheta\,\exp\left\{ -n{}^{{\rm (RS)}}\mathcal{H}\left(\vartheta,y_{\mu}\right)\right\} }}}\,.\end{eqnarray*}
 In the large \emph{n} limit, only the solutions $\tilde{\vartheta}_{\mu k}^{t}$
of ${\displaystyle \frac{\partial}{\partial\vartheta}}{}^{{\rm (RS)}}\mathcal{H}=0$,
that correspond to the minimum of $\mathcal{H}$ contribute to the
integral. The dominant term in the integral is obtained via saddle
point methods, which leads to the final expression for the message
\begin{equation}
^{{\rm (RS)}}\widehat{m}_{\mu k}^{t+1}=\varepsilon_{\mu k}\frac{\tilde{\vartheta}_{\mu k}^{t}-u_{\mu k}^{t}}{R^{t}}\,,\label{eq:mmrs}\end{equation}
 where $\tilde{\vartheta}_{\mu k}^{t}$ is given by equation~(\ref{eq:wrs}).

The 1RSB case is a little more delicate. The \cut{effective Hamiltonian
that appears as the argument of the} exponential is a sum over $L$
\cut{Hamiltonians} functions$^{{\rm (1RSB)}}\mathcal{H}\left(\vartheta^{0},\vartheta^{\ell},y_{\mu}\right)$.
Therefore, a Taylor expansion close to the saddle point of \cut{of
the total Hamiltonian near the ground state} equation~(\ref{eq:w1rsb})
is employed resulting in \begin{eqnarray*}
\sum_{\ell=1}^{L}{}^{{\rm (1RSB)}}\mathcal{H}\left(\vartheta^{0},\vartheta^{\ell},y_{\mu}\right) & \simeq & LE_{0}+\frac{L}{2}h_{0}\left(\upDelta\vartheta^{0}\right)^{2}+h_{1}\upDelta\vartheta^{0}\sum_{\ell=1}^{L}\upDelta\vartheta^{\ell}+\frac{1}{2}h_{2}\sum_{\ell=1}^{L}\left(\upDelta\vartheta^{\ell}\right)^{2}+\mathcal{O}\left(\upDelta\vartheta^{3}\right),\end{eqnarray*}
 where $E_{0}={}^{{\rm (1RSB)}}\mathcal{H}\left(\tilde{\vartheta}_{\mu k}^{0t},\tilde{\vartheta}_{\mu k}^{\ell t},y_{\mu}\right)$
is the energy of the ground state, $\upDelta\vartheta^{i}=\vartheta^{i}-\tilde{\vartheta}_{\mu k}^{it}\,\, i=0,\ell$
and the entries $h_{0}$, $h_{1}$ and $h_{2}$ satisfy the equation\[
\left(\begin{array}{cc}
h_{0} & h_{1}\\
h_{1} & h_{2}\end{array}\right)=\left(\begin{array}{cc}
\left(V^{t}-R^{t}\right)^{-1} & 0\\
0 & \left(V^{t}\right)^{-1}\end{array}\right)-\left.\frac{\partial\mathcal{P}}{\partial\vartheta}\right|_{\vartheta=\tilde{\vartheta}_{\mu k}^{t}}\left(\begin{array}{cc}
1 & 1\\
1 & 1\end{array}\right)\,,\]
 where $\tilde{\vartheta}_{\mu k}^{0\ell t}$ is the solution of equation~(\ref{eq:w1rsb}).
If $\mathbf{\Theta}^{\sf T}=\left(\vartheta^{0},\vartheta^{1},\dots,\vartheta^{L}\right)$
and $\left(\mathbf{H}_{\mathcal{H}}\right)_{ij}=\delta_{jk}\left[\delta_{j0}h_{0}+\left(1-\delta_{j0}\right)L^{-1}h_{2}\right]+\left(\delta_{j0}+\delta_{k0}\right)\left(1-\delta_{jk}\right)L^{-1}h_{1}$
is the Hessian of $\sum_{\ell=1}^{L}{}^{{\rm (1RSB)}}\mathcal{H}\left(\vartheta^{0},\vartheta^{\ell},y_{\mu}\right)$,
then \begin{eqnarray*}
\sum_{\ell=1}^{L}{}^{{\rm (1RSB)}}\mathcal{H}\left(\vartheta^{0},\vartheta^{\ell},y_{\mu}\right) & \simeq & LE_{0}+\frac{L}{2}\upDelta\mathbf{\Theta}^{\sf T}\mathbf{H}_{\mathcal{H}}\upDelta\mathbf{\Theta}\,.\end{eqnarray*}
 The matrix $\mathbf{H}_{\mathcal{H}}$ has the same structure as
$\mathbf{H}_{\Phi}$, therefore, the eigenvalues and eigenvectors
of $\mathbf{H}_{\mathcal{H}}$ can be obtained adapting equations~(\ref{eq:eigenvalues})
and (\ref{eq:eigenvectors}) by the substitutions $\alpha_{0}=h_{0}$,
$-\beta_{0}=h_{1}$ and $\gamma_{0}=h_{2}$. Expanding $\mathcal{P}\left(\vartheta,y_{\mu}\right)$
at the saddle point $\tilde{\vartheta}_{\mu k}^{0\ell t}$ one obtains
$\mathcal{P}\left(\vartheta_{\mu k}^{0\ell^{\prime}t},y_{\mu}\right)\simeq\mathcal{P}_{0}+\mathcal{P}_{1}\left(\upDelta\vartheta^{0}+\upDelta\vartheta^{\ell^{\prime}}\right)+\frac{1}{2}\mathcal{P}_{2}\left(\upDelta\vartheta^{0}+\upDelta\vartheta^{\ell^{\prime}}\right)^{2}$
where $\mathcal{P}_{j}\equiv\left.{\displaystyle \frac{\partial^{j}{\displaystyle \mathcal{P}}}{\partial\vartheta^{j}}}\right|_{\vartheta=\tilde{\vartheta}_{\mu k}^{t}}.$
The resulting messages are \begin{eqnarray*}
^{{\rm (1RSB)}}\widehat{m}_{\mu k}^{t+1} & = & \varepsilon_{\mu k}{\displaystyle \frac{{\displaystyle \int{\rm d}\mathbf{{\Theta}}\,\exp\left\{ -\frac{nL}{2}\upDelta\mathbf{\Theta}^{\sf T}\mathbf{H}_{\mathcal{H}}\upDelta\mathbf{\Theta}\right\} \left[\mathcal{P}_{0}+\mathcal{P}_{1}\left(\upDelta\vartheta^{0}+\upDelta\vartheta^{\ell^{\prime}}\right)+\frac{1}{2}\mathcal{P}_{2}\left(\upDelta\vartheta^{0}+\upDelta\vartheta^{\ell^{\prime}}\right)^{2}\right]}}{{\displaystyle \int{\rm d}\mathbf{{\Omega}}\,\exp\left\{ -\frac{nL}{2}\upDelta\mathbf{\Theta}^{\sf T}\mathbf{H}_{\mathcal{H}}\upDelta\mathbf{\Theta}\right\} }}}\end{eqnarray*}
 where the term proportional to $\mathcal{P}_{1}$ vanishes for parity
reasons. In the basis of eigenvectors of $\mathbf{H}_{\mathcal{H}}$,
i.e. $\mathbf{\Gamma}=\mathbf{U}^{\sf T}\mathbf{\Theta}=\left(\gamma_{0},\gamma_{1},\dots,\gamma_{L}\right)^{\sf T}$
where \textbf{U} is adapted from equation~(\ref{eq:rot}), the message
has the form:\[
^{{\rm (1RSB)}}\widehat{m}_{\mu k}^{t+1}\simeq\varepsilon_{\mu k}{\displaystyle \frac{{\displaystyle \int{\rm d}\mathbf{{\Gamma}}\,\exp\left\{ -\frac{nL}{2}\sum_{\ell=0}^{L}\lambda_{\ell}\left(\upDelta\gamma^{\ell}\right)^{2}\right\} }\left(\mathcal{P}_{0}+\frac{1}{2}\mathcal{P}_{2}\upDelta\mathbf{\Gamma}^{\sf T}\mathbf{M}_{0\ell^{\prime}}^{\prime}\upDelta\mathbf{\Gamma}\right)}{{\displaystyle \int{\rm d}\mathbf{{\Gamma}}\,\exp\left\{ -\frac{nL}{2}\sum_{\ell=0}^{L}\lambda_{\ell}\left(\upDelta\gamma^{\ell}\right)^{2}\right\} }}\,,}\]
 where $\lambda_{\ell}$ are the eigenvalues of $\mathbf{H}_{\mathcal{H}}$
and $\mathbf{M}_{0\ell^{\prime}}^{\prime}$ is adapted from equation~(\ref{eq:diagm}).

The expression for the message is reduced to \begin{eqnarray}
^{{\rm (1RSB)}}\widehat{m}_{\mu k}^{t+1} & \simeq & \varepsilon_{\mu k}{\displaystyle \left[\mathcal{P}_{0}+\frac{1}{n}\,\frac{\mathcal{P}_{2}}{2h_{2}}+\mathcal{O}\left(\frac{1}{nL}\right)\right]}\nonumber \\
 & \simeq & \varepsilon_{\mu k}\frac{\tilde{\vartheta}_{\mu k}^{t}-u_{\mu k}^{t}}{{2V}^{t}-R^{t}}+\frac{\varepsilon_{\mu k}}{2n}\,\frac{\mathcal{P}_{2}V^{t}}{1-\mathcal{P}_{1}V^{t}}\,.\label{eq:mm1rsb}\end{eqnarray}

The expression for the messages from \textbf{b}-nodes to \textbf{y}-nodes
is: \begin{eqnarray*}
m_{\mu k}^{t} & = & \sum_{\left\{ \mathbf{b}_{k}\right\} }\, b_{k}^{{\rm a}^{\prime}}P^{t}\left(\mathbf{b}_{k}|\left\{ y_{\nu\neq\mu}\right\} \right)\\
 & = & \frac{{\displaystyle \sum_{\left\{ \mathbf{b}_{k}\right\} }}\, b_{k}^{{\rm a}^{\prime}}{\displaystyle \prod_{\nu\neq\mu}\sum_{\left\{ \mathbf{b}_{l\neq k}\right\} }P\left(y_{\nu}|\mathbf{B}\right)}{\displaystyle \prod_{l\neq k}P^{t-1}}\left(\mathbf{b}_{l}|\left\{ y_{\sigma\neq\nu}\right\} \right)}{{\displaystyle \sum_{\left\{ \mathbf{b}_{k}\right\} }}\,{\displaystyle \prod_{\nu\neq\mu}\sum_{\left\{ \mathbf{b}_{l\neq k}\right\} }P\left(y_{\nu}|\mathbf{B}\right)}{\displaystyle \prod_{l\neq k}P^{t-1}}\left(\mathbf{b}_{l}|\left\{ y_{\sigma\neq\nu}\right\} \right)}\,,\end{eqnarray*}
 which can be approximated by \begin{eqnarray*}
m_{\mu k}^{t} & \simeq & \frac{{\displaystyle \sum_{\left\{ \mathbf{b}_{k}\right\} }}\, b_{k}^{{\rm a}^{\prime}}\int\mathrm{d}\mathbf{\Delta}_{\nu k}P\left(y_{\nu}|\mathbf{{\Delta}}_{\nu k};\mathbold{\gamma}\right)P\left(\mathbf{{\Delta}}_{\nu k}|\mathbf{B}\right)\left[1+\varepsilon_{\nu k}\mathbf{b}_{k}^{\sf T}\nabla_{\mathbf{{\Delta}}_{\nu k}}\ln P\left(y_{\nu}|\mathbf{{\Delta}}_{\nu k};\mathbold{\gamma}\right)\right]}{{\displaystyle \sum_{\left\{ \mathbf{b}_{k}\right\} }}\,\int\mathrm{d}\mathbf{\Delta}_{\nu k}P\left(y_{\nu}|\mathbf{{\Delta}}_{\nu k};\mathbold{\gamma}\right)P\left(\mathbf{{\Delta}}_{\nu k}|\mathbf{B}\right)\left[1+\varepsilon_{\nu k}\mathbf{b}_{k}^{\sf T}\nabla_{\mathbf{{\Delta}}_{\nu k}}\ln P\left(y_{\nu}|\mathbf{{\Delta}}_{\nu k};\mathbold{\gamma}\right)\right]}\\
 & = & \frac{{\displaystyle \sum_{b_{k}^{{\rm a}^{\prime}}=\pm1}}\, b_{k}^{{\rm a}^{\prime}}\int\mathrm{d}\mathbf{\Delta}_{\nu k}P\left(y_{\nu}|\mathbf{{\Delta}}_{\nu k};\mathbold{\gamma}\right)P\left(\mathbf{{\Delta}}_{\nu k}|b_{k}^{{\rm a}^{\prime}}\right)\left[1+\varepsilon_{\nu k}b_{k}^{{\rm a}^{\prime}}{\displaystyle \frac{\partial}{\partial\Delta_{\mu k}^{{\rm a}^{\prime}}}}\ln P\left(y_{\nu}|\mathbf{{\Delta}}_{\nu k};\mathbold{\gamma}\right)\right]}{{\displaystyle \sum_{b_{k}^{{\rm a}^{\prime}}=\pm1}}\,\int\mathrm{d}\mathbf{\Delta}_{\nu k}P\left(y_{\nu}|\mathbf{{\Delta}}_{\nu k};\mathbold{\gamma}\right)P\left(\mathbf{{\Delta}}_{\nu k}|b_{k}^{{\rm a}^{\prime}}\right)\left[1+\varepsilon_{\nu k}b_{k}^{{\rm a}^{\prime}}{\displaystyle \frac{\partial}{\partial\Delta_{\mu k}^{{\rm a}^{\prime}}}}\ln P\left(y_{\nu}|\mathbf{{\Delta}}_{\nu k};\mathbold{\gamma}\right)\right]}\\
 & = & \frac{{\displaystyle \sum_{b_{k}^{{\rm a}}=\pm1}b_{k}^{{\rm a}}}{\displaystyle \prod_{\nu\neq\mu}}{\displaystyle \frac{1+\widehat{m}_{\nu k}^{t}b_{k}^{{\rm a}}}{\mathscr N_{\nu k}^{t}}}}{{\displaystyle \sum_{b_{k}^{{\rm a}}=\pm1}}{\displaystyle \prod_{\nu\neq\mu}}{\displaystyle \frac{1+\widehat{m}_{\nu k}^{t}b_{k}^{{\rm a}}}{\mathscr N_{\nu k}^{t}}}}=\frac{{\displaystyle \prod_{\nu\neq\mu}}{\displaystyle \frac{1+\widehat{m}_{\nu k}^{t}}{\mathscr N_{\nu k}^{t}}}-{\displaystyle \prod_{\nu\neq\mu}}{\displaystyle \frac{1-\widehat{m}_{\nu k}^{t}}{\mathscr N_{\nu k}^{t}}}}{{\displaystyle \prod_{\nu\neq\mu}}{\displaystyle \frac{1+\widehat{m}_{\nu k}^{t}}{\mathscr N_{\nu k}^{t}}}+{\displaystyle \prod_{\nu\neq\mu}}{\displaystyle \frac{1-\widehat{m}_{\nu k}^{t}}{\mathscr N_{\nu k}^{t}}}}=\tanh\left(\sum_{\nu\neq\mu}{\rm arctanh}\left(\widehat{m}_{\nu k}^{t}\right)\right)\,,\end{eqnarray*}
 but since $\widehat{m}_{\nu k}^{t}\sim\mathcal{O}\left(\varepsilon_{\nu k}\right)$
we have that \begin{equation}
m_{\mu k}^{t}\simeq\tanh\left(\sum_{\nu\neq\mu}\widehat{m}_{\nu k}^{t}\right)\,.\label{eq:mmhatapprox}\end{equation}

\section{The saddle point of $\mathcal{H}$\label{sec:roots}}

For the RS case the equation to be solved is: \begin{eqnarray*}
\frac{\partial}{\partial\vartheta}{}^{{\rm (RS)}}\mathcal{H}\left(\vartheta,y_{\mu}\right) & = & \frac{\vartheta-u_{\mu k}^{t}}{R_{\mu k}^{t}}-\frac{\partial}{\partial\vartheta}\ln\mathcal{G}\left(\vartheta,y_{\mu}\right)\\
 & = & \frac{\vartheta-u_{\mu k}^{t}}{R_{\mu k}^{t}}-\mathcal{P}\left(\vartheta,y_{\mu}\right)\,,\end{eqnarray*}
 thus, the equation to be satisfied is:\begin{eqnarray}
\tilde{\vartheta}_{\mu k}^{t} & = & u_{\mu k}^{t}+R^{t}\mathcal{P}\left(\tilde{\vartheta}_{\mu k}^{t},y_{\mu}\right)\,.\label{eq:wrs}\end{eqnarray}

For the 1RSB case we have that ${\displaystyle \frac{\partial}{\partial\vartheta^{0}}{}^{{\rm (1RSB)}}\mathcal{H}}={\displaystyle \frac{\partial}{\partial\vartheta^{\ell}}{}^{{\rm (1RSB)}}\mathcal{H}}=0\,,$
resulting in the set of equations:\begin{eqnarray*}
0 & = & \tilde{\vartheta}_{\mu k}^{0t}-\left(V^{t}-R^{t}\right)\mathcal{P}\left(\tilde{\vartheta}_{\mu k}^{0\ell t},y_{\mu}\right)\\
0 & = & \tilde{\vartheta}_{\mu k}^{\ell t}-V^{t}\mathcal{P}\left(\tilde{\vartheta}_{\mu k}^{0\ell t},y_{\mu}\right)\,,\end{eqnarray*}
 which is equivalent to:\begin{eqnarray}
\tilde{\vartheta}_{\mu k}^{0\ell t} & = & u_{\mu k}^{t}+\left(2V^{t}-R^{t}\right)\mathcal{P}\left(\tilde{\vartheta}_{\mu k}^{0\ell t},y_{\mu}\right)\,,\label{eq:w1rsb}\end{eqnarray}
 where $\vartheta_{\mu k}^{0\ell t}=\vartheta^{0}+\vartheta^{\ell}+u_{\mu k}^{t}$.
Observed that equation~(\ref{eq:w1rsb}) is equivalent to equation~(\ref{eq:wrs})
and that the ground state $\tilde{\vartheta}_{\mu k}^{t}$ is independent
of the indices 0 and $\ell$.

\section{The optimisation condition \label{app:optimisation}}

Our goal is to devise an algorithm that returns a better estimate
of the message at each iteration; we therefore apply a variational
approach that optimises the free parameters of the model at each iteration.
We expect to find a suitable set of parameters $\mathbold\gamma^{c}$
that maximises the drop in error per bit rate.

The error function has the form \begin{equation}
\mathscr E^{t}\left(\mathbold{\gamma}\right)\equiv\lambda^{2}P_{b}^{t}-M^{t}/\sqrt{N^{t}}\,,\label{eq:gg}\end{equation}
 where $\lambda^{2}$ is a positive constant.

Observe that \[
M^{t}-N^{t}=\frac{1}{\sqrt{2\pi F^{t}}}\int{\rm d}z\,\exp\left[-\frac{z^{2}+\left(E^{t}\right)^{2}}{2F^{t}}-\ln\cosh(z)\right]\,\tanh(z)\sinh\left(\frac{E^{t}-F^{t}}{F^{t}}z\right)\,,\]
 and that ${\rm sgn}\left[\tanh(z)\sinh\left({\displaystyle \frac{E^{t}-F^{t}}{F^{t}}}z\right)\right]={\rm sgn}\left(E^{t}-F^{t}\right)\;\forall z.$
Therefore ${\rm sgn}\left(E^{t}-F^{t}\right)={\rm sgn}\left(M^{t}-N^{t}\right)$.

The second term of the right hand side of equation~(\ref{eq:gg})
is an implicit function of the parameters $\mathbold{\gamma}$ through
$E^{t}$ and $F^{t}$, therefore \begin{eqnarray}
\cut{{\rm d}\left(\frac{M^{t}}{\sqrt{N^{t}}}\right)}\frac{\partial}{\partial\gamma_{i}}\left(\frac{M^{t}}{\sqrt{N^{t}}}\right) & = & \frac{\partial}{\partial E^{t}}\left(\frac{M^{t}}{\sqrt{N^{t}}}\right)\frac{\partial E^{t}}{\partial\gamma_{i}}+\frac{\partial}{\partial F^{t}}\left(\frac{M^{t}}{\sqrt{N^{t}}}\right)\frac{\partial F^{t}}{\partial\gamma_{i}}\,,\label{eq:gammadiff}\end{eqnarray}
 where the partial derivatives with respect to $E^{t}$ and $F^{t}$
are \begin{eqnarray*}
\frac{\partial}{\partial E^{t}}\left(\frac{M^{t}}{\sqrt{N^{t}}}\right) & = & \left(N^{t}\right)^{-\frac{3}{2}}\int\mathcal{D}z\left[1-\tanh^{2}\left(\sqrt{F^{t}}z+E^{t}\right)\right]\left[N^{t}-M^{t}\tanh\left(\sqrt{F^{t}}z+E^{t}\right)\right]\\
\frac{\partial}{\partial F^{t}}\left(\frac{M^{t}}{\sqrt{N^{t}}}\right) & = & \left(N^{t}\right)^{-\frac{3}{2}}\int\mathcal{D}z\,\frac{z}{2\sqrt{F^{t}}}\,\left[1-\tanh^{2}\left(\sqrt{F^{t}}z+E^{t}\right)\right]\left[N^{t}-M^{t}\tanh\left(\sqrt{F^{t}}z+E^{t}\right)\right]\,.\end{eqnarray*}
 By the definition of the field $b_{k}h_{\mu k}^{t}$ we have that
${\rm sgn}\left(b_{k}h_{\mu k}^{t}\right)={\rm sgn}\left(b_{k}m_{\mu k}^{t}\right)={\rm sgn}\left(b_{k}m_{k}^{t}\right)$.
Exploiting Gaussian properties of the distribution of $h_{\mu k}^{t}$
(\ref{eq:EF}) \begin{eqnarray*}
P_{b}^{t} & \simeq & \frac{1}{2K}\sum_{k=1}^{K}\left(1-{\rm sgn}\left(b_{k}h_{\mu k}^{t}\right)\right)\\
 & \sim & \int_{-\infty}^{\infty}\frac{{\rm d}u}{\sqrt{2\pi F^{t}}}\,\exp\left\{ -\frac{\left(u-E^{t}\right)^{2}}{2F^{t}}\right\} \,\frac{1}{2}\left(1-{\rm sgn}(u)\right)\\
 & = & \int_{-\infty}^{-E^{t}/\sqrt{F^{t}}}\mathcal{D}u\,,\end{eqnarray*}
 and we suppose that $E^{t}$ and $F^{t}$ are both explicit functions
of the parameters $\mathbold{\gamma}$, therefore \[
\frac{\partial P_{b}^{t}}{\partial\gamma_{i}}=-\frac{1}{\sqrt{2\pi}F^{t}}\exp\left[-\frac{\left(E^{t}\right)^{2}}{2F^{t}}\right]\left\{ \frac{\partial E^{t}}{\partial\gamma_{i}}-\frac{1}{2}\,\frac{E^{t}}{F^{t}}\,\frac{\partial F^{t}}{\partial\gamma_{i}}\right\} \,.\]
 By differentiation equation~(\ref{eq:gg}) and using equation~(\ref{eq:gammadiff})
one obtains {\small\begin{eqnarray}
\frac{\partial}{\partial\gamma_{i}}\mathscr E^{t} & = & -\frac{\lambda^{2}}{\sqrt{2\pi}F^{t}}\exp{\textstyle \left[-{\displaystyle \frac{\left(E^{t}\right)^{2}}{2F^{t}}}\right]}\left(\frac{\partial E^{t}}{\partial\gamma_{i}}-\frac{1}{2}\,\frac{E^{t}}{F^{t}}\,\frac{\partial F^{t}}{\partial\gamma_{i}}\right)\nonumber \\
 &  & -\left(N^{t}\right)^{-\frac{3}{2}}\int\mathcal{D}z\,\frac{N^{t}-M^{t}\tanh\left(\sqrt{F^{t}}z+E^{t}\right)}{\cosh^{2}\left(\sqrt{F^{t}}z+E^{t}\right)}\left(\frac{\partial E^{t}}{\partial\gamma_{i}}+\frac{z}{2\sqrt{F^{t}}}\frac{\partial F^{t}}{\partial\gamma_{i}}\right)\nonumber \\
 & = & -\left(F^{t}N^{t}\right)^{-\frac{3}{2}}\int\frac{{\rm d}u}{\sqrt{2\pi}}\exp{\textstyle \left[-{\displaystyle \frac{\left(u-E^{t}\right)^{2}}{2F^{t}}}\right]}\,\frac{u}{2}\,\frac{N^{t}-M^{t}\tanh\left(u\right)}{\cosh^{2}\left(u\right)}\nonumber \\
 &  & -\left(\frac{\partial E^{t}}{\partial\gamma_{i}}-\frac{1}{2}\,\frac{E^{t}}{F^{t}}\,\frac{\partial F^{t}}{\partial\gamma_{i}}\right)\nonumber \\
 &  & \;\times\left\{ \frac{\lambda^{2}}{\sqrt{2\pi}F^{t}}\exp{\textstyle \left[-\!{\displaystyle \frac{\left(E^{t}\right)^{2}}{2F^{t}}}\right]}+\int\!\!\frac{{\rm d}u}{\sqrt{2\pi F^{t}\left(N^{t}\right)^{3}}}\exp{\textstyle \left[-\!{\displaystyle \frac{\left(u-E^{t}\right)^{2}}{2F^{t}}}\right]}\,\frac{N^{t}\!-\! M^{t}\tanh\left(u\right)}{\cosh^{2}\left(u\right)}\right\} \,.\label{eq:casi}\end{eqnarray}}
 To optimise $\mathscr E^{t}$ with respect to $\gamma_{i}$ one requires
$\frac{\partial}{\partial\gamma_{i}}\mathscr E^{t}=0$. The first
term of the right hand side of equation~(\ref{eq:casi}) is independent
of the index \emph{i} and is zero if and only if the integrand is
an odd function. This is true if $\tanh(u)={\displaystyle \frac{N^{t}}{M^{t}}}\,\tanh\left({\displaystyle \frac{uE^{t}}{F^{t}}}\right)\;\forall u\in\mathbb{R}$.
This condition is only satisfied if $E^{t}\left(\mathbold\gamma^{c}\right)=F^{t}\left(\mathbold\gamma^{c}\right)$
which automatically makes $M^{t}=N^{t}$. By the application of this
condition, the sum between curly brackets in the second term at the
right hand side of Eq.(\ref{eq:casi}) is always positive, which implies
$\left.{\displaystyle \frac{\partial E^{t}}{\partial\gamma_{i}}-\frac{1}{2}\,\frac{E^{t}}{F^{t}}\,\frac{\partial F^{t}}{\partial\gamma_{i}}}\right|_{\gamma_{i}^{c}}=0$.

The conditions $E^{t}\left(\mathbold\gamma^{c}\right)=F^{t}\left(\mathbold\gamma^{c}\right)$
and $\left.{\displaystyle \frac{\partial E^{t}}{\partial\gamma_{i}}-\frac{1}{2}\,\frac{E^{t}}{F^{t}}\,\frac{\partial F^{t}}{\partial\gamma_{i}}}\right|_{\gamma_{i}^{c}}=0$
imply that: \begin{eqnarray*}
\ln E^{t} & = & e_{0}+\mathbf{e}_{1}^{\sf T}\left(\mathbold{\gamma}-\mathbold\gamma^{c}\right)+\frac{1}{2}\left(\mathbold{\gamma}-\mathbold\gamma^{c}\right)^{\sf T}\mathbf{E}_{2}\left(\mathbold{\gamma}-\mathbold\gamma^{c}\right)+\dots\\
\ln F^{t} & = & e_{0}+2\mathbf{e}_{1}^{\sf T}\left(\mathbold{\gamma}-\mathbold\gamma^{c}\right)+\frac{1}{2}\left(\mathbold{\gamma}-\mathbold\gamma^{c}\right)^{\sf T}\mathbf{F}_{2}\left(\mathbold{\gamma}-\mathbold\gamma^{c}\right)+\dots\,,\end{eqnarray*}
 therefore, if the critical point is a minimum, then the expansion
$E^{t}/\sqrt{F^{t}}=\exp\left\{ \frac{1}{2}e_{0}+\frac{1}{2}\left(\mathbold{\gamma}-\mathbold\gamma^{c}\right)^{\sf T}\left(\mathbf{E}_{2}-\frac{1}{2}\mathbf{F}_{2}\right)\left(\mathbold{\gamma}-\mathbold\gamma^{c}\right)+\dots\right\} $
has a second term that satisfy the conditions: $\det\left(\mathbf{E}_{2}-\frac{1}{2}\mathbf{F}_{2}\right)>0$
and $\left(\mathbf{E}_{2}-\frac{1}{2}\mathbf{F}_{2}\right)_{ii}<0$,
validating the optimisation process.

\begin{thebibliography}{10}
\bibitem{MPV}M. M\'{e}zard, G. Parisi and M.A. Virasoro, \emph{Spin
Glass Theory and Beyond}, World Scientific, Singapore (1987).

\bibitem{MFAbook}M.~Opper and D.~Saad, \emph{Advanced Mean Field
Methods: Theory and Practice}, MIT Press, Cambridge, MA 2001

\bibitem{Pearl}J.~Pearl, \emph{Probabilistic Reasoning in Intelligent
Systems}, Morgan Kaufmann Publishers, San Francisco, CA (1988).

\bibitem{Jensen}F.V.~Jensen, \emph{An Introduction to Bayesian Networks},
UCL Press, London (1996).

\bibitem{macKay}D.J.C.~MacKay, \emph{Information Theory, Inference
and Learning Algorithms}, Cambridge University Press (2003).

\bibitem{KabashimaCDMA}Y.~Kabashima, J.~Phys.~A \textbf{36}, 11111
(2003).

\bibitem{Nishimoribook}H. Nishimori, \emph{Statistical Physics of
Spin Glasses and Information Processing}, Oxford University Press,
UK, (2001).

\bibitem{neirottisaad}J.P.~Neirotti and D.~Saad, Europhys.~Lett.~\textbf{71},
866 (2005).

\bibitem{footnote} Although we will be using the terms RS and RSB,
it should be clear that this is not directly related to the replica
approach~\cite{MPV,Nishimoribook}, but merely uses similar structures
for the cross-replica correlations.

\bibitem{CDMAbook}S.~Verd\'{u}, \emph{Multiuser Detection}, Cambridge
University Press UK (1998).

\bibitem{Seung}H. S. Seung, H. Sompolinsky and N. Tishby, Phys. Rev.
A \textbf{45}, 6056 (1992).

\bibitem{weiss}Y.~Weiss, \emph{Neural Computation} \textbf{12},
1 (2000).

\bibitem{TAPEPL}Y.~Kabashima, D.~Saad, Europhys.~Lett.~\textbf{44},
668 (1998).

\bibitem{YFW}J.S.~Yedidia, W.T.~Freeman and Y.~Weiss, in \emph{Advances
in Neural Information Processing Systems} \textbf{13}, 698 (2000).

\bibitem{MPZ}M.~M\'{e}zard, G.~Parisi and R.~Zecchina, Science
\textbf{297}, 812 (2002).

\bibitem{MZPRE}M.~M\'{e}zard and R.~Zecchina, Phys. Rev. E \textbf{66},
056126 (2002).

\bibitem{BZ} A.~Braunstein and R.~Zecchina, Phys.~Rev.~Lett.,
\textbf{96} 030201 (2006)

\bibitem{Kabashimanew}Y.~Kabashima, Jour.~of the Physical Society
of Japan \textbf{74} 2133(2005) \cut{ 

\bibitem{EM}A.P.~Dempster, N. M.~Laird and D.B.~Rubin, J. Roy.
Stat. Soc. B \textbf{39}, 1 (1977).

\bibitem{neirottisaad_long}J.P.~Neirotti and D.~Saad, in preparation
(2005).

\bibitem{kk}Comparison to the recently presented signal detection
algorithm of Ref.~\cite{Kabashimanew} cannot be carried out due
to absence of numerical data.} 

\end{thebibliography}
\end{document}